\documentclass[12pt]{article}
\usepackage{psfig,epsfig}

\begin{document}

\title{Many--body Correlation
Effects in the Ultrafast Nonlinear Optical 
Response of Confined Fermi Seas} 

\author{I. E. Perakis \& T. V. Shahbazyan\\ Department of 
Physics \& Astronomy\\Vanderbilt University  \\
Nashville, TN 37235}

\maketitle

\begin{abstract}
The dynamics of electrons and atoms interacting with intense
and ultrashort optical pulses 
presents an important problem 
in physics 
that cuts across 
different materials such as semiconductors and  metals.
The currently available laser pulses, 
as short as 5 fs, 
provide a  time resolution 
shorter than the dephasing and 
relaxation times in many materials. 
This allows for a systematic study 
of many--body effects using nonlinear optical 
spectroscopy. 
In this review article, we
discuss the role 
of Coulomb correlations in 
the ultrafast dynamics 
of  modulation--doped quantum wells and metal
nanoparticles. 
We focus in particular on the manifestations of 
non--Markovian 
memory effects induced by strong 
electron--hole and electron--plasmon  correlations. 
\end{abstract} 
\newpage
\tableofcontents
\newpage
\section*{List of acronyms}
\addcontentsline{toc}{section}{\numberline{}List of acronyms}
\begin{tabular}{ll}
FES & Fermi Edge Singularity \\
FS & Fermi sea \\
{\em e--h} & Electron--hole \\
{\em e--e} & Electron--electron \\
{\em h--h} & Hole--hole \\
{\em e--p} & Electron--phonon \\
{\em h--p} & Hole--phonon \\
HFA & Hartree--Fock approximation  \\
SBE's & Semiconductor Bloch Equations \\
SP &  Surface Plasmon \\
FWM & Four Wave Mixing \\
PP & Pump--probe \\
RPA & Random Phase Approximation \\
QW & Quantum Well\\
CCE & Coupled Cluster Expansion \\
fs & Femtosecond \\ 
ps & Picosecond \\ 
MDQW & Modulation--doped Quantum Well \\
TI--FWM & Time--integrated Four Wave Mixing 
\end{tabular}

\section{Introduction and Scope of this Review}
\label{sec:introduction}

One of the most  exciting thrusts 
in Physics focusses on
understanding  the properties of systems 
possessing many  degrees of freedom. 
The goal is to relate such properties to the 
interactions among the elementary excitations. 
Such many--body problems have  been a driving force 
in many different subfields of Physics and Chemistry: 
condensed matter physics, nuclear physics, quantum chemistry, 
high energy physics, $\cdots$ 
In condensed matter 
physics, the study of many--body phenomena has led to 
many exciting discoveries: superconductivity, quantum Hall effect, 
Kondo effect, Mott transition, quantum phase transitions,
Fermi Edge and X--ray Edge  Singularities, $\cdots$. 
In most cases, such fascinating  phenomena 
have been observed in equilibrium  systems at low temperatures.
In comparison, the role of many--body effects 
in the {\em dynamics} of non--equilibrium 
many--electron systems presents  a 
less explored frontier.

Ultrafast nonlinear optical 
 spectroscopy provides
unique and powerful tools for studying
the dynamics 
of many--body correlations. 
In such experiments,
an  intense 
optical pulse,
the pump,
with duration shorter 
than  
the scattering 
time
or the period of the elementary excitations
photoexcites the system 
from its ground state. 
The subsequent dynamics is then monitored 
by using  a second pulse, 
the probe. 
By measuring the amplitude and the phase of the 
optical field emerging from the 
photoexcited sample,
one can gain valuable 
insight into the interplay between 
many--body 
and quantum confinement effects
during very short time scales. 
A good understanding of the role of such 
fundamental 
physics 
can lead to  new ideas regarding the next generation 
of photonic and opto--electronic devices.
These 
operate under intense excitation conditions
in the technologically important Tbit/sec    regime
and are often based on semiconductor heterostructures 
and  novel  low--dimensional structures.
Such considerations have spurred intense efforts 
in recent years aimed at understanding 
how to control the optical 
dynamics
during sub--picosecond time scales
(1 Tbit/sec = (1 ps)$^{-1}$). 
In the highly non--equilibrium femtosecond regime, 
the dominant effects on the 
nonlinear optical dynamics 
come from the Pauli exclusion 
principle (Phase Space Filling effects) 
and the Coulomb interactions among the 
optically--excited and Fermi sea carriers. 
The phonons also play an 
important role, especially  
during picosecond time scales. 

Following their photoexcitation, 
the electrons and holes
undergo  a number of different relaxation 
stages as they scatter among themselves 
via the Coulomb interaction and with the phonons 
via the electron--phonon ({\em e--p}) interaction. 
Below we briefly discuss each of these relaxation
stages. 
Initially (stage {\bf I}, the {\em coherent regime})
the photoexcited electrons and holes 
have a well--defined phase relationship: 
since the photon momentum is negligible, 
they are restricted to have 
opposite momenta. Such a coherence 
is  described by an optical  polarization.
In the very beginning,
the photoexcited 
carriers interact via the bare Coulomb 
interaction.
After time intervals of the order of the inverse 
plasmon frequency corresponding to the carrier density, 
the interactions become screened. The problem can then 
be approached in terms of 
interacting Fermi liquid quasi---particles 
\cite{pines}. 
Even though  in typical metals 
such screening occurs within an extremely short time interval
$\sim$ \ 1 fs, 
in semiconductors the inverse plasma frequency 
is  of the order of tens of femtoseconds
and therefore the above regime 
is accessible experimentally by using 
$\sim$ \ 5--10 fs optical  pulses.
If one tunes  the pump frequency above the onset of 
linear absorption, 
{\em real} carriers are photoexcited, in which case 
the duration of stage {\bf I}
is determined by the dephasing time, $T_{2}$. 
In metals, the latter is typically 
of the order of a few femtoseconds, 
while in semiconductors it can be as long as a few picoseconds.
The dephasing processes are suppressed 
if one  tunes the pump  frequency {\em below} the  onset  of 
linear absorption. 
Under such below--resonance excitation conditions, 
the coherent effects become dominant 
since  only 
{\em virtual} carriers are photoexcited 
and thus 
{\em  dissipative} scattering processes 
cannot occur. 
The time--dependence of the 
optical polarization
is then determined by the 
pulse duration, which 
leads to extremely fast nonlinearities.

Immediately after  the {\em e--h} phase coherence 
is lost,
incoherent populations of  
electrons and holes are formed. These are initially 
described  by 
{\em non--thermal}  density distributions. Subsequently, these
evolve into hot thermal distributions  characterized 
by an electron temperature that can far exceed 
that of the lattice.
Such a time evolution  (stage {\bf II})
as the carriers equilibrate 
among themselves 
is mainly induced 
by the {\em e--e}, {\em e--h}, {\em h--h},  
and {\em e--p}
interactions. 
The duration of this 
stage {\bf II} is determined
by the population relaxation time, $T_{1}$, which  can be 
of the order of  picoseconds. 
Subsequently, the {\em e--p}
 interaction 
leads to the equilibration of the electronic and lattice 
temperatures on a time scale of   tens of
picoseconds, followed  
by the return to thermodynamic equilibrium 
after hundreds of 
picoseconds
via the  recombination of the photoexcited carriers 
or via the transport of heat 
to the surroundings (stage {\bf III}).

A large body of 
experimental and theoretical work has focussed 
on different aspects of the above relaxation. 
Such developments in  undoped semiconductors 
 have been extensively  described 
in a number of excellent review  articles and books. 
Among the  
 earlier reviews of interaction 
effects in highly--excited undoped semiconductors, 
we note those of Haug and Schmitt--Rink \cite{haugssr}, 
Stahl and Baslev \cite{stahl}, 
Zimmermann \cite{zimrev}, 
Schmitt--Rink et. al. \cite{ssrev}, 
and Haug and Koch \cite{haugkoch}. 
More recent reviews focussing 
on the femtosecond  {\em coherent} 
dynamics in undoped semiconductors
include those 
by Mukamel \cite{mukbook}, 
Binder and Koch \cite{bind95}, 
Shah \cite{shah96}, 
Haug and Jauho \cite{haugbook}, 
Axt and Mukamel \cite{axt98}, 
and Chemla \cite{dan99}. 
We also note  the collections of 
papers  in Refs. \cite{haug88,shah92,henn93,phil94}.
More recently, 
there has been a lot of interest in using ultrafast 
nonlinear spectroscopy to study the dynamics 
of many--body effects in metals. 
The charge carrier dynamics and the 
coherent ultrafast phenomena in 
bulk metals, metal interfaces, and metal surfaces 
were recently reviewed by Petek and Ogawa \cite{petek}. 

The purpose of this review article is to provide a comprehensive 
picture of recent experimental and theoretical 
advances in understanding femtosecond many--body phenomena 
in {\em Fermi sea} systems.
As has been extensively discussed in 
the review articles and books 
listed above, even though a  description of the ultrafast dynamics 
based on  few--level 
models \cite{yajima,eber,shen} has been very 
successful in atomic and molecular systems, 
it faces serious shortcomings when applied to 
semiconductors and metals. 
The development of microscopic 
non--equilibrium many--body theories has been found 
to be necessary for describing the femtosecond dynamics
and has made this field an exciting  area of research 
in condensed matter physics. 
In view of the large body of work, especially in undoped 
semiconductors, 
we will concentrate here on specific aspects 
of the many--body physics that have not 
been extensively reviewed  elsewhere. 
In particular, we will not touch on the impressive
advances in undoped semiconductors
and will  attempt to provide a unified picture of the ultrafast dynamics 
in different confined Fermi sea systems. 
We will mainly focus on coherent effects 
due to the Coulomb interaction and will only briefly 
discuss the electron--phonon ({\em e--p}) interaction effects. 
The main body of this article  
discusses the 
recent breakthroughs 
in understanding the 
sub--picosecond 
dynamics in modulation--doped quantum wells (MDQW's) and 
metal nanoparticles, 
with particular  emphasis  
on non--Markovian memory 
effects induced by carrier--carrier 
many--body  correlations. Such dynamical effects 
are beyond the scope of the dephasing and relaxation time approximations. 
One should note here that a detailed quantitative 
description of the experimental results presents 
a formidable task in the subpicosecond regime, which 
requires sophisticated numerical simulations 
as well as different approximations. This can make it 
difficult to discuss 
in a simple and intuitive way the important many--body 
physics conveyed by the experiment.
Here we will focus on the simplest 
possible models that can capture the most important 
 physics associated with the 
non--equilibrium Coulomb correlations
and  attempt to provide, to the degree possible, 
an intuitive explanation of the often  
delicate experimental features.

Let us now briefly discuss 
how many--body correlations enter into the ultrafast nonlinear 
optical response. 
The most commonly used theoretical 
approach 
starts with the Heisenberg equations of motion 
of the one--particle density matrix, 
$\rho_{ij}(t)$, where the indices $i,j=1,2 $ label the conduction 
and  valence band
\cite{mukbook}. 
The off--diagonal density matrix element, $\rho_{12}(t)=P(t)$,
is  the  optical polarization, 
while the diagonal matrix 
elements, $\rho_{ii}(t)$, correspond to 
the electron and hole 
occupation numbers (density distributions). 
Such equations of motion  must be  solved 
to obtain the polarization, which in turn acts as the source term 
in the Maxwell equations that determine the 
signal measured in the experiment \cite{mukbook}. 
The difficulty in treating the many--body interactions
comes from the coupling 
of  the one--particle density matrix 
to many--particle  correlation functions
(higher density matrices). As is typical for 
many--body systems, this  
leads  to an infinite hierarchy of
 coupled equations
which  cannot be solved 
exactly 
\cite{mukbook,axt98}.
One must therefore devise a controlled 
truncation scheme. 
One possibility is to adopt 
the time--dependent  Hartree--Fock approximation (HFA), which 
relies on the factorization of the many--particle 
correlation functions and only includes the two--particle 
interaction effects.
Such an approach 
developed 
into the very successful
Semiconductor Bloch Equations (SBE's)
formalism (see
e.g. Refs. \cite{stef86,stef88,lind88,basl89,bind91,haugkoch}). 
However, many recent experimental
observations, reviewed e.g. in Ref. \cite{dan99},
could not be explained within
such a 
mean field approach and were attributed to four--particle 
and higher order Coulomb correlations.
In order to develop 
a  many--body theory that can capture 
such effects,
it was noted by several different groups that, in undoped 
semiconductors, 
the above infinite density matrix hierarchy 
truncates if one adopts an expansion 
in terms of  the optical fields
\cite{axt98,axt94,ost95,sham98,che98,mukbook,bind95}. 
This is 
due to the fact that, in the undoped semiconductor ground state, 
the conduction band is empty while the valence band is full.
Such a  method for treating
the Coulomb correlations nonperturbatively 
is often referred to as 
the ``dynamic controlled truncation scheme''.
An alternative many--body approach
has been  based on diagrammatic expansions 
for Keldysh Green's functions,
whose multiple--time dependence is typically eliminated by using 
the Kadanoff--Baym ansatz \cite{haugbook}.

Let us now discuss how non--Markovian memory 
effects enter 
into  the density matrix equations of motion. 
Due to the many degrees of freedom that affect 
the dynamics, one usually distinguishes between 
a subsystem 
interacting  directly with the optical fields
(e.g. the photoexcited {\em e--h} pair) 
 and a reservoir/bath consisting 
of all the other degrees of freedom 
 (e.g. the phonons or the Fermi sea excitations).  
The polarization equation of motion can be  cast in the form 
\begin{equation} 
\frac{ \partial}{\partial t}  P(t) \ 
= \ \frac{ \partial}{\partial t} \left.  P(t) \right|_{\rm{coh}} 
 \ + \   \frac{ \partial}{\partial t} \left.  P(t) \right|_{\rm{scatt}} 
\label{poleom}
\end{equation} 
where the first term describes the Hartree--Fock  interactions 
among the coherent {\em e--h} pairs, 
while the second term describes the many--body scattering 
processes among the carriers as well 
as between the carriers and the bath excitations. 
The latter term can be expressed in the general form 
\begin{equation} 
\frac{ \partial}{\partial t}  \left. P(t) \right|_{\rm{scatt}} 
= \int_{-\infty}^{t} dt' \Gamma(t,t') P(t') \label{mem}
\end{equation} 
where 
$ \Gamma(t,t')$ is a memory kernel (similar equations 
can also be written for the carrier distribution functions). 
The importance of the many--body correlations 
of interest here 
is that they determine the dependence 
of  $ \Gamma(t,t')$ on the two times $t$ and $t'$
and therefore govern the memory effects. 
In systems 
 where 
only times $t' \sim t$ contribute to the above integral,
one  can apply the Markovian approximation, in which case 
\begin{equation} 
\frac{ \partial}{\partial t}  \left. P(t) \right|_{\rm{scatt}} 
\sim \Gamma(t) P(t).
\end{equation} 
In many cases, one can further approximate
$\Gamma(t)$ by a time--independent constant,
the dephasing energy width $\sim 1/T_{2}$.  
For example, such an approximation 
allows one to include, in a semi--phenomenological way, 
the effect of the scattering processes 
in the  SBE's. 
However, when 
such a Markovian approximation breaks down,  
the many--body correlations can affect the 
qualitative ultrafast dynamics 
by inducing strong memory effects and non--exponential 
polarization decay. 
Such 
effects in undoped semiconductors 
are known to arise e.g. from 
exciton--exciton
correlations or from 
 {\em e--p} and carrier--carrier 
interactions in the quantum kinetic regime  
and have been extensively reviewed e.g. 
in Refs. \cite{dan99,haugbook}. 
Here we  focuss on analogous 
effects whose origin lies 
in  the  non--perturbative Coulomb correlations  
between the photoexcited 
carriers and the cold electron Fermi sea
in MDQW's  and small metal nanoparticles.

This review is organized as follows.
First we present an  overview of the experimental 
results on the ultrafast nonlinear optical 
dynamics  in 
MDQW's (Section \ref{sec:MDQW}) and metal 
nanoparticles (Section \ref{sec:nano}). 
 In Section \ref{sec:formalism} we
outline the formalism necessary for 
describing the  {\em coherent} ultrafast 
nonlinear  optical response
of Fermi sea systems. 
In Section \ref{sec:FES}, the ultrafast nonlinear
dynamics of the Fermi Edge Singularity  (FES) 
in the coherent regime 
is discussed. In Section \ref{sec:decay}
the size-dependent dynamical screening of the 
Coulomb interaction in metal nanoparticles is discussed
and the results for the quasiparticle scattering 
rates are reviewed. In Section
\ref{sec:optics} these results are used to 
describe the ultrafast surface plasmon (SP) 
dynamics in small noble-metal nanoparticles. Section
\ref{sec:conclusion} concludes the  review.

\section{Nonlinear Optical Dynamics in 
Modulation--doped Quantum Wells} 
\label{sec:MDQW}

By separating the ionized impurity donors from the electrons, 
modulation doping can produce a high mobility two--dimensional 
electron gas.
In contrast to the case of a 
photoexcited electron gas, 
the temperature of such a system can be lowered, 
which enables 
the observation of fascinating 
many--body phenomena such as
the Quantum Hall Effect
\cite{qhe1,qhe2,qhe3,qhe4} 
and the Fermi Edge Singularity \cite{ssrev}. 

Recently there has been a lot of interest in studying 
the ultrafast dynamics in MDQW's. 
The role of the interactions 
between the photoexcited and the Fermi sea (FS) carriers 
during the  thermalization  stage {\bf II} 
has been reviewed e.g. in 
Refs. \cite{shah96,shah92,haugbook}
and is only briefly discussed here. 
In the section \ref{sec:SPUD} we will review 
corresponding measurements in metal films and nanoparticles 
and compare with MDQW's.
Much less is known about the role
of many--body effects in the {\em coherent} regime 
(stage {\bf I}). 
Even though 
some
aspects of the effects of the {\em e--e}
scattering  have been 
described within the dephasing time
approximation,  the role
 of the {\em e--h} correlations between the
photoexcited holes and the FS electrons
has only recently been studied. 
The main difficulty in treating such {\em e--h} 
correlation effects 
comes from the non--perturbative nature of the
Fermi edge singularity (FES), which dominates the
absorption spectrum close to the absorption onset,
and which cannot be described within the dephasing 
time approximation.
 Important here
is that the {\em e--h } interactions between a  heavy 
photoexcited hole and the FS-electrons lead to the
scattering of a macroscopic number of low--energy
FS-pairs, which readjusts  the entire FS into
a new orthogonal scattering  state
in the course of the optical excitation 
(Anderson orthogonality catastrophe \cite{ander67}).
Before 
we discuss 
the ultrafast dynamics 
due to such effects, 
we  briefly summarize in the next section the linear 
optical  properties of MDQW's.

\subsection{Fermi Edge Singularity in Linear Absorption} 
\label{sec:FESLA}

Close to the onset of linear absorption and for 
low--temperatures, the  optical properties 
of MDQW's are dominated by the FES. 
The  latter is a   many--body resonance that has been
observed in doped semiconductors
(see e.g 
\cite{fes1,fes2,fes3,fes4,fes5,ssrev}) 
as well as in 
metals \cite{mah88,ohtrev}, where 
it is referred to as the X--ray Edge Singularity. 
Despite the screening of the Coulomb interaction, 
the strength of this resonance is comparable to that 
of the undoped QW excitons. 
The  non--Lorentzian lineshape of the FES can be
viewed as originating from the decay of an excitonic bound state
caused by its interactions with the gapless FS 
excitations \cite{andrei}. 
In the case of  holes localized due to e.g. the disorder  \cite{fes1}
or for core holes in metals,
 the lineshape 
of the FES close to the onset of absorption 
can be approximately described by using the analytic 
power law expression \cite{andrei,mahbook,noz69,ssrev} 
\begin{equation}  
\label{feslin} 
\alpha(\omega) \propto  {\cal N}
\left( \frac{E_{F}}{\omega}\right)^{\beta}, 
\end{equation}
where ${\cal N}$ is  the density of states, 
$E_{F}$ is the Fermi energy, 
$\omega$ 
is the optical frequency measured  from the Fermi
edge, and $\beta=2 \delta/\pi- \left(\delta/\pi \right)^{2}$
is the FES exponent, where $\delta \sim  \tan^{-1}(\pi g)$ 
is the s--wave phaseshift of the screened {\em e--h}
potential $V$ 
evaluated at $E_F$, and $g= V {\cal N}$
is the dimensionless parameter characterizing the {\em e--h} 
scattering  
strength.
Such a non--Lorentzian 
lineshape results from the competition between the 
Mahan singularity,
 due to the attractive
interaction between the FS and the localized hole
(vertex correction effect,  \cite{mah67a,mah67b})
and the Anderson orthogonality
catastrophe due to the readjustment 
of the FS density profile during the optical 
transition (hole self--energy effect, \cite{ander67}). 
In the case of finite hole mass $m_{h}$, 
the FES is broadened by an additional energy width 
of the order of the hole recoil energy 
$\sim m_{e}/m_{h} E_{F}$, 
where $m_{e}$ is the electron mass \cite{andrei,gavor,sham90}.

In a first approximation, 
the excitonic effects in a MDQW  can
be described by extending the mean field approach (HFA)
to include  the effects of the screening and the Pauli
blocking due to the FS \cite{ssrev,andrei,stef83}. 
 However,
such a static treatment of the FS leads to 
a spurious bound state \cite{ssrev,andrei} 
with respect to the Fermi energy,
$E_{F}$, referred to in the following as 
the HFA bound state \cite{mah67a}.
This  discrete state, with binding energy $E_M$,
would appear at the energy $E_{F} - E_M$. Obviously,
for $E_M < E_{F}$ (as in typical MDQW's), 
such a state cannot exist since it overlaps with the FS continuum,
with which it interacts via the {\em e--h} potential
\cite{ssrev,andrei}.
The ``unbinding'' of this HFA bound state occurs via its
interactions  with the FS excitations,
 which are  {\em not} taken into account in the HFA 
 \cite{andrei}.
Note that, in two--dimensional systems, a static FS
allows for bound states even for arbitrarily small
attractive interactions and therefore such an unbinding
cannot arise from static screening or from Pauli blocking effects. 
This spurious bound state could be artificially merged
with the continuum by introducing a dephasing time
comparable to its binding energy.
Such an approximation,
however, neglects competely the 
dynamical correlations between
the photoexcited {\em e--h} pair and the FS excitations.
A microscopic description of the unbinding of the HFA bound state
presents a nontrivial problem due to the
non--perturbative nature of the {\em e--h} correlations
between the photoexcited hole and the FS excitations \cite{andrei}.
Even in linear absorption, 
within Green function techniques, this problem is rather
involved because vertex correction diagrams with arbitrarily
many crossed {\em e--h} interaction lines are as
divergent as the ladder diagrams and should be
treated on equal footing \cite{mahbook,mah67b}. 
 To perform
such a task, one must sum up at least the parquet
diagrams and  address the three--body correlations
between the photoexcited hole and a FS
excitation \cite{andrei,gavor}. 
Therefore, alternative methods
were developed for the case of linear absorption,
based on Fermi's golden rule with many--electron
eigenstates expressed in terms of Slater determinants
\cite{ohtrev}.
Such approaches become exact in the limit 
$m_{e}/m_{h} \ll 1$ and describe
quite accurately the FES lineshapes observed in
typical MDQW's 
\cite{ohtrev,oht89,sham90,hawr91}. 
Another approach to the FES problem is based on the
coupled cluster expansion (CCE) \cite{ccm1,ccm2,ccm3,schon}. This
general many--body technique provided exact results
in the limit
$m_{e}/m_{h} \ll 1$  \cite{ilias91a,ilias91b}
and was used to treat the hole recoil correlations in one
dimension \cite{ilias93}. In the latter case, an exact
solution was obtained for $m_{e}=m_{h}$. The CCE
has also been used to describe the
{\em e--e} correlation effects \cite{corr1,corr2}. More
importantly, however, this method is well--suited for
describing correlations in non--equilibrium systems,
where it retains the advantages of diagrammatic
expansions without resorting to the Kadanoff-Baym
ansatz or to the Markovian approximation \cite{neg}.

\subsection{Ultrafast Dynamics} 
\label{sec:UDMD}

In this section 
we summarize some recent studies of 
the dephasing and relaxation processes in
MDQW's. 
The earlier work focussed on the role of the 
interactions between the FS and the photoexcited 
carriers on the thermalization during 
relaxation stage  {\bf II}. 
In most cases, a non--thermal distribution of 
real electron and hole carriers was 
photoexcited well above the Fermi surface 
and the subsequent incoherent population dynamics was monitored 
with ultrafast pump--probe spectroscopy. 
Note that, in such experiments, 
the measured differential transmission is 
\begin{equation} \label{diff-T}
DST (\omega, \tau) =
\frac{\Delta T_s(\omega, \tau)} {T_0 (\omega)}
\, = \, \frac{ T_{s}({\cal E}_p)
-T_{s}({\cal E}_p=0)}
{T_{s}({\cal E}_p=0)} \, ,  
\end{equation}
where $T_{s}({\cal E}_p)$ is the transmission coefficient in the
probe direction in the presence of the pump field ${\cal E}_p$
and $\tau$ denotes the  time delay between the 
pump and probe optical pulses.
For samples with sufficiently small thickness $d$
 such that $\Delta \alpha \ d$ is small, 
the differential transmission 
reproduces the  change,
$\Delta \alpha(\omega ,\tau)$,
in the probe absorption coefficient
$\alpha (\omega , \tau)$ which is 
induced by the pump photoexcitation: 
$DST(\omega,\tau) \sim -\Delta \alpha(\omega ,\tau) d$.

For a simple intuitive interpretation 
of  the thermalization 
experiments in MDQW's, it is often  assumed that, 
for free carriers photoexcited well 
above the onset of absorption and for 
time scales much longer than the dephasing time, 
the differential absorption spectrum maps 
the carrier distribution functions  
at the probe photon energy
at  time $\tau$.
In particular, the following 
approximation is often  used: 
\begin{equation} 
\label{da} 
\Delta \alpha(\omega ,\tau) = ( 1 - f_{e} - f_{h} )  \
\alpha(\omega) 
\end{equation} 
where $\alpha(\omega)$ is the linear absorption coefficient
and $f_{e}$ and $f_{h}$ are the electron and 
hole distribution functions at the corresponding energies.  
Within such an approximation, the time evolution of the 
differential absorption is determined by the 
carrier distribution functions at time $\tau$.
There have been many 
theoretical calculations of the distribution 
function time evolution, 
mostly  based on numerical solutions of  the semiclassical 
Boltzmann equations (see e.g. Refs. \cite{haugbook,shah92,shah96}).
A detailed review of such an 
approach to thermalization 
can be found in  Ref. \cite{haugbook}. 
More recently, the dynamics in the 
quantum kinetic regime and the shortcomings of the Boltzmann 
equations have  been addressed, 
as reviewed e.g. in Ref. \cite{haugbook}.

Knox {\em et. al.} \cite{knox88}
studied the effect of the inelastic {\em e--e} scattering 
on the thermalization 
by photoexciting a non--thermal 
carrier distribution 
well above the Fermi surface 
and then monitoring its time evolution
to a thermal distribution at the bottom of the band. 
They observed that the {\em e--e} interactions
between the photoexcited and FS electrons 
significantly enhance the thermalization rates as compared 
to  undoped semiconductors. In particular, 
the thermalization time in MDQW's 
was found to be 
of the order of 
$\sim$10fs. 
The differential transmission lineshape 
mimicked  that of a Boltzmann distribution 
peaked at the bottom of the conduction band, 
which indicates a non--degenerate carrier distribution. 
This is expected  at room temperatures (as in the experiment
of Ref. \cite{knox88}), 
where the thermal energy exceeds 
the Fermi energy, and at high intensities, 
where the photoexcited carrier density
exceeds the FS density.
Knox {\em et. al.} also observed  an {\em instantaneous}
redshift due to the bandgap 
renormalization by the photoexcited carriers \cite{ssrev}. 
However, as we discuss later, the situation changes 
drastically at low temperatures, where the FS electrons 
have a sharp Fermi--Dirac distribution. 
Wang {\em et al.}
\cite{wang95} 
performed low temperature pump--probe measurements
and  found that, at low excitation densities, 
the differential transmission lineshape corresponded to a 
redshift of the FES due to 
the incoherent  bandgap renormalization.
 Importantly, they showed that 
such a redshift slowly builds up
with time as the hot electron temperature 
decreases due to the 
cooling to the lattice temperature.
This observation 
points out 
that the bandgap renormalization depends not only on the carrier 
density but also on the carrier distribution.
At high photoexcited densities, 
the heating of the electron gas becomes stronger 
and Wang {\em et. al.} observed a bleaching 
of the FES due to the smearing of the sharp Fermi surface. 

Both the above experiments were interpreted in terms 
of {\em e--e} interactions between the photoexcited and the 
FS carriers. 
Tomita {\em et. al.}
\cite{tomita93} 
performed 
ultrafast luminescence 
measurements at low temperatures to 
investigate the role of the {\em e--h} interactions 
on the thermalization in n--type MDQW's.
In particular, they 
 measured the time evolution of the luminescence 
intensity at the bandgap energy, where the 
optically--induced 
perturbation of the electron distribution
due to, e.g., heating 
 is minimal. 
Tomita {\em et. al.}
found that the number of holes that 
thermalized to the top of the valence band
as a function of time 
could not accounted for if one 
assumes a  thermalized hole distribution. 
This  result indicates that 
a substantial fraction of the 
photoexcited holes are non--thermal 
for time intervals as long as $\sim$ 800fs.
Furthermore, Monte--Carlo simulations \cite{tomita93} 
indicated that
the {\em e--h} scattering is the dominant hole thermalization 
mechanism.  
Similar measurements in n--doped bulk GaAs 
were performed by Chebira {\em et. al.} (at low temperatures) 
\cite{cheb92} and Zhou {\em et. al.} (at room temperature) 
\cite{zhou92}.  
Woerner {\em et. al.} \cite{woer94a,woer94b} studied the 
hole relaxation 
in p--doped semiconductors by 
photoexciting holes from the heavy--hole to
 the split--off valence band 
and then monitoring the time evolution 
of the inter--valence--band absorption spectrum
by using femtosecond pulses in the mid--infrared region. 
Due to the absence of conduction electrons, 
the dynamics is then solely determined  by the 
{\em h--h} and {\em h--p} 
scattering. 
They found that the inter--valence--band 
scattering via the emission of e.g. 
optical phonons  occurs within a short time interval 
$<$100fs, and that subsequently 
the phototexcited holes slowly thermalize via  {\em h--h} scattering
with the hole FS  within a time interval $\sim$ 700fs.

An important effect of the quantum confinement 
in QW's is the formation of discrete conduction and valence
subbands. 
 Recently, there has been much 
interest in exploring the role of 
the intersubband   excitations in the 
ultrafast dynamics, partly motivated by 
device applications such as the quantum cascade laser
\cite{faist}. 
Lutgen {\em et. al.} \cite{lutg96a,lutg96b} 
performed femtosecond pump--probe measurements and showed that the
intersubband nonlinear absorption is determined by both the 
intersubband and the intraband electron relaxation.
They measured the 
effects of {\em intersubband} pump excitation 
on the nonlinear absorption due to probe--induced
{\em interband} 
transitions from the valence to the conduction 
band. 
They observed a population 
relaxation that proceeds in two stages. 
First, the electrons photoexcited by the pump 
from the Fermi sea in the lowest conduction subband 
relax back  
via intersubband scattering 
within a time interval $\sim$ 1.3 ps. 
The latter is consistent with the 
intersubband relaxation times due to longitudinal 
optical  phonon emission \cite{phonon1,phonon2}. 
This process is followed by intraband {\em e--e} relaxation 
similar to the experiments discussed above. 
Lutgen {\em et. al.} 
also measured the time evolution of the 
nonlinear absorption spectrum
due to optical transitions between the conduction subbands. 
They observed  an intial decay of the differential transmission 
within a time interval $\sim$ 2--3 ps, which they 
attributed to the population relaxation 
due to the intersubband scattering.
This was followed by a slowly  decaying differential transmission signal,
whose time evolution depended strongly on
the photoexcitation  frequency (within the linear absorption linewidth). 
The latter regime was  attributed to the intraband 
population relaxation within the lowest subband
as the carriers cool to the lattice temperature.

We now come to the 
effect of a cold FS 
on the {\em e--e} scattering time. 
One would 
expect that the latter 
should become short as the FS density increases. 
This is indeed the case for  low FS densities,
high temperatures, or for large excess energies 
from the Fermi surface. 
Hawrylak {\em et. al.} \cite{hawr94}
obtained  the 
energy--dependent {\em e--e} 
scattering  times by calculating the 
equilibrium self--energies 
including  the effects of the electron--plasmon
and {\em e--p} interactions. They predicted step--like decreases
in the scattering time for  electron energy 
above the Fermi surface that exceeds 
the onset of plasmon and optical phonon emission.
They also concluded 
that short--range {\em e--e} correlations (not included in 
the RPA)  can significantly affect the 
scattering times at high energies. 
Such fast {\em e--e} scattering of 
photoexcited electrons well above the Fermi surface 
can explain for example the absence  of spectral 
hole burning in the pump--probe spectra 
of MDQW's  \cite{knox88,shah96}. 
However, the situation changes drastically 
for electron energies close to the Fermi surface
and temperatures smaller than the Fermi 
energy.
Under such conditions, the {\em e--e} scattering 
is in fact {\em suppressed} in the presence of a degenerate FS. 
Indeed, a photoexcited 
electron lowers its energy 
by interacting  with a FS electron while the
latter scatters above the 
Fermi surface. 
For electrons close to the Fermi surface, 
the drastic reduction 
of the phase space 
available for scattering 
due to the Pauli blocking 
by the FS electrons as well as the screening 
leads to the  strong suppression 
of the {\em e--e} scattering \cite{pines}. 
This is due to the sharp Fermi--Dirac 
distribution of the FS electrons 
at  low temperatures. 
In fact, at zero temperatures, the 
{\em e--e} scattering time $\tau_e(E)$ 
becomes infinite right at the Fermi surface, 
$E=E_{F}$ \cite{pines,hawr94}. 
According to Fermi liquid theory \cite{pines}, 
\begin{equation} 
\label{taue}
\tau_e(E)\propto \frac{E_{F}^{2}}{(E-E_F)^2 + (\pi k_{B} T)^{2}},
\end{equation}
where  $k_{B} T$ is the thermal energy.
Note that 
the above  result 
applies both to metals and  MDQW's 
due to the similar values of  $r_{s}$.
As the electron temperature increases, 
the smearing of the Fermi--Dirac distribution 
allows for the scattering of a photoexcited 
electron to states below $E_{F}$ 
and therefore $\tau_e(E)$ decreases as compared to the 
zero temperature limit. 
Even though 
the {\em e--e} scattering rate initially 
increases with electron density  for low densities,
it decreases at higher densities 
when the FS temperature is low 
and the Fermi edge is sharp \cite{levi}.
Therefore, one should 
expect delayed dephasing and thermalization
processes 
as the electrons approach the Fermi surface
(note  that the above 
phase space restriction does no apply to the 
{\em e--h} scattering of the photoexcited hole 
with the FS electrons).

The effect of a cold FS on the 
dephasing time of the interband optical polarization 
was investigated by 
Kim {\em et al.} \cite{kim92}
using  transient FWM  spectroscopy \cite{shah96}.
In such experiments, two optical 
pulses separated by a time delay 
$\tau$  propagate along two different directions,
${\bf k_{1}}$ and ${\bf k_{2}}$, and 
interfere within the sample,
thus generating a nonlinear polarization. 
In FWM spectroscopy, 
one measures the signal that emerges along the 
direction $2 {\bf k_{2}} - {\bf k_{1}}$. 
In the case of optically--thin samples, the 
time--resolved FWM signal corresponds to the square of the 
amplitude of the nonlinear polarization along the above direction
\cite{shah96}. 
The time--integrated FWM signal (TI--FWM)
is given by the integral over all times 
of the time--resolved FWM signal and is a function of the time delay 
$\tau$. 
To a first approximation, the FWM signal can be 
interpreted similar to atomic and molecular 
systems based on a two--level system \cite{yajima,shah96}. 
Within such a model, in the case of homogeneous broadening, 
the time--resolved FWM signal 
 is emitted immediately after the second pulse 
arrives and decays with a decay time 
$T_{2}/2$, where $T_{2}$ is the dephasing time. 
The TI--FWM signal also decays with a 
decay time 
$T_{2}/2$ and 
vanishes for negative time delays 
where pulse 1 comes after pulse 2. 
In the case of inhomogeneous broadening, the time--resolved 
FWM signal 
is delayed by a time interval $\tau$ after the second pulse arrives
and corresponds to a photon echo \cite{yajima,shah96,eber}. 
The TI--FWM signal decays with a decay time $T_{2}/4$, 
which  is a factor of 2 smaller than in the case of 
homogeneous broadening, and also vanishes for negative 
time delays. 
In undoped semiconductors, significant deviations 
from the above simple picture have been observed 
and have been attributed to Hartree--Fock 
exciton--exciton interactions 
and to exciton--exciton correlations \cite{dan99}. 
For example, such effects lead to a strong 
TI--FWM signal for negative time
delays as well as a delayed time--resolved FWM signal. 
For a detailed review of the 
exciton--exciton interaction  effects in undoped semiconductors, 
see e.g. Ref. \cite{dan99}. 

In MDQW's, Kim {\em et. al.} 
measured long dephasing times 
of a few picoseconds for energies within 
the frequency
range of the
FES (i.e. close to $E_{F}$). 
These decreased to less than 100fs 
as the excitation  frequency exceeded the FES peak 
by an energy comparable to the Fermi energy.
The above  result is consistent with the Fermi liquid 
energy dependence of the {\em e--e} scattering time, 
Eq. \ \ref{taue},
and points out 
that, unlike for the {\em e--h} correlations, 
in MDQW's the {\em e--e} scattering is suppressed within the
frequency range of the FES. 
This 
experiment also suggests that the hole dephasing times
are of the order of a few picoseconds. 
In the time--resolved FWM signal,
Kim {\em et. al.}   observed photon--echo--like behavior,  
which indicates that, 
despite the strong  FES peak in the linear absorption, 
the continuum of interband {\em e--h} states 
leads to behavior  similar to that of 
an inhomogeneously--broadened system.  
Finally, 
in MDQW's, the pair--pair interactions,
analogous to the exciton--exciton interactions
in undoped semiconductors, are screened out and the
negative time delay TI--FWM 
signal characteristic of exciton--exciton interactions 
was absent for high Fermi sea densities
\cite{kim92}. 
The  above experimental results were interpreted 
by Hawrylak {\em et. al.} \cite{hawr94}
within the two--level system approximation (which
neglects all FES  correlation effects) by 
introducing 
energy--dependent dephasing times
determined by the {\em e--e} and {\em e--p} scattering times. 

In order to provide further insight into the 
dephasing processes in MDQW's, 
Bar--Ad {\em et. al.}  \cite{barad94} performed  low temperature 
FWM measurements in the presence of a magnetic 
field parallel to the QW growth axis. 
For zero magnetic field, they observed a dephasing time 
$\sim$ 1ps. 
For magnetic fields
such that only the lowest Landau level is occupied, 
they observed  a suppression of  the {\em e--e} scattering. 
In particular, the 
dephasing times were then found to be longer, 
$\sim$1.5 ps, and,  unlike 
in the zero field case of Ref. \cite{kim92}, 
they did not depend significantly on the
frequency of the optical excitation. Bar Ad {\em et. al.} 
attributed the above observations to the suppression 
of the 
phase space available for scattering 
induced by the magnetic confinement. 
Indeed, the application of a magnetic field 
changes the continuous density of states into a 
series of discrete highly degenerate Landau level peaks, 
broadened by the  disorder. 
Finally, Bar--Ad {\em et. al.} 
observed 
a FES in the FWM spectrum at high magnetic fields, 
which they attributed 
to the  optical  transitions 
to the occupied lowest Landau level.

The dephasing  of the  intersubband optical  
polarization in MDQW's
was recently investigated by Kaindl {\em et. al.} 
\cite{kain98} using transient FWM in the midinfrared. 
They observed an exponential decay of the TI--FWM signal 
that was fairly
insensitive to the temperature. 
By modelling their data with a HFA calculation,
Kaindl {\em at. al.} 
deduced the values of the 
 dephasing times to be of the order of hundreds of femtoseconds
 and 
 concluded that the main dephasing mechanism is the {\em e--e} 
scattering. By increasing the Fermi sea density, they  observed 
faster dephasing.
In another experiment, 
Bonvalet {\em et. al.} \cite{bon96} 
used an extremely short, $\sim$ 12fs, pulse to excite 
a wavepacket consisting of states in the two lowest conduction 
subbands 
of a QW. By measuring the radiated  coherent electromagnetic 
field, they observed a quantum beat at room temperature 
oscillating 
at the inverse intersubband excitation energy.
 From the decay of the signal they deduced an
intersubband polarization  dephasing 
time of 110fs in the case of a MDQW, which was shorter 
than the dephasing time of 180fs observed in an undoped QW.
Such polarization dephasing 
is  determined by  the destructive 
interference effects as well as by the different scattering processes.

The above measurements of the polarization dephasing 
were performed
under resonant photoexcitation conditions. 
In contrast to undoped semiconductors, the
relaxation of such {\em real} pump--induced {\em e--h} pairs
due to their {\em e--e}  interactions with the FS
will obscure any coherent effects.
Fortunately, the dissipative processes 
can be suppressed by tuning the pump frequency 
{\em below} the onset of absorption. In this case, 
only {\em virtual} carriers are excited by the pump field 
and thus the coherent effects will dominate.
For well--below--resonance pump excitation, 
the {\em e--h} pair phase is primarily 
determined by the optical field 
and the time dependence of the 
interband polarization follows 
that of the pump pulse
(adiabatic following \cite{eber}). 
In this case, the 
pump creates a truly 
coherent {\em e--h} pair many--body state 
that lasts for the duration of the pulse.
Such pump--probe  experiments in atomic systems 
showed a resonance blueshift (
optical Stark effect) accompanied by 
a resonance bleaching due to the Phase Space Filling 
by the pump--induced {\em e--h} pairs 
\cite{eber,ssrev,atom}. 
In the case of excitons, 
it was shown that the Coulomb interaction can significantly 
alter the bleaching observed in atomic systems.
In particular, at low pump intensities
or for pulse duration longer than the dephasing time, 
the interactions 
lead to a pure exciton blueshift without significant  bleaching. 
For high intensities and  pulse duration 
shorter than the dephasing time,  the exciton bleaching 
is very strong. 
The exciton ac--Stark effect has been reviewed e.g. 
in Refs. \cite{ssrev,haugkoch,henn93,dan99,comb92}. 
Such interaction effects in undoped semiconductors
 suggest 
that the many--body {\em e--h} correlations 
between the photoexcited holes and the  FS electrons 
will significantly affect the coherent nonlinear
optical response of the FES. 
Brener {\em et. al.} \cite{bre95,per96} 
studied this issue 
by performing  pump--probe measurements
for pump detunings $\sim$ 50meV 
below the Fermi surface, 
much larger than the {\em e--h} 
Coulomb energy $E_{M}$ (which is much smaller than the 
Fermi energy $\sim$15meV).
They 
observed a qualitatively different bleaching 
of the FES
as compared to excitons
in undoped QW's, which they attributed  to the distinct 
 nature of the two resonances.
Sch\"{a}fer \cite{bre95} 
studied this issue by performing 
quantum kinetic 
calculations using Keldysh Green's functions.
Such an approach treats 
the HFA coherent effects and includes 
the  scattering contributions 
within the second order Born, Markovian, 
and static screening  approximations.
 The  numerical results for 
Fermi energy equal to the 
exciton binding energy 
showed a strikingly different bleaching 
between the 
FES and the exciton, which was attributed to the 
{\em e--e} scattering. 

As  we discuss at length in section \ref{sec:FES},
the   {\em e--h} correlations 
between the photoexcited hole  and the FS electrons 
lead to strong dephasing and  
affect qualitatively the coherent 
ultrafast dynamics of the FES. 
An important difference from undoped semiconductors 
is that, due to its {\em gapless} excitation spectrum,
a FS responds {\em unadiabatically} 
to time--dependent perturbations.
In contrast, 
because of its finite Coulomb binding energy, an
exciton can be polarized by the pump optical field
without being ionized.
A non--equilibrium treatment beyond the HFA 
is necessary in order to take into account the unadiabatic 
time--dependent {\em change} in the {\em e--h} pair--FS 
interactions and the {\em e--h} scattering processes
induced by the ultrafast pump excitation
in the coherent regime. 
As has already been noted in the case of photoexcitation 
of an electron gas 
within the continuum of states of an undoped 
semiconductor \cite{mycek}, 
the loss of coherence due to the many--body correlations 
cannot be fully described within the 
dephasing time approximation.  
In section \ref{sec:FES}   we show
that memory 
effects due to the {\em e--h} correlations 
lead to a time evolution of the pump--probe 
signal characterized by the inverse 
Coulomb energy $E_{M}$ rather than 
by the dephasing time (as in the case of exciton bound states). 

Concluding this section, let us briefly 
discuss  a very recent   FWM experiment
by Fromer {\em et. al.} \cite{from99}, which 
demonstrated for the first time 
non--Markovian memory effects in 
the Quantum Hall Effect regime.
Even though 
for magnetic fields between 
5.5 and 6.5 T the TI--FWM profile 
was found to be  a single exponential with an unusually 
long decay time, for magnetic fields 
that exceed 7.5 T such a 
profile was more complicated and characterized by a change of slope
indicating
memory effects in the polarization dynamics. 
Such effects were  also seen in the frequency domain:
the FWM spectrum 
profile changed from a Lorentzian lineshape 
to an asymmetric one corresponding 
to a frequency dependent width, $\Gamma (\omega)$.
To interpret such behavior, 
we note that, at high magnetic fields such that the cyclotron energy,
$\hbar\omega_c$, is large compared
to other characteristic energies of the system, the 
relaxation is dominated by intra-Landau-level processes. Such scattering
by collective excitations involves the matrix elements
of the dynamically screened interaction,
$U_{ij}^{<}(t,t')$, which in the lowest Landau level have the form:
\begin{eqnarray}\label{U1}
U_{ij}^{<}(t,t')&=&
\int \frac{d{\bf q}}{(2\pi)^2}
e^{-q^2l^2/2}v_q^2\,\bar{\chi}_q^<(t,t')
c_{ij}(q) \, ,
\end{eqnarray} 
where
$\bar{\chi}_q^<(t,t')=
\langle\bar{\rho}_{\bf q}(t')\bar{\rho}_{-\bf q}(t)\rangle$ 
is the density--density correlation function
projected onto the lowest Landau level \cite{gmp,haus96}, and
$\bar{\rho}_{\bf q}(t)$ is the corresponding
density operator. Here, $v_q$ is the unscreened
Coulomb interaction, $l = (\hbar/eB)^{1/2}$ 
is the magnetic length, and the coefficients 
$c_{ij}(q)$ with $i,j \rightarrow e,h$
model the asymmetry in the {\em e-e} and {\em e-h}
interaction matrix elements, which originates from the
difference between the 
electron and hole lowest Landau level wavefunctions.
Because of the breakdown of perturbation
theory due to the 
Landau level degeneracy in 2D systems, it
is incorrect to evaluate $\chi_q^<(t,t')$ 
within the
standard RPA \cite{haugbook}. Instead, one should account
for the true excitations of the interacting two--dimensional
electron liquid.
Several models can be found in the literature, and
we base our discussion on the magnetoroton model, 
which is the one best suited for the filling factors
$\nu$ in the experiment of Ref. \cite{from99}. 
 The most
salient features are, however, general and model
independent. The magnetoroton dephasing mechanism
is somewhat similar to that of acoustic phonon
scattering. Under the  experimental conditions of Ref. \cite{from99}, 
to a very good
approximation, the intra-Landau-level collective excitations
are not affected by the small density of photogenerated
carriers, so one can use the equilibrium density
correlation function\cite{gmp}. 
The equations for the density matrix elements then read 
\begin{eqnarray}\label{scatt1}
&&
\frac{\partial {\rho}_{ij}}{\partial t}
\Biggl|_{scatt}=i\sum_{k}\int_{-\infty}^{t}dt'
G_{i}^{r}(t-t')G_{j}^{a}(t'-t)
\nonumber\\&&\times
\Biggl(
[U_{ik}^<(t-t') -U_{kj}^<(t-t')]
\rho_{ik}^<(t') \rho_{kj}^>(t')
- (<\leftrightarrow >) \Biggr),
\end{eqnarray}
where $G_{i}^{r/a}(t)$ is the retarded/advanced 
Green function,
$\rho_{ij}^< = \rho_{ij}$, 
and $\rho_{ij}^> = \delta_{ij}-\rho_{ij}$. 
If all $U_{ij}$ are equal, i.e., $c_{ij}(q)=1$,
then the polarization scattering term 
vanishes\cite{lerner}. This corresponds 
to identical
electron and hole wavefunctions in the lowest Landau level. 
In practice, there is always an asymmetry between
electrons and holes, due to, e.g.,  
differing band offsets, lateral confinement, and disorder. 
Using the results of Ref.\ \cite{gmp},
Eq.\ (\ref{U1}) takes the form
\begin{eqnarray}\label{Umag}
U^{<}(t)=
-\frac{in}{2\pi}\int
&&
\frac{d{\bf q}}{(2\pi)^2}e^{-q^2l^2/2}v_q^2c_{ij}(q)
\nonumber\\&&\times
\bar{s}_q [(N_q+1)e^{i\omega_qt}+ N_qe^{-i\omega_qt}],
\end{eqnarray}
where $N_q$ is the Bose distribution function for 
magnetorotons of energy $\omega_q$, and $\bar{s}_q$ is 
the static stucture factor of the 2D electron liquid in the lowest 
Landau level. 
By comparing Eqs.\ (\ref{scatt1}) and (\ref{Umag}),
we see that the $\omega$ dependence of $\Gamma(\omega)$
is determined by the Fourier transform of $U^<(t)$, which
is governed by the $q$ dependence of $\bar{s}_q$. In the
lowest Landau level,  we have 
$\bar{s}_q =(2\nu^{-1} - 1)\tilde{s}_q$, where 
$\tilde{s}_q \sim (q l)^4$ for $ql\ll 1$,
$\sim \exp(-q^2l^2/2)$ for $ql\gg 1$, and  
$\tilde{s}_q$ displays a peak for $ql\sim 1$\cite{gmp} that
leads to the magnetoroton excitations. The corresponding
resonance in $\Gamma(\omega)$ near the magnetoroton
energy leads to non--Markovian behavior 
with a characteristic response time of approximately the inverse 
of this energy. The latter is
estimated from the gap $\Delta $ at the magnetoroton dispersion minimum, 
$\Delta\sim 0.1(e^2/\epsilon l)$ for the range of 
$\nu$ in this experiment  \cite{gmp}, 
which for $B=10$ T is $\approx 1.5$ meV.

This concludes our overview of the main ultrafast 
dynamical features observed in MDQW's. 
In the next section we discuss the ultrafast dynamics 
in the case of metal nanoparticles and compare to MDQW's.

\section{Nonlinear Optical Dynamics in   Metal Nanoparticles} 
\label{sec:nano}

The properties of small metal particles 
in the intermediate regime between 
bulk--like and molecular  behavior 
have been the subject of great interest 
lately. This was motivated in part by the need 
to understand 
how the  properties of matter evolve at the 
transition from atoms and molecules 
to  bulk solids. An additional motivation
comes from the technological 
trend towards 
electronic and optoelectronic devices 
based on   smaller and smaller solid state structures
 \cite{clustopt1,clustopt2,clustopt3,clustopt4,clustopt5}.
Metal clusters are also being 
used in a variety of applications, 
ranging 
from catalysis to biological and medical applications 
( see e.g. Refs.  \cite{clustmed1,clustmed2,clustmed3,clustmed4,clustmed5}).
The electronic and thermodynamic
properties of metal clusters have been 
extensively reviewed,
e.g. in Refs.  \cite{heer,brack,peren81,halp86}. 
The linear optical properties of metal nanoparticles 
have also been reviewed in detail by Kreibig and Vollmer 
 \cite{kreibig}. 

It has been known for a long time that 
surface collective excitations play an important role in the
absorption of light by metal nanoparticles.
In large particles with sizes
comparable to the wave--length of light $\lambda$ (but smaller than the
bulk mean free path), 
the lineshape of the surface plasmon (SP) 
resonance is determined by  the electromagnetic effects
 \cite{em1,em2,em3,kreibig}. 
However, as the size of the nanoparticle becomes 
smaller than the mean 
free path of electrons in the bulk metal, 
quantum confinement becomes important.
In small nanoparticles with radii
$R\ll \lambda$, the absorption spectrum is governed  by quantum 
confinement effects. For example, 
the momentum non--conservation due to the confining potential 
leads to the Landau damping of the SP and to a resonance linewidth
inversely proportional to the nanoparticle size
 \cite{kaw66}. 
A review of the extensive theoretical and experimental 
studies of this effect may be found in Ref.  \cite{kreibig}.   
More recently, it was established
that the static nonlinear optical properties of small nanoparticles 
are also affected by the confinement. 
In particular, a size--dependent  enhancement
of  the third--order nonlinear optical susceptibilities
for monochromatic photoexcitation, caused by the elastic 
surface scattering of single--particle excitations, 
was predicted by Flytzanis and collaborators 
 \cite{hac86,fly91}  
and observed experimentally 
by Yang {\em et. al.}   \cite{yan94}.  
Dielectric confinement 
also enhances the optical nonlinearities
close to the SP frequency  \cite{fly91,kreibig,aga88}. 
Before we proceed with the discussion of the ultrafast 
dynamics of such confined Fermi seas, 
let us briefly summarize in the next section 
the main features of the SP resonance in the linear absorption 
spectrum. 

\subsection{Surface Plasmon Resonance in Linear absorption} 
\label{sec:SPLA}

In this section we summarize the basic facts regarding the 
linear absorption by small 
metal particles embedded in a medium with
dielectric constant $\epsilon_m$.
We will focus primarily on
noble metal particles containing several hundreds of atoms; in this
case, the confinement affects the extended electronic states  
even though the bulk lattice structure has been established.
When the particle radii are small, $R\ll \lambda$, so that
only dipole surface modes can be 
optically  excited and non--local effects can be neglected, the
optical properties of this system are 
determined by the dielectric function  \cite{kreibig}   
\begin{equation}
\epsilon_{\rm col}(\omega)=
\epsilon_m+3p\epsilon_m\,
\frac{\epsilon(\omega)-\epsilon_m}{\epsilon(\omega)+2\epsilon_m},
\end{equation}
where $\epsilon(\omega)=\epsilon'(\omega)+i\epsilon''(\omega)$ is 
the dielectric function  of a metal particle
and $p\ll 1$ is the volume fraction
occupied by nanoparticles in the colloid.  Since the 
$d$--electrons play an important role in the optical properties of
noble metals, the dielectric function 
$\epsilon(\omega)$ includes also the interband contribution
$\epsilon_d(\omega)$ due to 
transitions  from the d--band to the s--p conduction band. 
For $p\ll 1$, the absorption  coefficient of such a system is
proportional to that of a single particle and is given by \cite{kreibig}  
\begin{equation}\label{absor}
\alpha(\omega)= -9p\,\epsilon_m^{3/2}\,\frac{\omega}{c}\,
\mbox{Im} {1\over \epsilon_{s}(\omega)},
\end{equation}
where
\begin{equation}\label{epseff}\epsilon_{s}(\omega)=
\epsilon_d(\omega)-\omega_p^2/\omega(\omega+i\gamma_s)+2\epsilon_m
\end{equation}
plays the role of an effective dielectric function of a particle in
the medium. Its zero, $\epsilon'_{s}(\omega_{s})=0$, determines the
frequency of the SP, $\omega_{s}$. In Eq.\ (\ref{epseff}),
$\omega_p$ is the bulk plasmon frequency of the conduction
electrons, and the width $\gamma_s$ characterizes the SP 
damping. 
The semiclassical result 
Eqs.\  (\ref{absor}) and  (\ref{epseff}) applies to nanoparticles
with radii $R\gg q_{_{TF}}^{-1}$, where  
$q_{_{TF}}$ is the Thomas--Fermi screening wave--vector
($q_{_{TF}}^{-1}\sim 1$ {\AA} in noble metals).
In this case, the electron density deviates from its 
classical shape only within a surface layer 
occupying  a small fraction of the total volume \cite{kre92}. 
Quantum mechanical corrections, arising from the discrete energy
spectrum, lead to a width $\gamma_s\sim v_{_F}/R$, where
$v_{_F}=k_{_F}/m$ is the Fermi velocity \cite{kreibig,kaw66}. 
Even though 
$\gamma_s/\omega_s\sim (q_{_{TF}}R)^{-1}\ll 1$, this damping
mechanism dominates for sizes 
$R < 10$ nm. Additional contributions 
to such a width come 
from the {\em e--p}, {\em e--e}, and electron--impurity 
interactions  as well as from the disorder.
On the other hand, in small clusters containing 
several dozens  of atoms, 
this semiclassical approximation breaks down and 
density functional or {\em ab initio} methods should
be used  \cite{kreibig,heer,brack,peren81,halp86,cluster}. 
In particular, discrete electronic levels whose  width
does not exceed their  spacing are expected 
for particle sizes smaller than a few angstroms. 
In the latter regime, the {\em e--e} interactions are similar  
to those  in atoms and molecules. 
One therefore  expects that the role of the 
many--body correlations
increases as we approach  the crossover from quasi--continuous  
to discrete nanoparticle energy levels.

It should be noted that, in contrast to the
scattering with  surface collective
excitations, the {\em e--e} scattering is not too  sensitive to the
nanoparticle size as long as  the condition 
$q_{_{TF}}R\gg 1$ holds \cite{siv94}. 
Indeed, for such sizes, the static screening is essentially 
bulk--like. At the same time, the energy dependence of the bulk 
{\em e--e} scattering rate (Eq. \ \ref{taue}) 
$\gamma_e\propto (E-E_F)^2$
comes from the phase--space restriction 
due to the energy and momentum conservation, and 
involves the exchange of typical  momenta $q\sim q_{_{TF}}$. 
If the size--induced momentum uncertainty 
$\delta q \sim R^{-1}$ is much smaller than $q_{_{TF}}$, the 
{\em e--e} scattering rate in a  nanoparticle is not 
significantly affected by the confinement \cite{alt97}.
Below we will see however that this is not the case for the 
electron--SP interactions.

\subsection{Ultrafast Dynamics} 
\label{sec:SPUD}

Even though the electronic, thermodynamic, 
 and optical properties of metal nanoparticles
have been extensively studied, the role of confinement in the
electron dynamics is much less understood.
Examples of outstanding issues include 
the role of {\em e--e} interactions 
in the process of cluster fragmentation, the role of surface 
lattice modes in providing additional
channels for intra-molecular energy relaxation, 
the influence of the electron and nuclear motion on the
superparamagnetic properties of clusters, 
and the effect of confinement on the nonlinear optical properties 
and  transient response under ultrafast excitation
 \cite{kreibig,cluster}.
These and other time--dependent phenomena can be studied with
femtosecond nonlinear optical spectroscopy, 
which in these structures provides time resolution shorter than the 
relaxation times. 
Similar experimental studies 
in bulk metals and metal interfaces
and surfaces  
 have been reviewed e.g.  in Ref.  \cite{petek}. 
Here we will focus on the more recent work 
in metal nanoparticles. 

Extensive experimental studies of 
the electron relaxation in 
noble--metal 
nanoparticles have  recently 
been performed using 
ultrafast pump--probe
spectroscopy.
In contrast to the situation in
semiconductors, 
the dephasing processes in metals
are very fast. 
In metal nanoparticles, 
the dephasing time
can be deduced, e.g., from the linear absorption 
SP width to be of the order of a few fs. 
Such measurements were performed
by Klar {\em et. al.}  \cite{kla98},  
who distinguished between 
homogeneous and inhomogeneous broadening
by measuring directly the SP lineshape 
of a {\em single} nanoparticle  using 
a near--field antenna effect.
The decay time of the SP resonance has also been studied with 
second and third harmonic generation  
measurements  \cite{hg1,hg2}. 

Within several femtoseconds 
after the photoexcitation, 
the optically--excited 
electrons and holes 
form
non--equilibrium
populations.
The pump--induced intraband transitions
create a broad non--thermal electron distribution 
that extends from the Fermi energy up to the pump 
photon energy, 
while the interband transitions 
create  additional electron and d-band hole populations 
when the pump frequency 
exceeds the d--band to conduction band transition 
threshold.
Within a few fs, 
the high energy 
electrons scatter to lower energies
due to {\em e--e} interactions 
with the FS electrons.
The latter interactions are  screened within  time intervals 
of the order of the inverse plasma frequency, typically 
$\sim$ 1fs. Note that, 
similar to MDQW's \cite{hawr94},
at {\em high} electron excess  energies 
the {\em e--e} 
 scattering time 
is of  the order of a fs, much shorter than the 
{\em e--p} scattering times (of the order of a ps).  
However, 
as such electrons scatter down to the Fermi surface, 
the {\em e--e} scattering times become 
much longer, of the order of several hundreds of fs. 
As discussed in the previous section, this is a consequence
of the Pauli blocking of the phase space available 
for scattering 
and the screening of the {\em e--e} 
interaction, which leads to energy--dependent 
scattering times given by Eq. \ \ref{taue} \cite{pines}. 
Therefore, one expects that thermalization 
should slow down as the electrons 
approach the Fermi surface. 
In the case of metal films, 
a non--thermal population consisting 
of a hot Fermi--Dirac distribution together with  a
tail of high energy electrons  right above the Fermi surface 
(and a tail of low energy holes below the Fermi surface)
was indeed observed 
in time--resolved photoemission experiments 
during time intervals of the order of hundreds of fs  \cite{fann1,fann2}.
Pump--probe experiments also showed 
a clear signature of delayed thermalization
(see e.g.  \cite{groen1,groen2,sun1,sun2,fly98}).
In particular, the rise time of the differential transmission and reflectance
signal was observed to be of the order of hundreds of fs, 
much longer than the pulse duration.

A similar slow  rise time of the differential transmission 
was also observed in the case of copper  \cite{big95},
gold  \cite{per97}, and silver  \cite{halt99} 
nanoparticles as well as in gold nanoshells  \cite{halas1,halas2}. 
It has been reported that, in small  nanoparticles with diameters
$\sim $5 nm,
such a delayed thermalization
regime (stage {\bf II}) lasts longer than in
metal films  \cite{big95,halt99}.
At the same  time, the {\em e--p} interaction for such sizes
was found to be weaker as compared to the bulk \cite{big95,halt99}.
A possible explanation of the reduced {\em e--p} 
interaction is that, for the small  sizes,  
the characteristic phonon energy,  given
by the Debye frequency, becomes smaller than the spacing between the
nanoparticle energy 
levels close to the Fermi surface, which is of the order of
$v_{F}/R$.
 As a result, the 
scattering
of an electron
via bulk--like phonons is suppressed  \cite{phonlong}. 
As the electrons approach the Fermi surface, 
the {\em e--e} scattering times become  comparable to 
the {\em e--p} scattering times and therefore 
the {\em e--p} interaction  also contributes 
to the thermalization of the electron gas.
The suppression of the latter could 
lead to delayed internal 
thermalization of the electron gas.

The main spectral feature in the 
differential  absorption of the copper 
and gold nanoparticles 
(where the onset of interband transitions is very close to 
SP frequency)
was a 
transient asymmetric broadening of the SP resonance  
( see e.g. Refs. \cite{tok94,say1,say2,zhang1,zhang2,
big95,per97,inou98,hart98}). 
Perner {\em et. al}  \cite{per97} 
observed a   build--up of 
such a pump--induced broadening 
during time intervals of the order of 1ps.
Bigot {\em et. al.}  \cite{big95} observed a similar 
buildup and also pointed out that
the differential 
absorption lineshape 
could not be understood without including the effects 
of the energy--dependent 
{\em e--e} scattering on the interband 
dielectric function. 
They attributed the SP broadening 
mainly to the smearing of the Fermi--Dirac 
electron distribution 
close to the Fermi surface, due to the 
heating of the electron gas, which affects 
the interband dielectric function. 
Inouye {\em et.al.}  \cite{inou98} 
arrived at similar conclusions and argued 
that, in  the case of resonant interband excitations, 
the 
contribution of the intraband
{\em e--e} scattering to the 
damping of the SP, which is described by the
width $\gamma_s$ in the intraband dielectric function
in Eq.\ (\ref{epseff})  \cite{fly98,gurzhi,kaveh},
plays a minor role. 

The time--delay
of the SP 
broadening can be understood as follows. 
For quasi--equilibrium conditions, the nonlinear absorption 
spectrum can be described using the linear absorption 
results but with dielectric constants determined by the time--dependent 
carrier populations. As discussed above, within the time resolution of the 
experiment, the electron distribution deviates from the
equilibrium  Fermi--Dirac 
distribution in the vicinity of the Fermi surface, 
due to the high electron temperature and the non--thermal 
population \cite{fann1,fann2,groen1,groen2}.
As can be seen from the linear absorption 
expressions, such a smearing of the FS distribution 
leads to the broadening  of the SP resonance. 
Therefore, 
the initial rise time of the differential 
transmission signal is determined by the 
non--equilibrium electrons ouside the 
equilibrium Fermi--Dirac distribution. 
Initially, such electrons occupy the broad photoexcited 
distribution and therefore 
their number 
is relatively small, determined by the 
pump intensity. However, their number 
increases with time as the 
{\em e--e} interaction leads to the 
scattering of   more electrons out of the 
FS. This is the origin of the time--dependent increase in the 
SP broadening.

Since the electron heat capacity is much smaller than that of the
lattice, an electron temperature much higher than that of the lattice 
can be reached during 
subpicosecond  time scales.
Subsequently, the electron and phonon baths
equilibrate  through the {\em e--p} interactions 
over  time intervals of  a few picoseconds.
During this incoherent stage, 
the hot electrons
can be characterized by a 
thermalized Fermi--Dirac distribution
with time--dependent temperature $T(t)$, while 
the phonons can be characterized by a 
 Bose--Einstein distribution 
with 
time--dependent temperature
$T_{l}(t)$. 
During the cooling of the hot electron gas, 
the SP width 
decreases \cite{tok94,say1,say2,zhang1,zhang2,
big95,per97,inou98,hart98}.
As soon as  the electrons have equilibrated
among themselves,
one can study the 
subsequent 
time 
evolution by using  a set of coupled differential 
equations for 
the temperatures 
$T(t)$ and $T_{l}(t)$,
referred 
to as the two--temperature model  \cite{tt,anis75,met1,met2,met3,met4}:
\begin{eqnarray}
\label{TT1}
C_{e}(T)\frac{\partial T}{\partial t} =
 &&\mbox{\hspace{-6mm}}
\nabla \cdot \left( \kappa_{e} \nabla  T \right) 
-G(T-T_l) + P({\bf r}, t), 
\nonumber\\
C_l\frac{\partial T_l}{\partial t} = 
&&\mbox{\hspace{-6mm}}
G(T-T_l),
\end{eqnarray}
where $C_{e}(T)=\gamma T$ and $C_l$ are the electron and lattice heat
capacities, respectively, $G$ is 
the energy transfer coefficient 
between the electrons and the lattice,
which is proportional to the {\em e--p} coupling constant
 \cite{allen},  
$\kappa_{e}$ is the electronic thermal conductivity,
and the source term 
 $P({\bf r},t)$
describes  the local energy density 
per unit time absorbed by the 
electron system from the pump optical  pulse.
The first term on the rhs of Eq.\ (\ref{TT1})
describes the thermal diffusion   from the nanoparticle 
to the surrounding matrix,
which occurs on a time scale 
of several tens of ps. 
The second  term 
on the rhs of Eq.\ (\ref{TT1})
determines the tranfer of  heat from  the 
electron gas to the lattice as the former cools down 
during a time interval of several ps. 
The above model assumes  electron and lattice 
temperatures larger than the Debye temperature  \cite{anis75}
and  that the interactions can maintain a quasi--equilibrium 
for both the electron and the phonon populations at all times. 
The cooling of the electron gas to the lattice 
manifests
itself via an exponential decay of the differential 
transmission.
According to Eq.\ (\ref{TT1}),
the decay time of the differential tranmission signal 
is inversely proportional to the strength of the 
{\em e--p} 
interaction  \cite{sun1,sun2}.
Stella {\em et. al.}  \cite{stella96} performed transient 
reflectivity measurements in metallic tin nanoparticles 
with radii ranging from 2 nm to 6nm
and saw a decay of the signal  as a function
of pump--probe delay that became faster 
by decreasing the nanoparticle  radius.
They deduced from their data a contribution to the energy 
relaxation time during the thermalization of the hot electron Fermi
sea with the lattice 
that was inversely proportional 
to the nanoparticle radius.
In a subsequent paper, Nisoli {\em et. al.}  \cite{nis97} 
performed femtosecond pump--probe measurements 
in solid and liquid gallium nanoparticles 
with radii ranging from 5 nm to 9 nm
and observed energy relaxation time constants that  varied from 
1.6 ps to 600 fs
with decreasing  nanoparticle size.
They observed similar electron relaxation dynamics in the 
solid and liquid nanoparticle phases, 
which indicates that in this system, the
scattering of electrons with 
 bulk phonons   plays only a minor
role in the relaxation
because of the reduction in the available
phase space  due to the quantum confinement.
Instead, the hot electrons 
transfer their excess energy  to the lattice through the 
generation of surface vibrational waves (capillary waves), 
which leads to an electron--surface phonon interaction $G$ 
inversely proportional to the nanoparticle radius
 \cite{phonshort}.
An additional contribution to this effect 
may come from the
fact that the heat diffusion to the matrix 
becomes  faster with decreasing nanoparticle size due to 
a thermal buildup that occurs for larger particles. 
A similar effect was observed by Halte {\em et. al.} 
in the case of very small silver nanoparticles  \cite{halte1}.
The role of surface effects on the dynamics was also
discussed  in Ref.  \cite{zhang2}.

In the gold and copper nanoparticles, condidered above,
the SP frequency is very close to the onset 
of interband transitions.
In contrast, in silver nanoparticles,
the SP and interband 
transitions are well separated in energy.
Halte {\em et. al.}  \cite{halt99} compared the electron dynamics 
between silver thin films and silver nanoparticles embedded in
glass  in the same
spectral range,
and studied its  dependence on the photoexcitation
intensity. In thin films, they found that the RPA 
dielectric function corresponding to the instantaneous 
temperature determined by the two--temperature model reproduced
their  results quite well. In the case of the silver nanoparticles, 
the main feature in the differential absorption spectrum was an
apparent SP redshift, which was attributed to the changes in the
real  part of the interband dielectric function due to the
thermal broadening of the electron distribution
close to the Fermi level. It was also shown that the finite 
{\em e--e} 
lifetime of the  carriers photoexcited via interband 
transitions 
was essential for obtaining such a redshift.
Finally, the strong 
non--parabolicity of the conduction band 
at the energies corresponding to the optical frequencies 
was shown to be important for interpreting the 
data.
Averitt {\em et. al.}  \cite{halas1,halas2} performed ultrafast 
pump--probe measurements in gold nanoshell structures, which consist
of a dielectric core surrounded by a thin metallic shell of
nanometer dimensions. In such structures, 
the SP frequency can be tuned by changing the 
ratio of the core diameter to the  shell thickness, and the 
experiment was performed at frequencies 
far off the onset of the interband transitions.
The delayed pump--induced broadening  of the SP resonance 
suggests that a nonthermal electron population is present 
in these structures 
during the initial $\sim$100 fs following the pump photoexcitation.
The decay of the pump--probe signal
was found to be somewhat slower than in the bulk, 
which was attributed to a weaker 
{\em e--p} interaction due to the reduction 
of phase space available for scattering
with bulk phonons induced by the confinement.
Finally, they observed a transient blueshift of the SP 
resonance,  attributed to the off--resonant interband transitions. 

Recent experimental results in the case of {\em small}
noble metal particles \cite{big95,halt99} 
indicate that 
 many--body correlation effects play an important role 
during the cooling of the electron gas 
to the lattice (relaxation stage  {\bf III}). 
Despite the similarities to the bulk--like behavior,
observed, e.g., in metal films, 
certain  aspects of the optical
dynamics in nanoparticles are 
significantly different  \cite{big95,sha98,sha99}.
For example, the experimental studies of copper 
nanoparticles by Bigot {\em et. al.} \cite{big95}
revealed that, 
for sizes  smaller than $\sim$ 5nm,  
the decay times  of the
pump--probe signal {\em depend strongly on the 
probe frequency} in the immediate vicinity of the SP resonance.
In particular, 
the relaxation is considerably slower
at the SP resonance, and becomes 
faster right above and right below the SP 
frequency \cite{big95,sha98,sha99}.
This and other  observations suggest that 
collective surface excitations
play an important role in the electron dynamics
in small metal particles. This important 
issue will be discussed at length 
in section \ref{sec:optics}.

\section{Coherent Ultrafast Response of the Fermi Edge Singularity: 
Formalism}
\label{sec:formalism}
\subsection{Basic equations}
\label{sec:basic}

Before we proceed with our discussion of the 
time evolution of the FES in the coherent regime, we outline 
the main points of a
formalism  \cite{per99,per94,per96,chemphys,pri99} recently developed 
to account for the non--Markovian 
dynamics  due to the dynamical 
FS response  ({\em e--h} correlations).
More details may be found in Appendix A.

As discussed in section \ref{sec:FESLA}, 
even in linear absorption, the HFA and the 
dephasing time approximation 
have serious shortcomings when used to describe the FES.
In the case of a photoexcited  electron gas, 
the  correlation effects have been treated 
within the second Born approximation by using 
Keldysh Green functions (see e.g. Ref. \cite{haugbook}). 
However, at low  temperatures and within the frequency
range of the FES, the effects of the {\em e--h} interaction 
must be consistently accounted for to arbitrary order. 
For example, as discussed in section \ref{sec:FESLA}, 
even in linear absorption 
the vertex correction diagrams with arbitrarily
many crossed {\em e--h} interaction lines are as
divergent as the HFA ladder diagrams 
and should therefore be
treated on equal footing.
With Green functions,
this requires  summing up at least the parquet
diagrams \cite{andrei,gavor},  a  formidable task  
especially in the non--equilibrium femtosecond 
regime. Alternative methods are therefore 
desirable. 
In undoped semiconductors, the ``dynamic controlled 
truncation scheme'' has been used to treat 
the  exciton--exciton correlations \cite{axt98,axt94}
(see section \ref{sec:introduction}). 
However, such a hierarchy of density matrix equations 
no longer truncates if the ground state of the semiconductor 
includes a FS. 
Furthermore, as we demonstrate  in section 
\ref{sec:monochr}, 
an expansion in terms of the optical fields 
breaks down for frequencies within the FES range. 
Finally, 
in view of the significant 
complexity of the problem, it is highly desirable 
to use a method that also allows for  a physically 
intuitive interpretation of the results. 
The purpose of this section is to outline the main 
points of such a method, which will be used in section 
\ref{sec:FES} to describe the coherent nonlinear 
response of the FES. 
 
We first consider the case of one conduction subband
and later extend the formalism to include the 
Coulomb coupling to a second subband.
We also consider spinless electrons for simplicity. 
In the rotating frame \cite{eber}, the Hamiltonian 
describing this system is 
\begin{equation} \label{H-tot}
H_{{\rm tot}}(t) = H + H_{p}(t) + H_{s}(t) \, .
\end{equation}
The first term is the ``bare'' Hamiltonian,
\begin{equation} \label{H0} 
H =  \sum_{\bf {\bf q}}\varepsilon_{{\bf {\bf q}}}^{v}
b_{-{\bf {\bf q}}}^{\dag} b_{-{\bf {\bf q}}} + 
\sum_{{\bf {\bf q}}} (\varepsilon_{{\bf {\bf q}}}^{c}
+\Omega) a^{\dag}_{{\bf {\bf q}}}a_{{\bf {\bf q}}}
+V_{ee} +V_{hh}+V_{eh},
\end{equation}
where  $a^{\dag}_{{\bf {\bf q}}}$ is the creation
operator of a conduction electron with energy
$\varepsilon_{{\bf {\bf q}}}^{c}$ and mass $m_{e}$,
$b^{\dag}_{-{\bf {\bf q}}}$ is the creation operator
of a valence hole with energy
$\varepsilon_{{\bf {\bf q}}}^{v}$ and mass $m_{h}$,  
$V_{ee}, V_{eh}$, and $V_{hh}$ describe the
{\em e--e}, {\em e--h}, and {\em h--h} interactions,
respectively, and
$\Omega= E_{g} + E_{F}(1 + m_{e}/m_{h}) - \omega_{p} $
is the detuning of the central frequency of the
optical fields, $\omega_p$, from the Fermi level, $E_{g}$ being
the bandgap (we set $\hbar=1$ everywhere).
The second and third
terms describe the coupling of  the pump optical field,
${\cal E}_p(t) e^{i {\bf k}_p \cdot {\bf r}  - i\omega_{p} t}$,
and  the probe optical field,
${\cal E}_s(t)e^{i{\bf k}_s \cdot {\bf r} - i\omega_{p} (t-\tau)}$, 
respectively: 
\begin{eqnarray}\label{Hp}
H_{p}(t)
\mbox{\hspace{-6mm}}&&
= -\mu {\cal E}_p(t) \left[ e^{i {\bf k}_{p} \cdot
 {\bf r }}U^{\dag}+ {\rm H.c.} \right], 
\nonumber\\
H_s(t) =
\mbox{\hspace{-6mm}}&& 
-\mu {\cal E}_s(t) \left[ e^{i {\bf k}_{s} \cdot
 {\bf r } + i \omega_{p} \tau}U^{\dag}+ {\rm H.c.} \right],
\end{eqnarray} 
where the pump amplitude ${\cal E}_p(t)$ is
centered at time $t=0$ and the probe amplitude
${\cal E}_s (t)$ is centered at the time
delay $t=\tau$,  $\mu$ is the interband
transition matrix element, and
\begin{equation}
U^{\dag} = \sum_{{\bf {\bf q}}}
a^{\dag}_{{\bf {\bf q}}}
b^{\dag}_{-{\bf {\bf q}}} \, 
\label{trans-op}
\end{equation} 
is the optical transition operator. 
The conventions for the time delay $\tau$ are clarified in 
Appendix B.

In many experiments, the amplitude of the probe field is much
smaller than that of the pump, 
$|{\cal E}_p(t)| \gg |{\cal E}_s (t)|$. 
In that case, as was shown in Ref.\  \cite{per99} 
(see Appendix A),
the experimentally--measurable nonlinear optical
polarization can be obtained in terms of the
linear response functions of a ``pump--dressed''
semiconductor to a probe field
(note that, within $\chi^{(3)}$, this is true even for comparable 
pulse amplitudes). This ``dressed'' system is 
described by a time--dependent effective Hamiltonian
$\tilde{H}(t)$, which is obtained by performing a
time--dependent Schrieffer--Wolff/Van Vleck  
canonical transformation on the Hamiltonian $H+H_{p}(t)$
\cite{s-w,s-w2,cohen,per93}.
As we shall see 
later, 
in all the cases
of interest, the effective Hamiltonian $ \tilde{H}(t)$
has the same operator form as the bare Hamiltonian 
$H$, with the important difference that the band
dispersions (effective masses) and interaction
potentials are {\em time--dependent} through
${\cal E}_{p}(t)$. Thus, the calculation of the nonlinear 
absorption spectrum reduces to that of the linear 
absorption spectrum of the ``pump--dressed'' semiconductor 
with uncoupled ``effective bands'' --- a great
simplification that allows us to use 
straightforward generalizations of 
well established theoretical tools in order to treat the 
correlations. 
In fact, such an approach 
mimics nicely the spirit of the pump--probe experiments
and allows for a physically intuitive interpetation 
of the results.
It is important to note here that
this ``pump--dressed semiconductor'' 
approach is not restricted to monochromatic pulses 
and is valid for {\em any}  pulse duration \cite{per99}.

The pump--probe
nonlinear polarization has the following form (see Appendix A):
\begin{equation} 
P_{{\bf k}_{s}}(t) =
i\mu^2   e^{i {\bf k}_{s} \cdot {\bf r}- i\omega_{p} (t - \tau)}
\int_{-\infty}^{t} dt' {\cal E}_{s}(t')\langle \Phi_0(t)| 
\tilde{U}(t){\cal K}(t,t')\tilde{U}^{\dag}(t')
|\Phi_0(t') \rangle \, .
\label{spec1}
\end{equation}
Here, $|\Phi_0(t)\rangle$ is the state evolved with 
$\tilde{H}(t)$ from the semiconductor ground state 
$| 0 \rangle$ of $H$, $\tilde{U}^{\dag}(t )$
is the effective optical transition operator
describing the  probability amplitude 
for the photoexcitation  of an {\em e--h} pair  
by the probe field in the presence of the pump
excitation, and ${\cal K}(t,t')$ is the 
time--evolution operator satisfying  the
Schr\"{o}dinger equation
\begin{equation}\label{eigen}
i\frac{ \partial}{\partial t}{\cal K} (t,t') 
=\tilde{H}(t){\cal K} (t,t') \, .
\end{equation}
The above equation  describes the time evolution of a 
probe--photoexcited {\em e--h} pair
in the presence of the pump excitation. The
effective Hamiltonian and effective transition
operator are given by (see Appendix A)
\begin{equation} 
\tilde{H}(t) = H_{0} +\frac{\mu}{2}\, 
\left({\cal E}_p(t)\left[\hat{\cal P}(t), U^{\dag} \right]
+ {\rm H.c.} \right),
\label{eff}
\end{equation} 
and 
\begin{equation} 
\tilde{U}^{\dag}(t) 
= U^{\dag} + \frac{1}{2} 
\left[ \hat{\cal P}(t), \left[U^{\dag}, \hat{\cal P}^{\dag}(t)\right]
\right] + \frac{1}{2} 
\left[\hat{\cal P}^{\dag}(t), \left[ U^{\dag}, \hat{\cal P}(t)\right]
\right], 
\label{me}
\end{equation} 
where the  operator 
$\hat{\cal P}^{\dag}(t)$, which generates the canonical 
transformation, satisfies the equation 
\begin{equation}
i   \frac{\partial
\hat{\cal P}^{\dag}(t)}{\partial t} =
\left[ H,\hat{\cal P}^{\dag}(t)
\right] + \mu {\cal E}_p(t) U^{\dag},
\label{sig}
\end{equation}
with the initial condition $
\hat{\cal P}^{\dag}=0$ before
the pump arrives.
Eqs. (\ref{eff}) and  (\ref{me}) include  all the
pump--induced corrections to $\tilde{U}^{\dag}(t)$ 
and $\tilde{H}(t)$ up to the second order in the pump
optical field and are valid when
$\left( \mu E_{p}/\Omega \right)^{2} < 1$ (for
off--resonant excitation) or 
$\left( \mu E_{p} t_{p}\right)^{2}  < 1$ (for
resonant excitation), where $t_{p}$ is the pump
duration.

It should be emphasized that, although
 Eq.\ (\ref{eff}) 
gives the effective Hamiltonian 
up to the  second order in the pump field,
${\cal E}_p{\cal E}_p^{\ast}$,  
the polarization 
expression Eq.\ (\ref{spec1}) describes the effects of 
$\tilde{H}$
in {\em all} orders.
For example, as we  
shall see in section \ref{sec:hamilt}, the pump--induced term in
$\tilde{H}$ contains  self--energy
corrections to the electron/hole energies, 
which describe 
(among other effects) the resonance blueshift due to
the ac--Stark effect. Although the {\em magnitude} of
these self--energy corrections, calculated 
using Eq.\ (\ref{eff}),  
is quadratic in
${\cal E}_p$, the
correct position of the resonance 
can only be obtained
by evaluating the PP polarization (\ref{spec1})
nonperturbatively (beyond $\chi^{(3)}$), i.e., without
resorting to the expansion of the time--evolution
operator ${\cal K}(t,t')$ in the pump field. 
Importantly, 
as we demonstrate in \ref{sec:monochr},
the same is 
true when calculating the effects of the 
self--energy corrections 
on the {\em e--h} correlations. 
As we shall see in section \ref{sec:FES}, such a nonperturbative (in
the pump field) treatment of the nonlinear response of
the FES is crucial for the adequate description of
the PP spectrum at negative time delays.
In Sections \ref{sec:cce} and \ref{sec:e-h},
we will describe the corresponding procedure, which
accounts  for the FS dynamical response. 
In contrast, the third--order polarization, $\chi^{(3)}$,
can be simply obtained from Eq.\ (\ref{spec1}) by
expanding ${\cal K}(t,t')$ to the {\em first} order
in the pump--induced term in $\tilde{H}$ [second term
in Eq.\ (\ref{eff})]. We did not include in Eq.\ (\ref{spec1}) 
the  biexcitonic contribution  
(coming from the excitation of two {\em e--h} pairs by the pump 
{\em and} the probe pulses)
since it vanishes for the negative time delays ($\tau<0$)
considered below where 
the coherent effects dominate \cite{per99}.

In addition  to including important 
contributions beyond $\chi^{(3)}$
via the solution of Eq. \ (\ref{eigen}) 
as discussed above,  
the advantage of  Eq.\ (\ref{spec1}), as
compared to the  equations of motion for the
polarization, comes from its similarity to the
linear polarization that determines the linear
absorption spectrum  \cite{mahbook}. This can be seen
by setting ${\cal E}_p(t) =0$ in the Eqs. (\ref{eff})
and  (\ref{me}), in which case the effective
time--evolution and optical transition operators
transform into their ``bare'' counterparts:
${\cal K}(t,t') \rightarrow  e^{-i  H (t - t' )}$ and
$\tilde{U}^{\dag}(t) \rightarrow U^{\dag}$. Moreover, like
$U^{\dag}$, the effective transition operator
$\tilde{U}^{\dag}(t)$ creates a single {\em e--h} pair,
while, as we shall see in section \ref{sec:hamilt},
the effective Hamiltonian
$\tilde{H}(t)$ can be cast in a form similar to
$H$. This allows one to interpret the Fourier transform
of  Eq.\ (\ref{spec1}) as the linear absorption spectrum 
of a ``pump--dressed'' semiconductor with two uncoupled
 but time--dependent effective bands. This mapping  
simplifies significantly the  calculation of the FES 
ultrafast nonlinear optical response by allowing a
straightforward generalization of the CCE.
It also allows one to interpret the various  dynamical 
features in the nonlinear absorption spectra, 
originating from the correlation effects, 
within the familiar framework 
developed for linear spectroscopy.

\subsection{Overview of the Coupled Cluster Expansion} 
\label{sec:cce}

In this section, 
we show how the time--dependent 
CCE can be used to study the effects of the
{\em e--h} correlations (dynamical FS response) on the
time evolution of  the {\em e--h} pair photoexcited
by the probe. Our goal is to evaluate the many--body
state $| \Psi(t) \rangle = {\cal K}(t,t')
\tilde{U}^{\dag}(t') |\Phi_0(t') \rangle$ that enters
in  Eq.\ (\ref{spec1}). This state satisfies the  Schr\"{o}dinger
equation
\begin{equation}
i   \frac{ \partial}{\partial t} | \Psi(t) \rangle 
=\tilde{H}(t) | \Psi(t) \rangle.
\label{eigen1}
\end{equation}
As already mentioned, $\tilde{H}(t)$ has the same
form as the bare Hamiltonian $H$. This allows us to obtain
$| \Psi(t) \rangle$ through a straightforward
generalization of the linear absorption calculations
 \cite{schon,ilias91a,ilias91b,ilias93}. After eliminating the 
valence hole degrees of freedom  \cite{llp,ilias93},  
$| \Psi(t) \rangle$ is expressed in the CCE form \cite{ccm1} 
\begin{equation}\label{ccw}
| \Psi(t) \rangle =  e^{S(t)} |\Phi(t) \rangle,
\end{equation}
where  the time--dependent operator $S(t)$
creates FS--pairs and is given by 
\begin{eqnarray} 
S(t) = 
\mbox{\hspace{-6mm}}&&
\sum_{p > k_{F},  k <k_{F}} 
s({\bf p,k},t) a^{\dag}_{{\bf p}}   
a_{{\bf k}} 
\nonumber\\&&
+ \sum_{p_{1}, p_{2} >k_{F},
k_{1}, k_{2}<k_{F} } s_{2}({\bf p_{1},p_{2},k_{1},k_{2}},t) 
a^{\dag}_{{\bf p_{1}}} a^{\dag}_{{\bf p_{2}}} a_{{\bf k_{2}}}
a_{{\bf k_{1}} }
+ \cdots, \label{S}
\end{eqnarray}
while the  state $ |\Phi(t) \rangle$, discussed in
Section \ref{sec:e-h}, describes the time evolution of the
probe--induced {\em e--h} pair. In Eq.\ (\ref{S}),
the amplitude $s({\bf p}, {\bf k}, t) $ describes
the {\em e--h} correlations which, in particular,
are responsible for the unbinding of the HFA bound state; 
the two--pair amplitude $s_{2}$
describes the {\em e--e} interaction effects at the 
RPA level and beyond.
>From a physical point of view, the operator 
$e^{S(t)}$ describes the readjustment 
of the FS density profile 
during the optical transition 
in response to the FS  
interactions 
with the photoexcited {\em e--h} pair.

Substituting  Eq.\ (\ref{ccw}) into the Schr\"{o}dinger
equation Eq.\ (\ref{eigen1}), multiplying by the
operator $e^{-S(t)}$ from the lhs, and using the
fact that $[S(t), S(t')]=[\dot{S}(t), S(t')]=0$, one obtains
\begin{equation} \label{phi}
i\frac{\partial}{\partial t}|\Phi(t) \rangle + 
i\dot{S}(t)|\Phi(t) \rangle 
=  e^{-S(t)} \tilde{H}(t) e^{S(t)}| \Phi(t) \rangle,  
\end{equation}
where the transformed Hamiltonian on the rhs 
can be expressed in terms of the   commutator series
(Baker--Campbell--Hausdorff expansion)
\begin{equation} \label{trH}
e^{-S(t)}  \tilde{H}(t) e^{S(t)}
= \tilde{H}(t) 
+ [ \tilde{H}(t), S(t) ] 
+ \frac{1}{2} [[H, S(t)],S(t) ] + \cdots  
\end{equation}
An important advantage of the CCE is that, 
due to the FS momentum  restrictions in  Eq.\ (\ref{S}), 
the above series {\em terminates}
after the first few terms 
(three for quartic interactions) and a closed--form  expression of
the transformed Hamiltonian (\ref{trH}) can be obtained  in terms of
$S(t)$. By requiring that all FS--pair creation processes are
eliminated from the above equation, one obtains the CCE
equations \cite{ccm1,schon} for $S(t)$. Before proceeding
with such a calculation however, one needs to derive
explicit expressions for $\tilde{H}(t)$ and $\tilde{U}(t)$.

%
\subsection{Effective Hamiltonian and Transition Matrix Elements}
\label{sec:hamilt}

\subsubsection{Discussion for Doped Semiconductors} 
\label{sec:discuss}

In the general case, the effective Hamiltonian $\tilde{H}(t)$
 is given 
by Eq. \ (\ref{eff}) and the effective 
optical transition operator $\tilde{U}(t)$ by 
Eq. \ (\ref{me}). 
For calculations, it is useful to re--express such 
equations in second--quantized
form. 
In the case of undoped semiconductors, 
it was shown in Ref. \cite{per99} that the effective Hamiltonian 
$\tilde{H}(t)$ describes the same interactions 
as the bare Hamiltonian $H$, however
among quasiparticles with time--dependent properties 
determined by the pump polarization.
Similarly, the effective interband transition 
matrix element in $\tilde{U}(t)$ was shown to have 
 an imaginary part that describes  the 
dephasing due to the exciton--exciton interactions. 
The purpose of this section is to present similar 
results in the case of a doped semiconductor
\cite{pri99,chemphys,per96}. 

We start with Eqs.\ (\ref{eff})
and\ (\ref{me}), which express $\tilde{H}(t)$
and $\tilde{U}(t)$ in terms of the canonical
transformation operator $\hat{\cal P}^{\dag}(t)$.
The latter is given by Eq.\ (\ref{sig}), which 
includes the effects of the Coulomb interactions
on the {\em pump} photoexcitation. It is important to realize that
the effect of the  {\em e--h} interactions on the pump and
the probe
photoexcitations is very different. 
For an adequate description
of the FES, the {\em e--h} interactions should be taken into account
{\em non--perturbatively} for the {\em probe}--photoexcited
pair. Indeed, the nonlinear  absorption spectrum 
at a given frequency $\omega$ close to the 
FES resonance is determined by the time--evolution
for {\em long} times (of the order of
the dephasing time $T_2$)
of an {\em e--h} 
pair photoexcited 
by the {\em probe} at 
the  energy
$\omega$ \cite{mahan2}. 
Since the characteristic 
``{\em e--h} interaction time''
$E_M^{-1}$ (inverse HFA bound state energy)
that determines  the non--exponential polarization decay 
of the FES is much shorter that the 
dephasing time,  
the long--time asymptotics of the response function 
(to the {\em probe}) depends non--perturbatively on the 
{\em e--h} interactions.  
In contrast,  a short {\em pump}
optical pulse excites a wavepacket 
of {\em continuum} 
{\em e--h} pair states
(unlike in the discrete exciton case)
with energy width $\sim t_{p}^{-1}$,
which thus evolves during timescales 
comparable to the pulse duration $t_{p}$.
Also, the 
corrections to the effective Hamiltonian 
are determined by the time 
evolution of the 
pump--induced carriers 
only up to times  $\sim t_{p}$
[see Eq.\ (\ref{eff})].
Therefore, if the 
``{\em e--h} interaction time'' 
$E_M^{-1}$ is larger than 
$t_{p}$, $t_p E_M < 1$, 
(i.e., if the pump pulse
frequency width exceeds $E_M$), 
the {\em e--h} interactions can be
treated perturbatively when describing the time--evolution of the 
{\em pump}--photoexcited pairs. 
This   can also be  shown explicitly for 
the third--order nonlinear polarization. 
In the general expression for  $\chi^{(3)}$, 
all contributions that 
depend on the pump are integrated over the
width of the pump pulse; therefore, any resonant enhancement
of $\chi^{(3)}$  that depends on the  pump frequency
will be broadened  out for sufficiently short pulses 
with frequency width that  exceeds $E_M$. 
In other words, when deriving the
pump--renormalized parameters, one can
treat Coulomb interactions perturbatively if the above condition is
fulfilled. 
In fact, the above 
situation  is somewhat similar to the calculation
of the linear absorption spectrum close to the indirect
transition threshold, 
where perturbation theory can be used \cite{andrei,gavor}.
Thus the
above consideration applies  even for long  pulse  durations
provided that  the detuning $\Omega$ exceeds $E_M$.
However, in order to obtain the full absorption spectrum,  
the time--evolution of the {\em probe}--photoexcited
pair with such effective Hamiltonian (with perturbatively calculated
time--dependent parameters) should be treated 
{\em non--perturbatively}.
As can be seen from the above discussion, 
an important advantage of this formalism is that it 
seprates naturally between the perturbative and the 
non--perturbative interaction effects.

\subsubsection{Second Quantization Expressions} 
\label{sec:sq}

We now proceed with the derivation of the effective Hamiltonian.
To lowest order in the interactions,
$\hat{\cal P}^{\dag}(t)$ can be presented as 
\begin{eqnarray} \label{op1}
\hat{\cal P}^{\dag}(t) 
=  
&&\mbox{\hspace{-6mm}}
\sum_{{\bf q}} {\cal P}_{eh}({\bf q},t)
a^{\dag}_{{\bf q}} b^{\dag}_{-{\bf q}}
+ \frac{1}{2} \sum_{{\bf p,p',k}} 
{\cal P}_{eh}^{e}({\bf p p';k;}t)b^{\dag}_{{\bf k-p-p'}}  
a^{\dag}_ {{\bf p}} a^{\dag}_ {{\bf p'}}a_ {{\bf k}} 
\nonumber\\&&\mbox{\hspace{-6mm}}
+\frac{1}{2} \sum_{{\bf p, p',k}} 
{\cal P}_{eh}^{h}({\bf p p';k};t) 
a^{\dag}_ {{\bf p + p' - k}} 
b^{\dag}_{-{\bf p}} b^{\dag}_{-{\bf p'}}
b_{-{\bf k}},
\end{eqnarray}
where ${\cal P}_{eh}$ is the  probability amplitude 
for excitation of an e--h pair with zero momentum 
satisfying 
\begin{equation}
\label{dir} 
i\frac{\partial}{\partial t}{\cal P}_{eh}({\bf q},t) 
=\left[ \Omega + \varepsilon_{\bf{q}}^{(c)} +
\varepsilon_{\bf{-q}}^{(v)}  \right] 
{\cal P}_{eh}({\bf q},t)+ \mu {\cal E}_p(t)-
\sum_{{\bf q}'} v({\bf q}-{\bf q}'){\cal P}_{eh}({\bf q'},t).
\end{equation} 
In Eq.\ (\ref{op1}), 
\begin{eqnarray}\label{eeh}
\mbox{\hspace{-6mm}}
{\cal P}_{eh}^{e}({\bf pp';k};t)
\mbox{\hspace{-6mm}}&&
=i\int_{-\infty}^{t}dt'
e^{-i(t - t') ( \Omega + \varepsilon_{{\bf p}}^{c} 
+ \varepsilon_{{\bf p'}}^{c} 
- \varepsilon_{{\bf k}}^{c}
+ \varepsilon_{{\bf p+p'-k}}^{v})} 
\nonumber \\ \times \mbox{\hspace{-6mm}}&&
\left[
 v({\bf p}-{\bf k})
\left[{\cal P}_{eh}({\bf p'},t') -
{\cal P}_{eh}({\bf p+p'-k},t') \right]
- ({\bf p} \leftrightarrow {\bf p'})\right]
\end{eqnarray} 
describes the scattering of the photoexcited
{\em e--h} pair with an electron, and 
\begin{eqnarray} \label{ehh}
\mbox{\hspace{-6mm}}
{\cal P}_{eh}^{h}({\bf pp';k};t) 
\mbox{\hspace{-6mm}}&& 
= -i\int_{-\infty}^{t}dt'e^{-i(t-t') 
\left( \Omega + \varepsilon_{{\bf p+p'-k}}^{c} 
+ \varepsilon_{{\bf p'}}^{v} + \varepsilon_{{\bf p}}^{v} 
- \varepsilon_{{\bf k}}^{v} \right)} 
\nonumber \\ \times \mbox{\hspace{-6mm}}&&
\left[
 v({\bf p}-{\bf k})
\left[{\cal P}_{eh}({\bf p'},t') -
{\cal P}_{eh}({\bf p+p'-k},t') \right] 
- ({\bf p} \leftrightarrow {\bf p'})
\right]
\end{eqnarray} 
describes the scattering of the photoexcited
{\em e--h} pair with a hole. 
The above expressions 
describe in the lowest order in the screened
interaction \cite{ander1}
$\upsilon({\bf p} - {\bf k}) $ 
the coherent pump--induced processes, 
the effects of the Hartree--Fock 
pair--pair and pair--FS interactions, and the  dynamical FS
response  to the {\em  pump} photoexcitation.

By substituting Eq.\ (\ref{op1}) into Eq.\
(\ref{me}), we obtain the following expression
for the effective optical transition operator: 
\begin{eqnarray} \label{tme}
\tilde{U}^{\dag}(t) |\Phi_0(t) \rangle =
\mbox{\hspace{-6mm}}&& 
\sum_{p > k_{F}} M_{{\bf p}}(t)  a^{\dag}_{{\bf p}} \,
b^{\dag}_{-{\bf p}} |0 \rangle 
\nonumber \\ \mbox{\hspace{-6mm}}&&
+ \frac{1}{4} \sum_{p,p'>k_{F},k<k_{F}}
M_{{\bf pp'k}}(t) a^{\dag}_{{\bf p}} \, a^{\dag}_{{\bf p}'} \, 
b^{\dag}_{{\bf k - p - p}'} \, a_{{\bf k}}  |0\rangle,
\end{eqnarray} 
where the effective matrix element $M_{{\bf p}}(t)$ 
includes corrections due to phase space filling and 
Hartree--Fock  interactions, and $M_{{\bf pp'k}}(t)$
is the probability amplitude for indirect optical
transitions  \cite{andrei} induced by the pump optical
field, which contribute to the pump--probe
polarization in the second order in the
interactions. The explicit expressions for
$M_{{\bf p}}(t)$ and $M_{{\bf pp'k}}(t)$ are
given in Appendix C. 

We turn now to the effective Hamiltonian
$\tilde{H}(t)$. After substituting Eq.\
(\ref{op1}) into Eq.\ (\ref{eff}) 
we obtain that 
\begin{eqnarray} \label{Heff}
\tilde{H}(t)= 
\sum_{\bf q} \varepsilon^{v}_{{\bf q}}(t) 
b_{-{\bf q}}^{\dag}b_{-{\bf q}}
+\sum_{{\bf q}} \varepsilon^{c}_{{\bf q}}(t)
a^{\dag}_{{\bf q}}a_{{\bf q}} 
+ V_{eh}(t) + V_{ee}(t),
\end{eqnarray} 
where 
\begin{eqnarray}  \label{ed} 
\varepsilon^{c}_{{\bf q}}(t)=
\varepsilon_{{\bf q}}^{c} - \mu {\cal E}_p(t)
{\rm Re}\left[ {\cal P}_{eh}({\bf q},t) - 
\sum_{{\bf q}'} {\cal P}_{eh}^{e}({\bf qq';q};t)\right]
\end{eqnarray} 
is the effective conduction electron energy; 
\begin{eqnarray} \label{hd}
\varepsilon^{v}_{{\bf q}}(t)= E_{g} 
+ \varepsilon^{v}_{{\bf q}} -\mu {\cal E}_p(t) 
{\rm Re}\left[ {\cal P}_{eh}({\bf q},t) - 
\sum_{{\bf q}'} {\cal P}_{eh}^{h}({\bf qq';q};t)\right],
\end{eqnarray} 
is the effective  valence hole energy; 
\begin{eqnarray}
V_{eh}(t)= 
-\sum_{{\bf kk'q} }
\upsilon_{eh}({\bf q; kk'};t)
a^{\dag}_{{\bf k+q}}
a_{{\bf k}}b^{\dag}_{-{\bf k'-q}}b_{-{\bf k}'}, 
\end{eqnarray} 
is the effective {\em e--h} interaction; and 
\begin{eqnarray}
V_{ee}(t)= \frac{1}{2} \sum_{{\bf kk'q}}  
\upsilon_{ee}({\bf q; kk'};t) 
a^{\dag}_{{\bf k+q}}a^{\dag}_{{\bf k'-q}}a_{{\bf k'}}a_{{\bf k}},
\end{eqnarray} 
is the effective {\em e--e} interaction. The explicit
expressions for $\upsilon_{eh}({\bf q; kk'};t)$
and $\upsilon_{ee}({\bf q; kk'};t)$ are given
in Appendix C. As can be seen, $\tilde{H}(t)$ has the
same operator form as the bare Hamiltonian $H$. However,
both the effective band dispersions and the effective
interaction potentials are now {\em dependent on time}.
Note here that the above  pump--induced 
renormalizations  only last for the pulse duration 
$t_{p}$.
As discussed above, they are therefore perturbative 
in the screened interactions for $
t_p E_M < 1$ or for $\Omega > E_{M}$.

\begin{figure}
\vspace{-10mm}
\begin{center}
\epsfxsize=5.0in
\epsffile{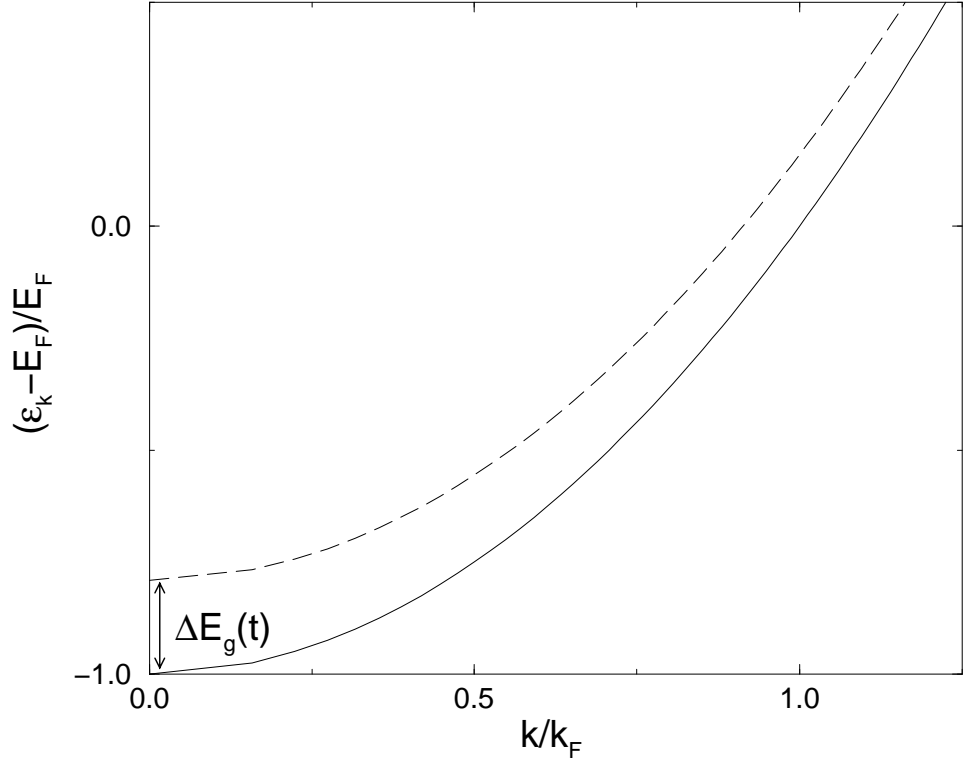}
\vspace{5mm}
\end{center}
\caption{
The effect of the pump pulse on the band dispersion,
$\varepsilon_{{\bf k}}(t) 
=\varepsilon^{c}_{{\bf k}}(t)  + \varepsilon^{v}_{-{\bf k}}(t)$, 
of the ``pump--dressed'' system. 
Solid line: ``bare'' dispersion ($t=-\infty$). Dashed line:
pump--renormalized dispersion ($t=0$). The bands become ``heavier''
for the duration of the pump.
}
\label{FES-fig1}
\end{figure}

Let us first discuss the effect of the pump--induced
self--energy corrections to the conduction and
valence band energies, given by the last terms in Eqs.
(\ref{ed}) and (\ref{hd}). The dispersion of the
effective band is shown in Fig.\ \ref{FES-fig1}. As can be seen,
the pump pulse leads to a bandgap increase as well as
a change in the momentum dependence (band dispersion) 
that last {\em as long as the pump pulse}. The
magnitude of the bandgap increase  is of the order of 
$(\mu{\cal E}_{p})^{2}/\Omega $ (for off--resonant
excitation) and $(\mu{\cal E}_{p})^{2} t_p$ (for
resonant excitation) and leads to, e.g., 
the ac--Stark blueshift. 
As we shall see, for pulse duration shorter than the
dephasing time, it also leads to bleaching and gain
right below the onset of absorption, analogous to
the case of excitons or two--level systems.
It should be emphasized that these are {\em coherent}
effects that should not be confused with the incoherent
bandgap redshift due to the {\em e--e} interactions
among  real photoexcited carriers \cite{wang95}. In particular,
the
above bandgap renormalization is induced by the
{\em transverse} EM--field of the laser, as compared to the usual 
bandgap renormalization due to a {\em longitudinal}
EM--field, i.e., Coulomb screening.
The change in the band
dispersion, whose relative magnitude is of
the order of $(\mu{\cal E}_{p}/\Omega)^{2}$
(for off--resonant excitation) or
$(\mu{\cal E}_{p}t_{p})^{2}$ (for resonant
excitation), can be viewed as an increase in
the effective density of states and, to the
first approximation, in the effective mass.
This is important in doped semiconductors 
because, 
as we shall see later, it leads to an {\em optically--induced}
time--dependent  enhancement 
of  the {\em e--h} interactions and
scattering processes with the FS electrons.

The effective Hamiltonian $\tilde{H}(t)$ also includes
pump--induced corrections in the effective interaction
potentials, determined by the pair--pair and pair--FS
interactions during the pump photoexcitation. By expanding 
Eqs.\ (\ref{veh}) and (\ref{vee}) for carrier energies 
close to the Fermi surface using Eqs.\ (\ref{eeh}) 
and (\ref{ehh}), one can show that these corrections
vanish at the Fermi surface; for the typical FS excitation
energies $\Delta \varepsilon \sim E_M$ that contribute
to the FES, their order of magnitude is 
$(\mu{\cal E}_{p}\,\Delta \varepsilon /\Omega^{2})^{2}$ 
(for off--resonant  excitation) or 
$ (\mu{\cal E}_{p}\,\Delta\varepsilon\, t_{p}^{2})^{2}$ 
(for resonant excitation). Thus the corrections to the
interaction potentials are suppressed for below--resonant
excitation by a factor of $(E_M /\Omega)^{2}$, or for
short pulses by a factor of $( E_M t_{p})^{2}$, 
as compared to the self--energy corrections. Such a
suppression is due to the Pauli blocking effect and the 
screening, which leads to the  vanishing of the pump--induced
corrections to the interaction
potentials at the Fermi surface. Similarly, the
pump--induced indirect optical transition matrix
elements $M_{{\bf pp'k}}(t)$ are suppressed by the
same factor as compared to the  direct transition
matrix element $M_{\bf{p}}(t)$ [first term
in  Eq.\ (\ref{tme})], while they  contribute
to the pump--probe polarization only in 
the second order in the screened interactions.
Therefore, in the doped case,
the screened Coulomb interaction 
leads to subdominant parameter 
renormalizations to the effective Hamiltonian 
$\tilde{H}(t)$ 
and transition operator $\tilde{U}^{\dag}(t)$
for sufficiently 
short pump pulses or for off--resonant excitation.
In contrast, in undoped semiconductors, 
the exciton--exciton interactions 
lead to non--perturbative contributions 
discussed in Ref. \cite{per99}.

\subsection{Description of the Electron--Hole Pair Dynamics}  
\label{sec:e-h}

\begin{figure}
\vspace{-10mm}
\begin{center}
\epsfxsize=3.0in
\epsffile{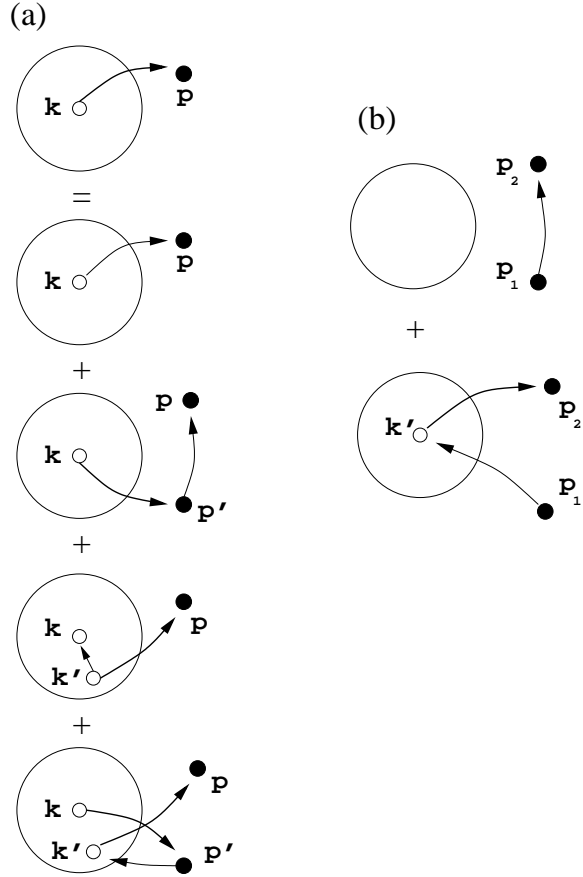}
\end{center}
\vspace{5mm}
\caption{
(a) The {\em e--h} scattering processes that contribute to the rhs of
Eq.\ (\ref{de}). Full lines correspond to $s({\bf p,k},t)$ and thin lines
to $V$. The diagrams describe (from top to bottom) (i) Born
scattering of a FS electron,  
(ii) FS electron ladder diagrams,  
(iii) FS hole ladder diagrams, and 
(iv) Nonlinear vertex corrections due to the dynamical FS response.
(b) Scattering processes state that
determine the time-- and momentum--dependence 
of the  effective {\em e--h} potential  
$\tilde{V}({\bf p},t) $ and 
lead to the unbinding of the HFA bound
state.}
\label{FES-fig2}
\end{figure}

In this section, we present the final formulae for
the nonlinear pump--probe polarization of the
interacting system by applying the CCE to the
effective Hamiltonian $\tilde{H}(t)$
in order to treat the dynamical FS response.
The CCE equation Eq.\ (\ref{phi}) contains the operator
$S(t)$ described by a hierarchy of coupled
equations for the amplitudes $s, s_{2}, \cdots$,
defined by Eq.\ (\ref{S}). As discussed 
at length above, in the coherent regime  of 
interest here, the {\em e--e} scattering effects 
are suppressed, and the nonlinear absorption spectrum 
is dominated by the {\em e--h} interactions. This allows
us to use the dephasing time approximation for treating the  
probe--induced {\em e--e} scattering 
processes  \cite{sham90,oht89,hawr91},
in which case the above 
hierarchy terminates after $s$. Importantly, the  
{\em e--h} correlations (dynamical FS response) are 
still treated 
non-perturbatively, since they are 
determined by $s$ \cite{schon,ilias91a,ilias93}. 
Then all FS--pair creation processes can be eliminated
explicitely from the rhs of Eq.\ (\ref{phi}), leading
to the following nonlinear differential equation
for the one--FS--pair scattering amplitude \cite{schon,ilias91a,ilias93},
$s({\bf p},{\bf k},t)$:
\begin{eqnarray} \label{de} 
i\frac{\partial s({\bf p,k},t)}{\partial t}-
\mbox{\hspace{-6mm}}&&
\left[\varepsilon_{{\bf p}}^{c}(t) - 
\varepsilon_{{\bf k}}^{c}(t)\right] 
s({\bf p,k},t)= 
\nonumber \\ \mbox{\hspace{-6mm}}&&
V\left[ 1 + \sum_{p' > k_{F}}
s({\bf p',k},t) \right]  
\left[ 1 - \sum_{k' < k_{F}}s({\bf p,k'},t) \right].
\end{eqnarray}
The {\em e--h} scattering processes described by
the above equation are sketched in Fig.\ \ref{FES-fig2}a. 
Here $V$ is the s--wave component  \cite{ssrev,stef83} 
of the screened  
interaction  \cite{ander1}, 
 approximated for simplicity  by its value 
at the Fermi energy \cite{andrei,oht89,sham90}.
This neglects plasmon effects, which are
however small within the  frequency range of the FES 
\cite{hawrpl,hawr94}.
Although this approximation is standard for the linear
absorption case, its justification for the transient spectra
requires more attention. Indeed, the
characteristic time for screening buildup is of the order 
$\Omega_p^{-1}$, where  $\Omega_p$ is the typical
plasma frequency corresponding to the FS 
 \cite{el-sayed94,banyai98,mahan2}.
This time is however shorter than the
typical pump duration $\sim$ 100fs  and dephasing time (which is of the order
of ps near the Fermi energy \cite{kim92}), 
so the Coulomb interactions can be  considered 
screened also for the nonlinear absorption case.
In Eq.\ (\ref{de}), we neglected the hole recoil energy
contribution to the excitation energy  \cite{ilias93}  since 
$\varepsilon^{v}_{k_{F}}(t)-\varepsilon^{v}_{0}(t)< E_M$
due to a  sufficiently heavy hole mass \cite{andrei}.
Note that, by increasing the hole effective mass, 
the pump--induced hole self--energy, Eq.\ (\ref{hd}), 
reduces the hole recoil energy and thus the  corresponding
broadening \cite{andrei}. In real samples, the
relaxation of the momentum conservation condition
due to the disorder will also suppress the hole
recoil broadening effects. 
In fact, the disorder can even lead to localized 
hole states \cite{fes1}, which corresponds to the infinite hole 
mass limit.

From  Eq.\ (\ref{phi}) we then easily obtain the
following expression for $|\Phi(t) \rangle$:
\begin{equation} \label{photo}
|\Phi(t) \rangle = \sum_{p>k_{F}} \Phi_{{\bf p}}(t,t') 
 a^{\dag}_{{\bf p}}b^{\dag} | 0 \rangle,
\end{equation}
where  $b^{\dag}$ is the creation operator of the 
zero--momentum hole state and the {\em e--h} pair
wavefunction $ \Phi_{{\bf p}}(t;t')$ satisfies the
``Wannier--like'' equation of motion,
\begin{eqnarray}\label{phip} 
 i \frac{\partial \Phi_{{\bf p}}(t,t')}{\partial t} 
= 
\biggl[\Omega + \varepsilon_{{\bf p}}^{c}(t)
+\varepsilon_{-{\bf p}}^{v}(t) 
\mbox{\hspace{-6mm}}&&
- \epsilon_A (t) 
- i \Gamma_{{\bf p}}\biggr] \Phi_{{\bf p}}(t,t')
 \nonumber \\ \mbox{\hspace{+6mm}}&&
 - \tilde{V}({\bf p}, t)\sum_{p' > k_{F}}
\Phi_{{\bf p'}}(t,t'),
\end{eqnarray}
where
\begin{equation}\label{Veff}
\tilde{V}({\bf p}, t) =
V\left[ 1 - \sum_{k'<k_{F}} s({\bf p,k'},t) \right],
\end{equation}
is the  effective    {\em e--h} 
potential whose 
{\em time-- and momentum--dependence}
is determined by the
response of the FS electrons to their 
 interactions with  the probe--induced
{\em e--h} pair (vertex corrections)
 [sketched schematically
in Fig.\ \ref{FES-fig2}(b), 
 responsible for the
unbinding of the HFA bound state], 
\begin{equation} 
\epsilon_A (t)=V \sum_{k'<k_{F}} \left[1+
\sum_{p'>k_{F}} s({\bf p',k'},t)\right]
\end{equation} 
is the self--energy due the to the sudden appearance
of the photoexcited hole, 
which leads to non--exponential polarization
decay [described by ${\rm Im}\,\epsilon_A(t)$] due
to the  Anderson orthogonality catastrophe  \cite{ander67} 
and a dynamical resonance redshift [described by
${\rm Re}\,\epsilon_A(t)$], and
the dephasing width  $\Gamma_{{\bf p}}$
describes all additional  dephasing processes (due
to {\em e--e} interactions, hole recoil, and phonons).
Eq.\ (\ref{phip}) should be solved with the initial
condition $\Phi_{{\bf p}}(t',t') = M_{{\bf p}}(t')$, 
where the effective matrix element 
$M_{{\bf p}}(t')$ is defined by Eq.\ (\ref{tme}). 

It is worth stressing here the analogy between Eqs.
(\ref{photo}) and (\ref{phip}) and the corresponding
problem in undoped semiconductors. Indeed, Eq. (\ref{photo})
is the direct analog of an exciton state, whereas
again Eq. (\ref{phip}) is very similar to a Wannier
equation. However, Eqs. (\ref{de}) and (\ref{phip})
include the effects of the interactions between the
probe--photoexcited 
{\em e--h} pair and the FS--excitations, and  the 
wavefunction $\Phi_{{\bf p}}(t,t')$ describes
the propagation of the photoexcited pair
``dressed'' by the FS excitations. Such a ``dressing'' 
is due to the dynamical FS response, which leads to  
the dynamical screening of the effective {\em e--h}
interaction 
Eq.\ (\ref{Veff}). The time--dependence of the latter  is determined
by the FS scattering amplitude $s({\bf p,k},t)$
and is affected by the pump excitation as described by Eq.\ (\ref{de}).
One can easily verify that
by setting $s({\bf p,k},t)=0$ in Eq.\ (\ref{phip}), we
recover the results of the Hartree--Fock (ladder diagram,
static FS \cite{andrei}) approximation. If one neglects
the nonlinear (quadratic) term in  Eq.\ (\ref{de}),
one recovers the three--body (Fadeev)
equations \cite{andrei,fadeev}.
Note that the coupled equations
for $\Phi_{{\bf p}}(t,t')$ and $s({\bf p,k},t)$,
obtained by neglecting the multipair excitations
in Eq.\ (\ref{S}), can  be extended to include the
hole recoil--induced corrections \cite{ilias93}. 

Using Eqs. (\ref{ccw}), (\ref{S}), and (\ref{photo}),
we now can express the pump--probe polarization Eq.\
(\ref{spec1}) in terms of the {\em e--h} wavefunction
$\Phi$ and the effective transition matrix element
$M_{{\bf p}}$. Assuming, for simplicity, a
delta--function probe pulse centered at time
delay $\tau$, 
${\cal E}_s(t)={\cal E}_s\delta(t-\tau)$, we obtain
a simple    final expression:
\begin{eqnarray}\label{p-p} 
P_{{\bf k}_{s}}(t)= i\theta(t-\tau)\mu^2{\cal E}_s
e^{i{\bf k}_{s}\cdot {\bf r}- i\omega_{p} (t - \tau)} 
\sum_{p > k_{F} } M_{{\bf p}}(t)  \Phi_{{\bf p}}(t,\tau). 
\end{eqnarray}
Eq.\ (\ref{p-p}) expresses the pump--probe
polarization in terms of two physically distinct
contributions. First is the effective transition
matrix element $M_{{\bf p}}(t)$, which includes
the effects of pair--pair and pair--FS interactions
and Phase space filling effects due to the
pump--induced carriers present during the probe
photoexcitation. Second is the  wavefunction
$\Phi_{{\bf p}}(t)$ of the {\em e--h} pair
photoexcited by the probe, whose time dependence, 
determined by Eqs.\ (\ref{phip}) and (\ref{de}), describes the
formation of the absorption resonance.
 Despite the formal similarities,
there are two important differences between the
doped and the undoped cases. First, in the doped
case, the time evolution of the {\em e--h}
wavefunction $\Phi_{{\bf p}}(t)$ is strongly
affected by the interplay between the {\em e--h}
correlations and the pump--induced 
transient changes in
the bandgap and band dispersion relations.
As we shall see later, this can be viewed as 
an excitation--induced dephasing. Second,
unlike in the undoped case \cite{per99}, the
pump--induced corrections in the effective
matrix element $M_{{\bf p}}(t)$ are
perturbative in the screened interactions
if the pump detuning or the pump frequency
width exceed the Coulomb energy $E_M$.

\subsection{The Case of Two Coupled Subbands} 
\label{sec:hybrid}
 
In this section, the above results are extended to the 
multisubband case in order to 
investigate the ultrafast PP dynamics of the
FES--exciton hybrid formed in asymmetric QW's with
partially occupied subbands 
\cite{chen90,chen91,fes3,mue90,mue91,hawr2sub,sko91}. 
 In such one--sided MDQW  structures, 
interband optical transitions 
from the  valence band to several conduction subbands are allowed
due to the finite overlap between the hole and electron envelope
wave--functions. 
When the exciton Fano resonance from a higher empty  
conduction subband is nearly resonant 
(within a few meV) 
to the Fermi level, 
 the FES coming from the 
lowest occupied subband 
is  enhanced by over 
two orders of magnitude 
\cite{fes3,chen90,chen91}. 
Such many--body effects on the linear 
spectrum have been described by using the simple two--subband
Hamiltonian \cite{mue90,mue91,hawr2sub} 
\begin{equation}
\label{H0-h}
H=
\sum_{i\bf k}\epsilon_{i\bf k}^ca_{i\bf k}^{\dagger}a_{i\bf k}+
\sum_{\bf k} (\epsilon_{\bf k}^v+E_g)b_{\bf -k}^{\dagger}b_{\bf -k}
-\sum_{ij}\sum_{\bf pkq}v_{ij}({\bf q}) a_{i\bf p+q}^{\dagger}a_{j\bf p}
b_{\bf k-q}^{\dagger}b_{\bf k},
\end{equation}
where $a_{i\bf k}^{\dagger}$ and $\epsilon_{i\bf k}^c$
are the creation operator and the energy of a conduction
electron in the $i$th subband, $b_{\bf -k}^{\dagger}$ and 
$\epsilon_{\bf k}^v$ are those of a valence hole
($E_g$ is the bandgap), and $v_{ij}({\bf q})$ is the {\em screened}
{\em e--h} interaction matrix with diagonal (off--diagonal)
elements describing the intrasubband (intersubband)  scattering. Due
to the screening, the interaction potential is short--ranged
and can be replaced by its s--wave 
component  \cite{ssrev,stef83}; close to the Fermi surface, 
$v_{ij}({\bf q})\simeq v_{ij}$  \cite{mue90,mue91,mahbook}.
Here we consider the case where only the first subband is occupied, but
the Fermi level is close to the exciton level (with binding energy $E_B$) 
below the bottom of the second subband.
For large values of the FES--exciton splitting
$\Delta-E_F-E_B$, where $\Delta$ is the subband separation, 
the linear absorption spectrum consists of two well separated peaks,
the lower corresponding to the FES from subband 1, and the higher
corresponding to the Fano resonance from the exciton of subband 2
broadened by its coupling to the continuum of states in subband 1. With
decreasing  $\Delta-E_F-E_B$, the FES and the exciton become hybridized 
due to the intersubband scattering arising from the Coulomb interaction.
This results in the transfer of oscillator strength from the exciton
to the FES and a strong enhancement of the absorption peak near the
Fermi level due to the resonant scattering of the photoexcited
electron by the exciton level \cite{chen90,chen91,fes3,mue90,mue91,hawr2sub}.
Since the  PP signal is linear in the probe field, 
the essential physics can be captured by assuming
a $\delta$--function probe pulse,  
${\cal E}_{\tau}(t)={\cal E}_{\tau}e^{i\omega_{p}\tau}\delta(t-\tau)$,
and a Gaussian pump pulse. 
The PP polarization has the form ($t>\tau$)
\begin{equation}
\label{pol-h}
P(t)=i{\cal E}_{s}e^{-i\omega_pt}
\sum_{ij}\mu_i\mu_j
\langle 0|\tilde{U}_{i}(t){\cal K}(t,\tau)
\tilde{U}_{j}^{\dag}(\tau)|0 \rangle,
\end{equation}
where ${\cal K}(t,\tau)$ is the time--evolution operator for the
effective Hamiltonian,
\begin{equation}
\label{eff1-h}
\tilde{H}(t)=\sum_{ij\bf k} \epsilon^c_{ij\bf k}(t)
a_{i\bf k}^{\dagger}a_{j\bf k}+
\sum_{\bf k} \epsilon_{\bf k}^v(t)b_{\bf -k}^{\dagger}b_{\bf -k}
+V_{eh}(t)+
V_{ee}(t),
\end{equation}
where $V_{eh}$ and  $V_{ee}$ are the effective 
{\em e-h} and {\em e-e} interactions and $\tilde{U}_{i}^{\dag}(t)$
is the effective transition operator given below. 
Here

\begin{eqnarray}  
\epsilon^c_{ij\bf k}(t)=\delta_{ij}\epsilon_{i\bf k}^c
+\Delta\epsilon_{ij\bf k}^c(t),
~~~
\epsilon_{\bf k}^v(t)=\epsilon_{\bf k}^v+\Omega+\Delta\epsilon_{\bf k}^v(t)
\end{eqnarray}
are the band dispersions with pump--induced self--energies
(to lowest order in the interactions) 
\begin{eqnarray}
\Delta\epsilon_{ij\bf k}^c(t)=
\mbox{\hspace{-6mm}}&&
-{\cal E}_p(t) [\mu_i p_{j{\bf k}}^{\ast}(t)+
\mu_j p_{i{\bf k}}(t)]/2,
\nonumber\\
\Delta\epsilon_{\bf k}^v(t)=
\mbox{\hspace{-6mm}}&&
-{\cal E}_p(t){\rm Re}\sum_i\mu_ip_{i{\bf k}}(t),
\end{eqnarray}
with $p_{i{\bf k}}(t)$ satisfying
\begin{equation}
\label{sigmai}
i\frac{\partial p_{i{\bf k}}(t)}{\partial t}=
(\epsilon_{i\bf k}^c+\epsilon_{\bf k}^v+\Omega)
p_{i{\bf k}}(t)
-\sum_{j{\bf q}}v_{ij}p_{j{\bf q}}(t)+\mu_i{\cal E}(t).
\end{equation}
Note that the pump induces additional
intersubband scattering, described by $\Delta\epsilon_{12\bf k}^c(t)$. 
To lowest order in the interactions, 
the effective transition operator appearing in Eq.\ (\ref{pol-h}) is
given by 

\begin{equation}
\tilde{U}_i^{\dag}(t)=\sum_{j{\bf k}}\phi_{ij{\bf k}}(t)
a_{j\bf k}^{\dagger}b_{\bf -k}^{\dagger},
\end{equation}
with 
\begin{equation}
\label{phiij}
\phi_{ij{\bf k}}(t)
=\delta_{ij}\left[1-\frac{1}{2}\sum_l|p_{l{\bf k}}(t)|^2\right]
-\frac{1}{2}p_{i{\bf k}}(t)p_{j{\bf k}}^{\ast}(t).
\end{equation}
In the one--subband case and in the coherent limit, 
Eq.\ (\ref{phiij}) takes a
familiar form, $\phi_{{\bf k}}(t)=1-|p_{{\bf k}}(t)|^2$ ---
the usual Pauli blocking factor \cite{pri99}; 
in a multi--subband case, the latter is a matrix.

Similar to the one--subband case, the state
${\cal K}(t,\tau)\tilde{U}_{i}^{\dag}(\tau)|0 \rangle$,
entering into (\ref{pol-h}), can be viewed as describing
the propagation of the {\em e--h} pair (with wavefunction 
$\Phi_{ij}({\bf k},t)$) excited by the probe pulse at time $\tau$,
 dressed by the scattering of the FS excitations (dynamical FS response). 
The latter leads to a dynamical broadening
described by the amplitude $s_{ij}({\bf p},{\bf k},t)$ 
that satisfies the differential equation \cite{schon,pri99}  
\begin{eqnarray}
\label{for s}
i\frac{\partial s_{ij}({\bf p},{\bf k},t)}{\partial t}=
\mbox{\hspace{-6mm}}&&
(\epsilon_{i\bf p}^c-\epsilon_{j\bf k}^c)s_{ij}({\bf p},{\bf k},t)
\nonumber\\ \mbox{\hspace{-6mm}}&&
+\sum_{l} [\Delta\epsilon_{il\bf p}^c(t)s_{lj}({\bf p},{\bf k},t)
-\Delta\epsilon_{lj\bf k}^c(t)s_{il}({\bf p},{\bf k},t)]
\nonumber\\ \mbox{\hspace{-6mm}}&&
-\sum_{l}\tilde{v}_{il}({\bf p},t)[\delta_{lj}
+\sum_{q>k_F}s_{lj}({\bf q},{\bf k},t)],
\end{eqnarray}
with initial condition $ s_{ij}({\bf p},{\bf k},\tau)=0$,
and {\bf p} and {\bf k} labeling respectively the ($i$th
subband) FS electron and the ($j$th subband) FS hole. Since
only the first subband is  occupied, the only non--zero
components of $s_{ij}$ are $s_{11}({\bf p},{\bf k},t)$ and
$s_{21}({\bf p},{\bf k},t)$, which describe the intra and
intersubband FS excitations 
respectively. The photoexcited 
 {\em e--h} pair wavefunction
$\Phi_{ij}({\bf k},t,\tau)$ satisfies the
Wannier--like equation
\begin{eqnarray}
\label{for phi}
i\frac{\partial\Phi_{ij}({\bf k},t)}{\partial t}=
\mbox{\hspace{-6mm}}&&
\sum_{l} [\epsilon_{il\bf k}^c(t)
+\delta_{lj}[\epsilon_{\bf k}^v(t)
+\epsilon_{A}(t)]-i\Gamma]\Phi_{lj}({\bf k},t)
\nonumber\\ \mbox{\hspace{-6mm}}&&
-\sum_{l,q>k_F}\tilde{v}_{il}({\bf k},t)\Phi_{lj}({\bf q},t)
\end{eqnarray}
with initial condition
$\Phi_{ij}({\bf k},\tau)=\phi_{ij{\bf k}}(\tau)$, where

\begin{equation}
\epsilon_{A}(t)=-\sum_{k'<k_F} [v_{11}
+\sum_{p'>k_F}s_{11}({\bf p}',{\bf k}',t)v_{11}]
\end{equation}
is the self--energy due to the readjustment
of the FS to the photoexcitation of a hole \cite{pri99,schon} 
and $\Gamma$ is the inverse dephasing time due to 
all the processes not included in $H$.
In Eqs.\ (\ref{for s})\ and\ (\ref{for phi}),

\begin{equation}
\tilde{v}_{ij}({\bf k},t) =v_{ij}-
\sum_{l,k'<k_F}s_{il}({\bf k},{\bf k}',t)v_{lj}
\end{equation}
is the effective {\em e--h} potential 
whose time--dependence is due to the 
dynamical FS response  \cite{pri99}.
Note that it is the interplay between this effective potential and
the pump-induced self-energies that gives rise to the unadiabatic FS
response to the pump field.
In terms  of $\Phi_{ij}({\bf k},t)$, the polarization
(\ref{pol-h}) takes the simple form ($t>\tau$)
\begin{equation}
\label{pol1-h}
P(t)=-i
{\cal E}_{s}e^{-i\omega_pt}
\sum_{ijl}\mu_i\mu_j\sum_{k>k_F}
\Phi_{il}({\bf k},t)\phi_{ij{\bf k}}^{\ast}(t),
\end{equation}
with $\phi_{ij}({\bf k},t)$ given by (\ref{phiij}). The
nonlinear absorption spectrum is then proportional to
${\rm Im}P(\omega)$, where $P(\omega)$ is the Fourier
transform of the rhs of (\ref{pol1-h}).

\section{Coherent Ultrafast Dynamics of the Fermi Edge Singularity}
\label{sec:FES} 
\subsection{Discussion of the Physics: Monochromatic Excitation}
\label{sec:monochr}

In this section, the 
role of the {\em e-h}
correlations in the nonlinear optical response  of the FES
is  illustrated in a simple way 
for monochromatic
excitation \cite{chemphys,per96,per94,pri99}. 
In the latter case, the theory 
discussed
in the previous section applies for  pump detunings
larger than the characteristic Coulomb energy, 
$\Omega> E_M$. As discussed 
above, close to the Fermi edge, the
linear absorption spectrum of the FES can be
approximately described using  the simple power law  expression
in Eq.\ (\ref{feslin}). The monochromatic
pump excitation leads to a resonance blueshift,
originating from the shift in the effective band
energies [see Fig.\ \ref{FES-fig1}], and to a bleaching mainly
due to the Pauli blocking (Phase Space Filling) which reduces the effective
transition matrix element (analogous to the dressed
atom picture  \cite{cohen}). More importantly, however,
the pump--induced change in the band dispersion
increases the density of states 
 ${\cal N}$  close to the Fermi
surface and thus also increases both the {\em e--h}
{\em scattering} strength $g=V {\cal N} $ and the phaseshift 
$\delta \sim  \tan^{-1}(\pi g)$ that determine the FES lineshape 
(see Eq.\ (\ref{feslin})). 
This, in turn, leads to an increase in the FES
{\em exponent} $\beta$ that determines the resonance
{\em width}. In contrast, in the case
of a bound excitonic state of dimensionality $D$
and Bohr radius $a_{B}$, the resonance width remains
unchanged, while the oscillator strength,
$\propto a_{B}^{-D}$, increases by a factor
$ \sim (1 + D\,\Delta m)$, where $\Delta m$ is the
pump--induced change in the effective
mass  \cite{per94,chemphys,per96}. Such an optically--induced enhancement
of the exciton strength competes with the bleaching
due to the Pauli blocking and the exciton--exciton
interactions. This
results in an almost rigid exciton blueshift, consistent  
with experiment  \cite{knox1,chemla89} and previous
theoretical results  \cite{ell,comb91}. 

However, in the case of a FES resonance, the
pump--induced change in the exponent $\delta$ leads
to a stronger oscillator strength enhancement than
for a bound exciton state. Obviously, such an
enhancement cannot be described perturbatively,
i.e., with an expansion in terms of  the optical
field, since, as can be seen 
from Eq.\ (\ref{feslin}),
the corresponding corrections to the
absorption spectrum {\em diverge logarithmically} 
for frequencies $\omega$
at the
Fermi edge.
Eq.\ (\ref{feslin}) shows that 
the effect of the pump on the FES can
be thought of as  an excitation--induced dephasing
that affects the {\em frequency dependence} of
the resonance; again, this is in contrast to the
case of the exciton. In the time domain,
this also implies a memory structure 
related to the response--time of the FS-excitations.
Therefore, the qualitative differences between
the nonlinear optical response of the FES and 
the exciton
originate from the fact that an exciton is a discrete
bound state, while the FES is a {\em continuum}
many--body resonance. The FS responds {\em
unadiabatically} to the pump--induced change in
the density of states via an {\em increase} in
the {\em e--h}
scattering of {\em low--energy} pair excitations.
Such  scattering processes, which determine the
response of the Fermi sea to the hole potential
in the course of the optical excitation,
are responsible for the unbinding 
and broadening of the HFA bound state \cite{mahan2}.
Therefore, the pump field 
changes the broadening and dephasing effects
even for  below--resonant photoexcitation. 
On the other hand, due
to the finite Coulomb binding energy of the exciton, 
the pump optical field can polarize such a bound
state and change its Bohr radius without ionizing
it.

\subsection{Results for Short--pulse Excitation} 
\label{sec:short}

In this section,  we discuss the nonlinear
absorption of the FES in the case of short pulse
excitation. The results presented 
here  
were  obtained by solving
numerically the differential equations
(\ref{phip}) and (\ref{de}), using the Runge--Kutta
method, for Gaussian pulses with duration
$t_p = 2.0 E_F ^{-1}$ \cite{pri99}.
In order
to suppress the incoherent effects due to the
{\em e--e} scattering of real 
carrier populations with the FS,
below--resonant pump excitation 
and negative time delays were considered.
Under such 
excitation conditions, the {\em coherent}
effects in which we are interested dominate, 
and the Coulomb--induced corrections 
to the effective  parameters, discussed
in section \ref{sec:hamilt}, are perturbative. 
The goal of this section is to study the role of the dynamical FS
response ({\em e--h} correlations) on the
pump--probe dynamics. For this reason, we compare
the results of the theory outlined in section \ref{sec:formalism} 
to those of the HFA,
obtained by setting $s({\bf p},{\bf k},t)=0$
in Eq.\ (\ref{phip}). As mentioned above, in
the latter case, 
the (spurious) HFA bound state does
not interact with the FS pair excitations, even
though it can merge with the continuum when one
introduces a very short dephasing time. 

\begin{figure}
\vspace{-10mm}
\begin{center}
\epsfxsize=5.0in
\epsffile{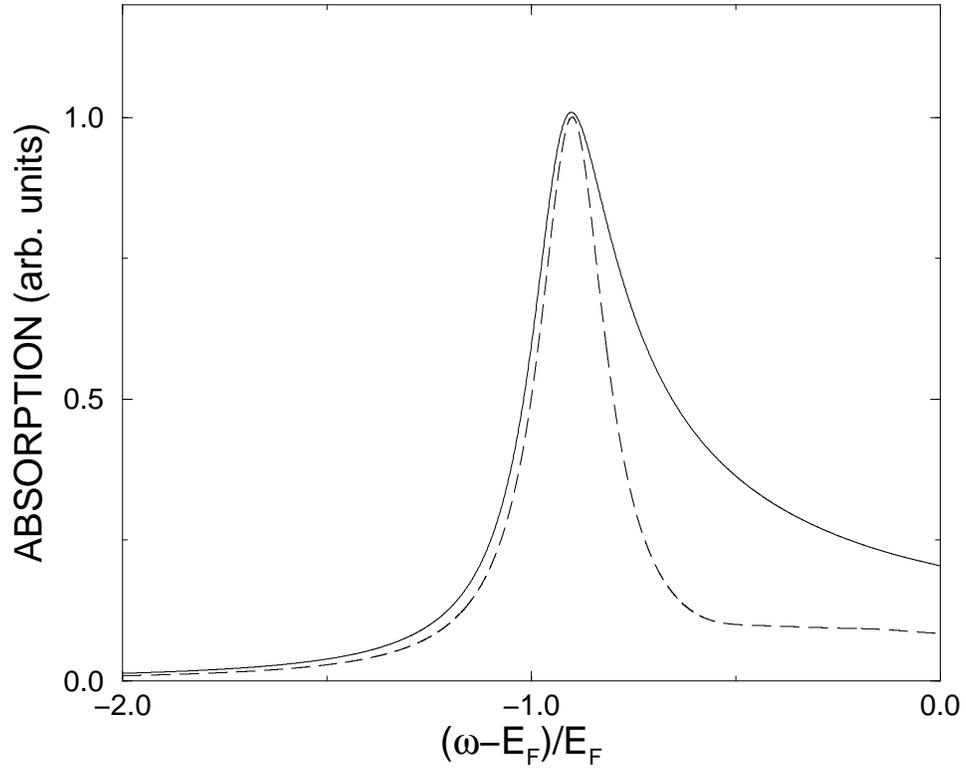}
\end{center}
\vspace{5mm}
\caption{
Linear absorption resonance lineshape 
for FES (solid curve) compared to the HFA (dashed curve), 
calculated with $g=0.4$ and $\Gamma=0.1 E_{F}$. 
The HFA resonance position was shifted for better visibility.}
\label{FES-fig3}
\end{figure}

In Fig.\ \ref{FES-fig3}, the linear absorption lineshape 
(i.e., in the absence of pump) of the FES is compared
to the HFA (without the dynamical FS response). 
The parameter values $g=0.4$ and 
$\Gamma=0.1 E_{F}$ were used, which were 
previously used to fit the experimental spectra 
in  modulation doped quantum wells \cite{oht89,ohtrev}.
For better visibility, we shifted
the
curves in order to compare their lineshapes.
The linear absorption FES lineshape is consistent with
that obtained in Ref.  \cite{oht89}.
On the other hand, the HFA
spectrum is characterized by the coexistence of
the bound state and a continuum contribution
due to the fact that, in 2D, a bound state exists
even for an arbitrary weak attractive potential.
We note that if one limits oneself to linear absorption,
it is possible to artificially shorten the dephasing
time $T_2 = \Gamma^{-1}$, mainly determined by the hole
recoil effects, by taking $\Gamma \simeq E_M$. Then the
spurious discrete state and the continuum merge, and
the discrepancy between the two linear absorption
lineshapes decreases. This trick has been
used for phenomenological fits of linear absorption 
experimental data. However, as we discuss below, 
in the nonlinear absorption case the differences in the
transient spectra are significant so that the processes 
beyond HFA can be observed experimentally.

\begin{figure}
\vspace{-10mm}
\epsfxsize=5.0in
\epsffile{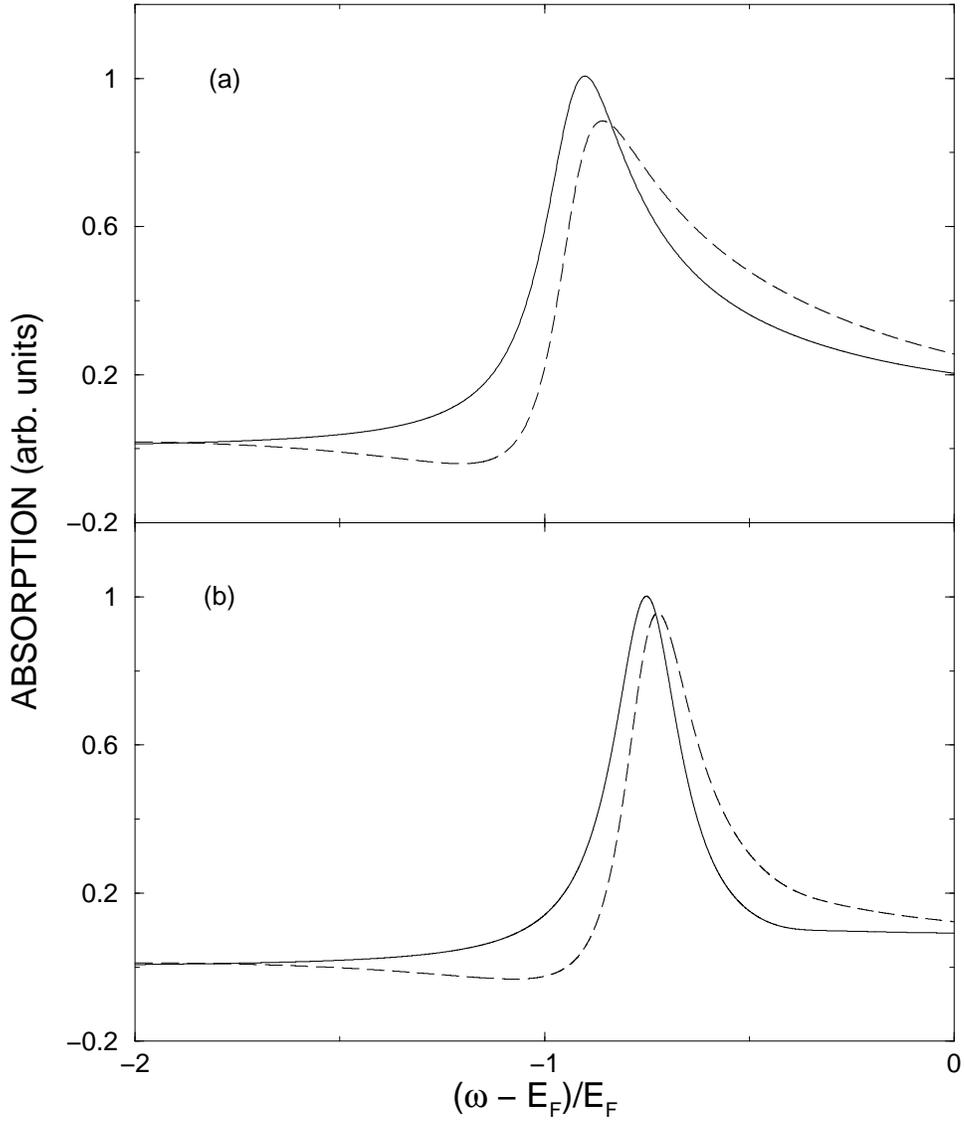}
\vspace{5mm}
\caption{The absorption spectra for the FES (a) compared 
to the HFA (b). Solid curves: linear absorption. 
Dashed curves: nonlinear  absorption.
The curves were calculated with 
$g=0.4$ and $\Gamma =0.1 E_{F}$ at time 
delay $\tau=-0.1\Gamma^{-1}=-t_{p}/2$ 
and pulse duration $t_{p} = 2.0E_{F}^{-1}$.
The nonlinear absorption lineshapes exhibit bleaching, 
resonance blueshift, and gain below
the absorption onset
that differ in the two cases.}
\label{FES-fig4}
\end{figure}

Let us turn to the time evolution of the pump--probe
spectra. In Fig.\ \ref{FES-fig4} we show the nonlinear absorption
spectra calculated by including the dynamical FS response 
[Fig.\ \ref{FES-fig4}(a)] and within the HFA [Fig.\ \ref{FES-fig4}(b)]
at a short time delay $\tau = - t_p/2$.
The main features of the spectrum 
are a pump--induced resonance bleaching, blueshift,
and gain right below the onset of absorption. For
off--resonant pump, these transient effects vanish
for positive time delays after the pump is gone, and
persist for negative time delays shorter than the
dephasing time $T_2 = \Gamma^{-1}$. Similar features
were also obtained for different values of the pump
amplitude, duration, and  detuning. They are mainly due to
the broadening induced by the transient renormalization
of the energy band dispersion [Eqs.\ (\ref{ed}) and
(\ref{hd})] when its duration $\sim t_{p}$ is
shorter than the dephasing time (analogous to
excitons and  two--level systems).

\begin{figure}
\vspace{-10mm}
\epsfxsize=5.0in
\epsffile{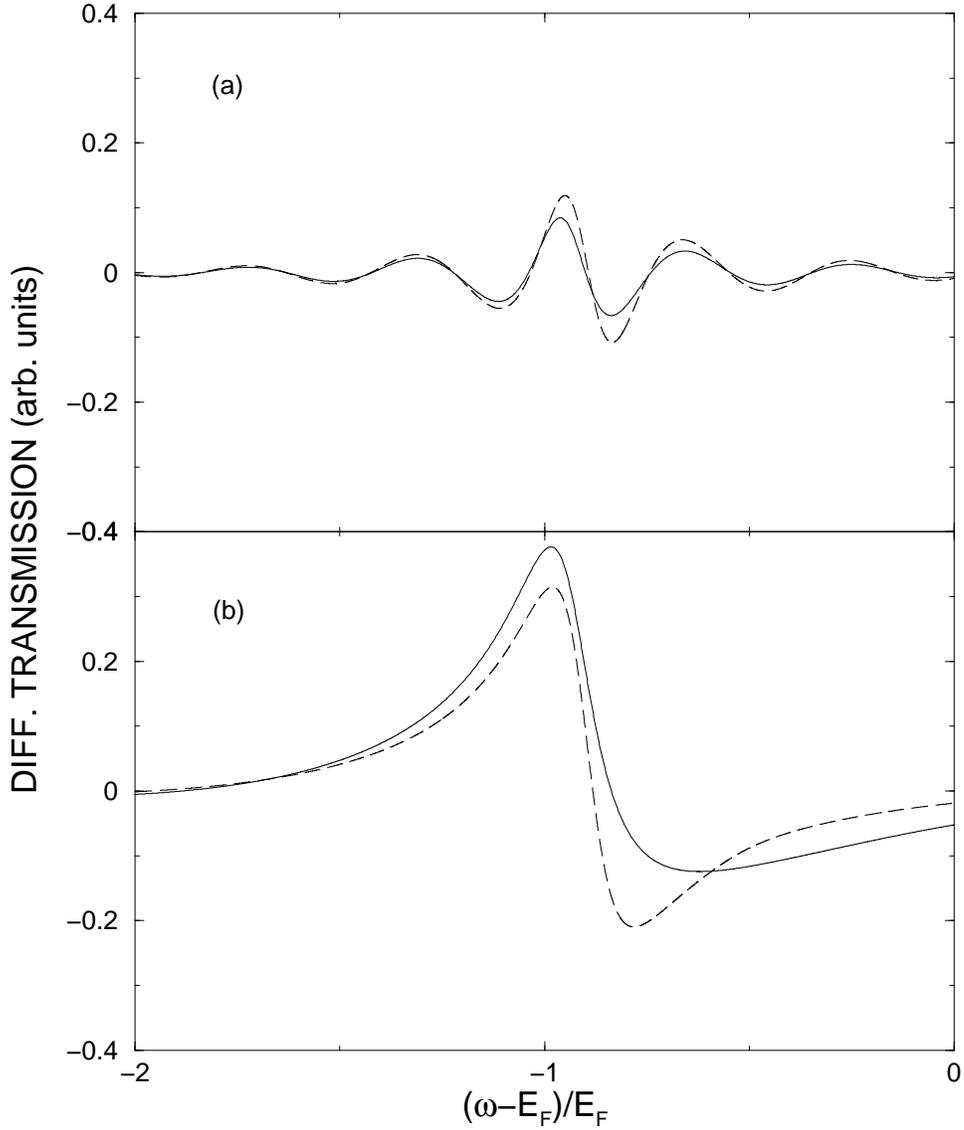}
\vspace{5mm}
\caption{
Differential transmission lineshape for various time delays calculated with 
$g=0.4$, $\Gamma =0.1 E_{F}$, and $t_{p} = 2.0E_{F}^{-1}$.
(a) Long time delay, $\tau = - 1.5\Gamma^{-1}=-15.0 E_F^{-1}$. 
For the FES, the oscillations in
the differential transmission spectra are reduced (solid curve)
as compared to the HFA (dashed curve). 
(b) Short time delay, 
$\tau = - 0.1 T_2 = - t_p/2 = - 1.0 E_F^{-1}$.
For the FES, the differential transmission spectrum 
is asymmetric (solid curve) as compared to the
symmetric lineshape for the HFA (dashed curve).
The above curves were shifted for better visibility.
}
\label{FES-fig5}
\end{figure}

\begin{figure}
\vspace{-10mm}
\epsfxsize=5.0in
\epsffile{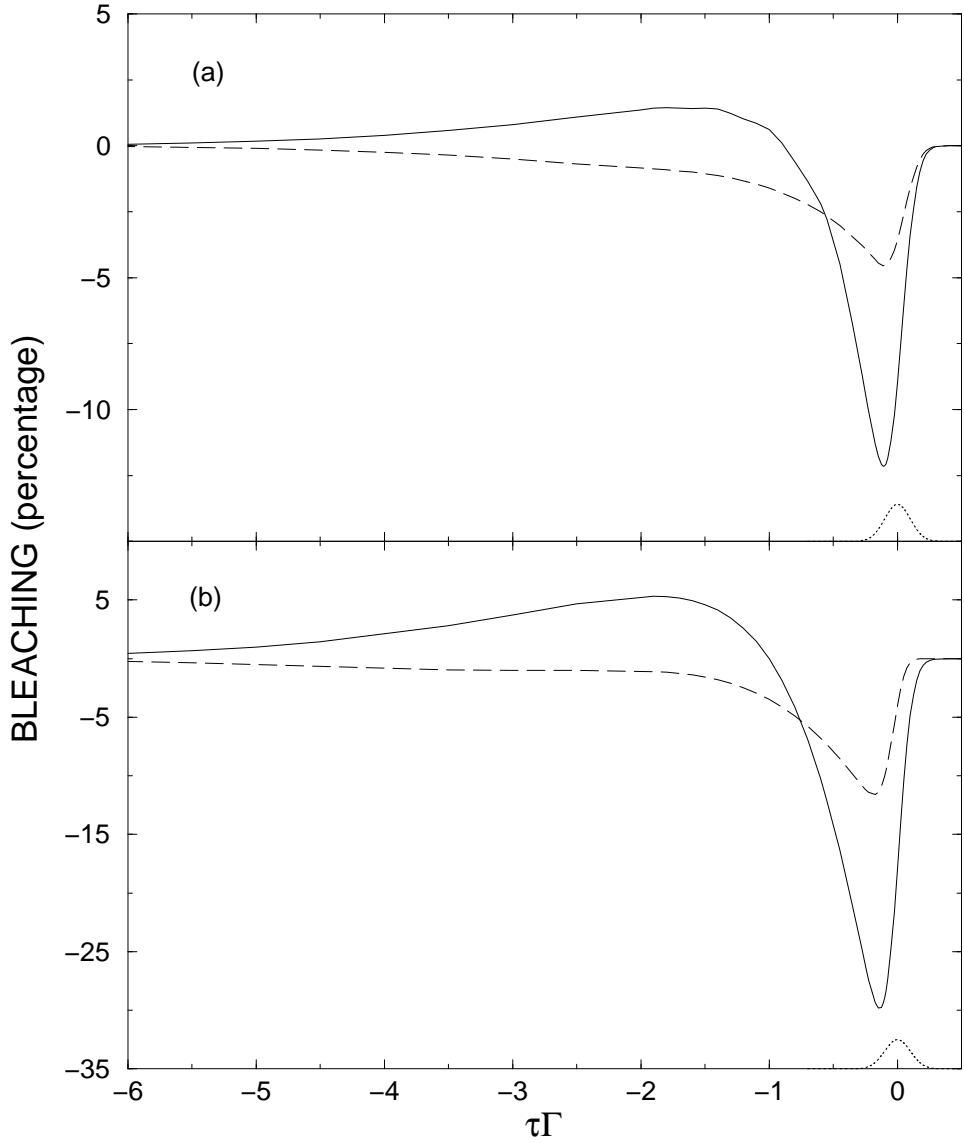}
\vspace{5mm}
\caption{
(a) Nonlinear absorption resonance bleaching 
evaluated at the instantaneous peak frequency 
as function of time delay 
for the FES (solid curve) compared to the HFA (dashed curve). The
curves were calculated with 
$g=0.4$, $\Gamma =0.1 E_{F}$, and $t_{p} = 2.0E_{F}^{-1}$.
The time--dependence of the pump pulse 
is also presented for comparison (dotted curve).
(b) Same calculated using the rigid band shift model, Eq.\ (\ref{nla}).
}
\label{FES-fig6}
\end{figure}

\begin{figure}
\vspace{-10mm}
\epsfxsize=5.2in
\epsffile{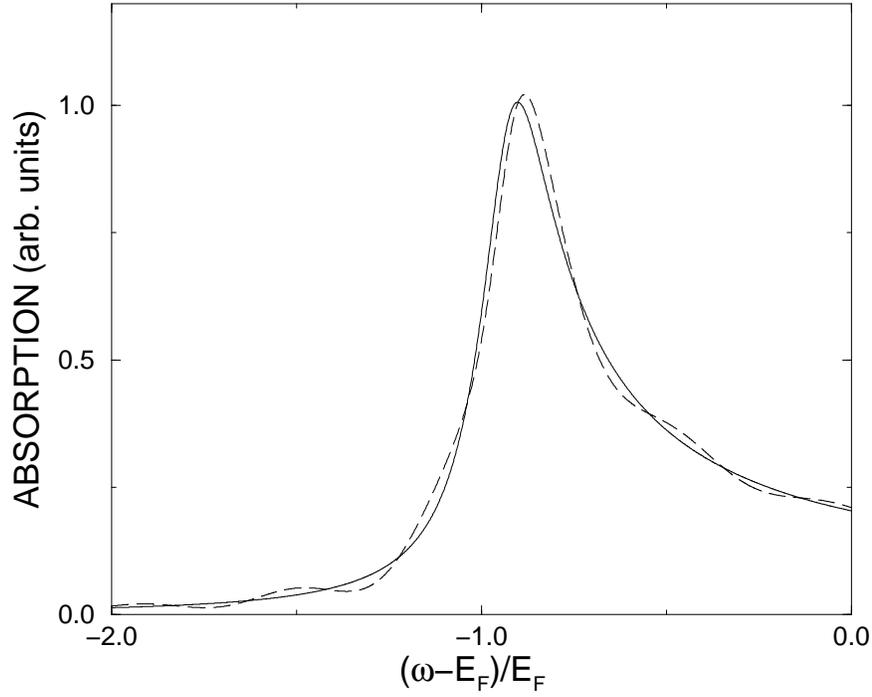}
\vspace{5mm}
\caption{
The enhancement of the nonlinear
absorption resonance of the  FES (dashed curve) vs. linear
absorption (solid curve) at long time delay $\tau=-1.5 \Gamma^{-1}$.
The curves were calculated with 
$g=0.4$, $\Gamma =0.1 E_{F}$, and $t_{p} = 2.0E_{F}^{-1}$.
}
\label{FES-fig7}
\end{figure}

\begin{figure}
\vspace{-10mm}
\epsfxsize=5.2in
\epsffile{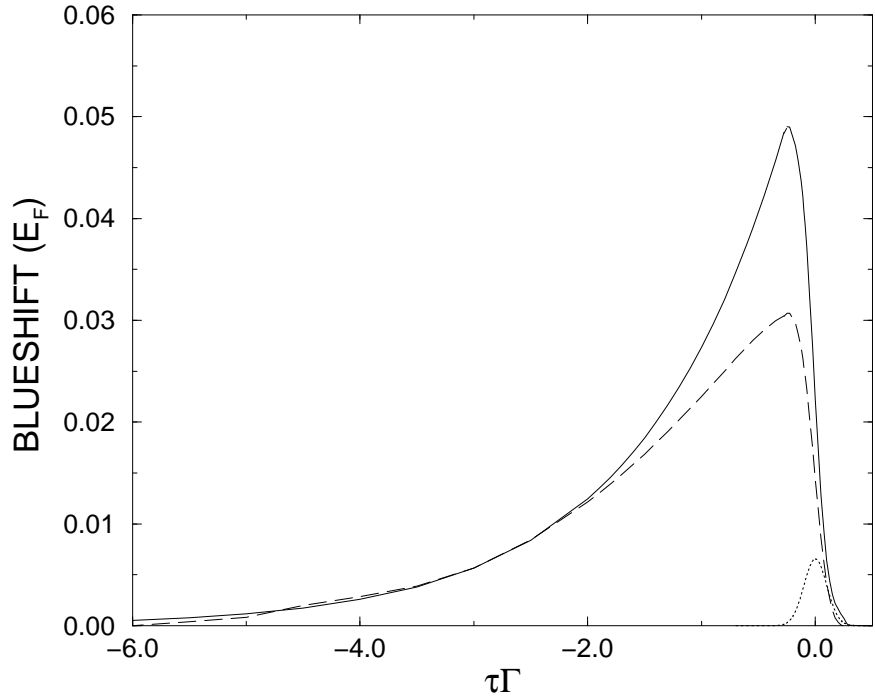}
\vspace{5mm}
\caption{
Nonlinear absorption resonance blueshift as function of time delay 
for the FES (solid curve) compared to the HFA (dashed curve).
The blueshift is significantly weaker for the HFA.
The curves were calculated with 
$g=0.4$, $\Gamma =0.1 E_{F}$, and $t_{p} = 2.0E_{F}^{-1}$.
The time--dependence of the pump pulse 
is also presented for comparison (dotted curve). 
}
\label{FES-fig8}
\end{figure}

Let us now turn to the role of the {\em e--h}
correlations. 
In Fig.\ \ref{FES-fig5} we compare the differential
transmission spectrum calculated by including the dynamical
FS response or within the HFA for long and short 
negative time delays. Note that, in  PP
spectroscopy,
the experimentally measured differential transmission is 
given by Eq.\ (\ref{diff-T})
and in 
the weak signal regime, it reproduces the pump--induced 
changes
in the probe absorption coefficient
$\alpha (\omega , \tau)$:
$DST(\omega,\tau)\propto-\Delta \alpha(\omega ,\tau)$.
Fig.\ \ref{FES-fig5}(a) shows 
the results obtained for a  long time delay,
$\tau = -1.5T_2= - 15.0E_F^{-1}$,
in which case frequency
domain oscillations are observed. 
These oscillations are
similar to those seen in undoped semiconductors
and two--level systems \cite{fluegel87}; however, their amplitude
in the FES case is reduced.  On the other hand, 
as shown in Fig.\ \ref{FES-fig5}(b),
for time delays comparable to the pulse duration,
$\tau = - 0.1 T_2 = - t_p/2 = - 1.0 E_F^{-1}$,
the main features are a blueshift and bleaching. 
In this case the {\em e--h} correlations lead to
a substantially larger width and asymmetric lineshape
of the differential transmission spectrum. This comes
from the different
response of the FES 
to the pump--induced dispersion renormalizations 
when the {\em e--h} correlations are accounted for. 
This is
more clearly seen in Fig.\ \ref{FES-fig6}, where the
magnitude of the resonance decrease, evaluated at
the peak frequency, 
is plotted
as a function of $\tau$.
Clearly, the bleaching of the FES peak
is substantially stronger when
the dynamical FS response is included 
than in the HFA case. Note  that for 
$|\tau| \sim \Gamma^{-1}$ the
FES resonance is actually enhanced by the pump, 
as can be seen more clearly in Fig.\ \ref{FES-fig7}. 
The time dependence of the resonance bleaching is
strikingly different in the two cases. In the HFA
case,  the $ |DST(\omega,\tau)|$ evaluated at the 
instantaneous  peak frequency decays over
a time scale $|\tau| \sim \Gamma^{-1}$, i.e. during
the dephasing time. This is similar to results
obtained for a two--level system with the same
effective parameters. On the other hand, the decay
of $|DST(\omega,\tau)|$ at the peak frequency
is much faster when we take into account the {\em e--h}
correlations. Note that the above results were obtained for
off--resonant  excitation. Under
resonant conditions, a spectral hole is produced. 
In Fig.\ \ref{FES-fig8} we compare the resonance blueshifts, 
evaluated at the peak frequency, as a function of
$\tau$. Again, a larger blueshift is predicted when 
the dynamical FS response 
is included. This suggests that
in the experiment of Ref.\  \cite{bre95}, where
similar blueshifts were observed in two quantum well
samples (one MDQW with a FES and one
undoped sample with a 2D-exciton) the effective
parameters were larger in the latter case, due to
the absence of screening and exciton--exciton
interaction effects. 

In order to gain qualitative understanding of the role of the 
{\em e--h} correlations, let us
for a moment neglect the momentum dependence of the
pump--renormalization of the band dispersion and
the phase space filling effects and  consider the bleaching 
caused by a  rigid semiconductor band shift $\Delta E_{g}(t)$,
obtained from the
pump--induced self--energies, Eqs. (\ref{ed})
and (\ref{hd}), evaluated  at the bottom of the
band (note that $\Delta E_{g}(t)$ lasts for the duration of pulse). 
Within this model, the pump
excitation has no effect on the {\em e--h}
scattering amplitude $s({\bf p},{\bf k},t)$
[see Eq.\ (\ref{de})]. It is thus convenient to factorize 
the effects of the rigid band shift on the {\em e--h} 
wavefunction $\Phi_{{\bf p}}(t,t')$:
\begin{equation}
\label{factor}
\Phi_{{\bf p}}(t,t')=
e^{- i \int_{t'}^{t} \Delta E_{g}(t'') dt''}
 \, \tilde{\Phi}_{\bf p}(t,t') \, .
\end{equation}
This relation is general and defines 
$\tilde{\Phi}_{{\bf p}}(t,t')$, which does {\em not} 
depend on $\Delta E_{g}(t)$. In the special 
case of a rigid shift, 
$\tilde{\Phi}_{{\bf p}}(t,t')$ coincides
with $\Phi_{{\bf p}}^0(t-t')$ describing  the
propagation of the probe--photoexcited {\em e--h} pair 
in the {\em absence} of the pump pulse.
By substituting into Eq.\ (\ref{factor}) the long--time
asymptotic expression \cite{mahan2} 
$\tilde{\Phi}_{{\bf p}}(t,t') 
=\Phi_{{\bf p}}^0(t-t')\propto[i(t-t') E_{F}]^{\beta-1}$ that
gives the linear absorption spectrum of the FES at 
$\omega \simeq E_F$, and substituting the resulting 
$\Phi_{{\bf p}}(t,t')$ into Eq.\ (\ref{p-p}), 
we obtain a simple  analytic expression
for the effect of a pump--induced
 rigid band shift on the nonlinear 
absorption spectrum: 
\begin{equation}\label{nla} 
\alpha(\omega) \propto {\rm Re}\,
\int_{\tau}^{\infty} dt e^{i(\omega +i\Gamma)
(t-\tau)- i \int_{\tau}^{t}
\Delta E_{g}(t') dt' } 
[i(t-\tau) E_{F}]^{\beta-1}.
\end{equation}
For $\Delta E_{g}(t)=0$ one, of course, recovers
the linear FES absorption in the vicinity of the 
Fermi edge \cite{mahan2}.
For $\beta=0$, 
 Eq.\ (\ref{nla}) gives the absorption of 
the non--interacting continuum. 

The physics of the FES can be seen from Eq.\ (\ref{nla}). 
For $\beta=1$, this gives a discrete 
Lorentzian peak corresponding to the HFA bound state.  
However, during the  optical transition, the {\em e--h} 
pair interacts with the FS electrons, leading to the 
readjustment of the FS density profile 
via the scattering of FS pairs.
This results in the  broadening of the discrete HFA bound state,  
which is governed  by the time evolution of the FS. 
Such time evolution is unadiabatic due to the low--energy FS pairs,
which leads to the characteristic power--law time dependence of the
broadening factor in Eq.\ (\ref{nla}). The interaction with the
FS-pairs determines the exponent, $0 \leq \beta \leq 1$,
of the latter, which leads to a non--Lorentzian
lineshape in the frequency domain and a
non--exponential decay in the time domain. 
A detailed discussion of the above physics and the analogy 
to phonon sidebands and collision broadening may be found in 
Ref.  \cite{mahan2}. 
However, the CCE must be used in order to calculate the spectrum at
{\em all} frequencies 
(and not just asymptotically close to $E_{F}$ as 
with Eq.\ (\ref{nla})) 
and, most importantly, 
to describe the non--equilibrium FS and {\em e--h} pair response to the
time--dependent increase in the 
effective mass/density of states, not included in 
Eq.\ (\ref{nla}).

\begin{figure}
\vspace{-10mm}
\epsfxsize=5.0in
\epsffile{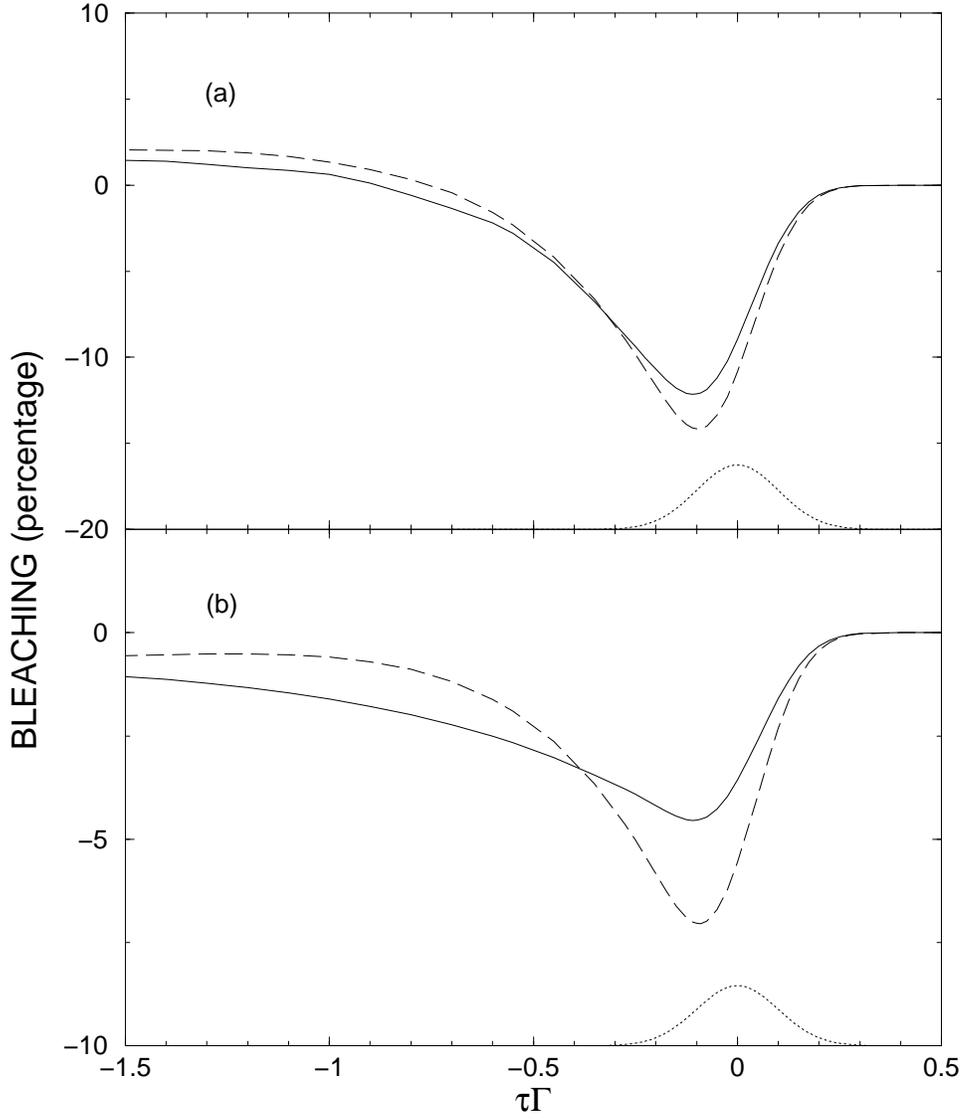}
\vspace{5mm}
\caption{
Resonance bleaching as function of time delay 
for the FES (a) compared to the HFA (b) for 
different strengths of the {\em e--h} interaction: 
g=0.4 (solid curve) and g=0.3 (dashed curve).
The curves were  calculated with
$\Gamma =0.1 E_{F}$, and $t_{p} = 2.0E_{F}^{-1}$.
The time--dependence of the pump pulse 
is also presented for comparison (dotted curve). 
}
\label{FES-fig9}
\end{figure}

The resonance bleaching obtained from 
Eq.\ (\ref{nla}) as a function of $\tau$, is shown in 
Fig.\ \ref{FES-fig6}(b) 
for $\beta=0.6$, corresponding to the value of the parameters used 
in Fig.\ \ref{FES-fig6}(a) ($g=0.4$), together with the HFA
result ($\beta =1$). Comparing Figs.\ \ref{FES-fig6}(a) and (b),
one can see that the rigid band shift approximation
qualitatively accounts for the dynamics, but that
there are strong  discrepancies (see vertical scales), 
whose origin is discussed below. Both the magnitude and the
time--dependence of the bleaching depends critically
on the value of $\beta$, which characterizes the 
interaction of the photoexcited {\em e--h} pair 
with the FS excitations. Because of such
coupling, many polarization components are
excited in the case of the continuum FES resonance, 
and it is their interference that governs the 
dynamics of the PP signal. 
Such interference 
is  also responsible for the resonance enhancement and
differential transmission oscillations at $\tau <0$
shown in Figs.\ \ref{FES-fig5} and\ \ref{FES-fig7}. 
As $\beta$ increases the
interference effects are suppressed because the  
energy  width of the continuum states contributing
to the FES narrows. In fact, this energy width is 
directly related to that of the linear absorption resonance.
This is clearly seen in Fig.\ \ref{FES-fig9} where we show
the effect of increasing $g$ on the dynamics of the
bleaching. It becomes more bound--state--like as, 
with increasing $g$, the FES resonance becomes narrower. 
On the other hand, in the HFA case,
the decay rate is $\sim T_2 = \Gamma^{-1}$,
i. e. it is independent on $g$, when $E_M$ becomes
smaller than $\Gamma$, while for $E_M \simeq \Gamma$, 
the contribution of the continuum states produce
a faster decay.

\begin{figure}
\vspace{-10mm}
\epsfxsize=5.0in
\epsffile{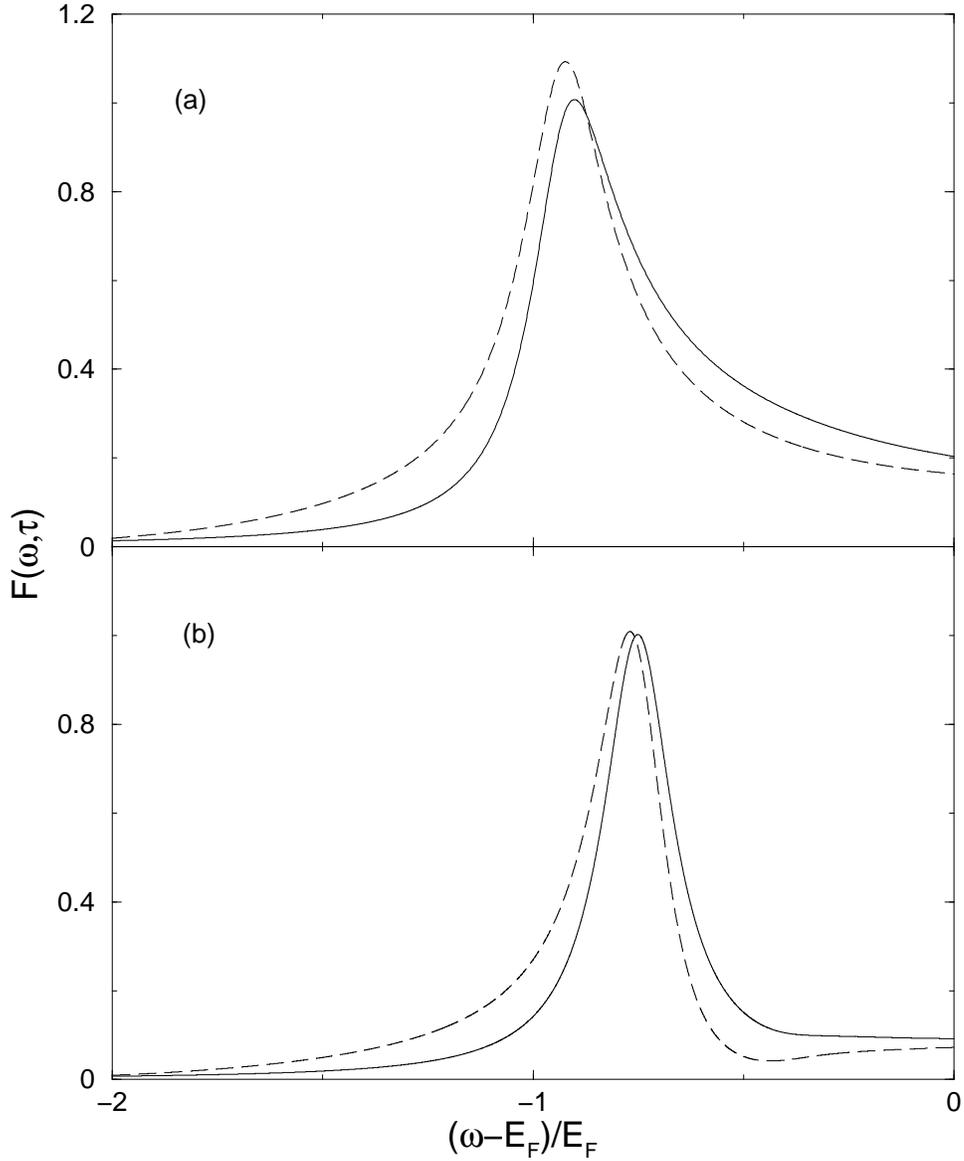}
\vspace{5mm}
\caption{The effect of the pump--induced renormalization of the band
dispersions on the {\em e--h} interactions. The 
function $F(\omega,\tau)$, given by Eq.\ (\ref{F}), for the
FES (a) compared to the HFA (b) in the presence (dashed curve)
and absence (solid curve) of the pump pulse.
The curves were calculated with 
$g=0.4$, $\Gamma =0.1 E_{F}$, and $t_{p} = 2.0E_{F}^{-1}$.
}
\label{FES-fig10}
\end{figure}

\begin{figure}
\vspace{-10mm}
\epsfxsize=5.0in
\epsffile{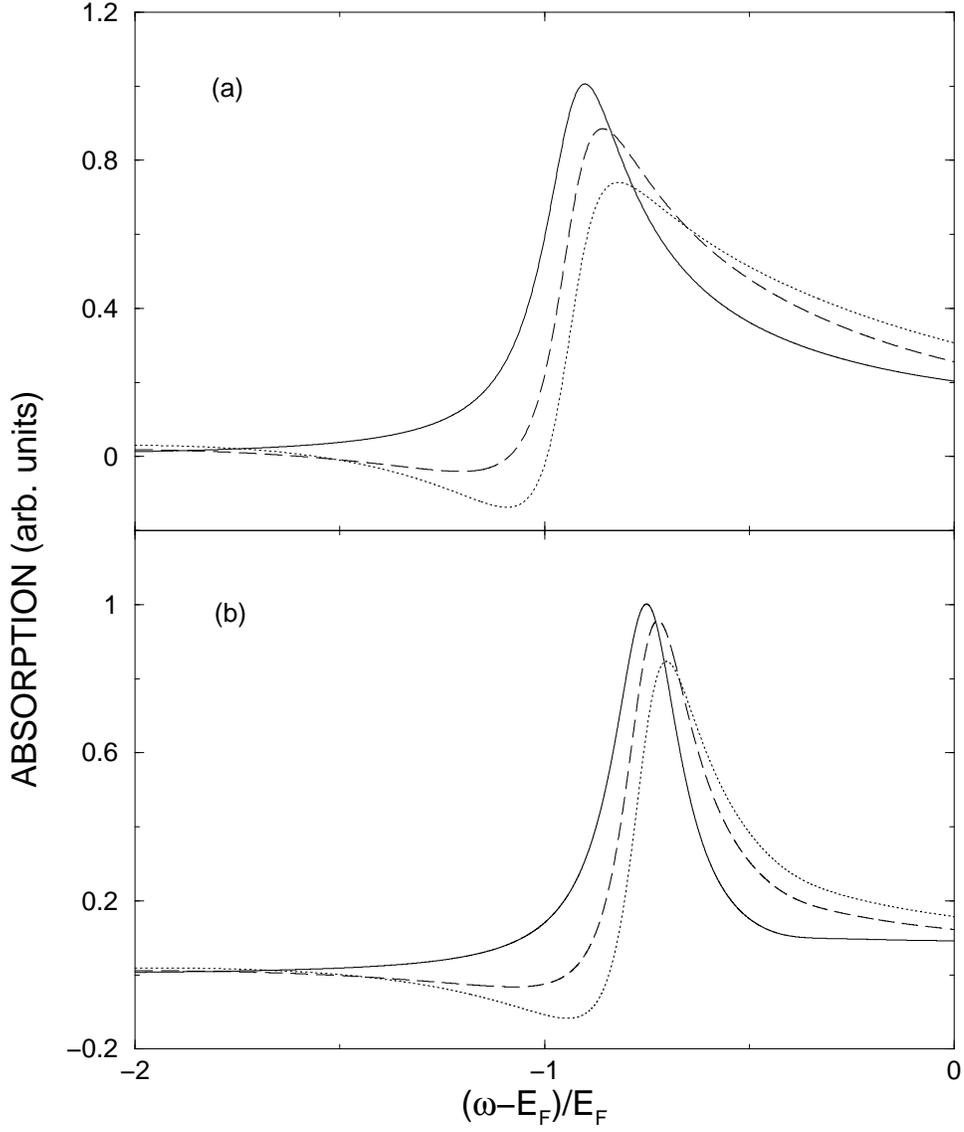}
\vspace{5mm}
\caption{
The effect of the pump--induced 
renormalization of the band dispersions on the 
resonance strength.
The nonlinear absorption spectrum for the 
FES (a) compared to the HFA (b). 
Solid curves: Linear absorption. 
Dashed curves: Nonlinear absorption. 
Dotted curves: 
Nonlinear absorption for a rigid band shift only. 
The curves were  calculated with 
$g=0.4$, $\Gamma =0.1 E_{F}$, and $t_{p} = 2.0E_{F}^{-1}$.
}
\label{FES-fig11}
\end{figure}

Although the transient rigid band shift approximation,
Eq. (\ref{nla}), explains some of the features of
the dynamics of the bleaching, it strongly overestimates
its magnitude. This is because Eq.\ (\ref{nla})
neglects the response of the many--body system
to the pump--induced renormalization of the 
band's dispersion.
 Such a  transient change in the 
dispersion, which can be viewed as an increase in the
density of states/effective mass
 for the duration of the pump,
is important because it results in an 
enhancement of  the {\em e--h} scattering. 
For example, 
in the case of monochromatic excitation, 
this leads to the change in the exponent $\beta$ 
of the broadening prefactor in the integrand of 
Eq.\ (\ref{nla}), as discussed Section \ref{sec:monochr}.  
For the short--pulse case, 
it is not possible to describe analytically
the effect of the pump on the {\em e--h} 
scattering processes, 
due to the non--equilibrium unadiabatic FS response.
The latter can however be described
in a simple way  with the numerical 
solution of Eqs.\ (\ref{de}) and (\ref{phip}), 
which is
presented in Figs.\ \ref{FES-fig10} 
and \ \ref{FES-fig11} and discussed below.

In order to show  the role 
of the pump--induced renormalization of the band 
dispersion in the presense of the 
dynamical FS response, 
we plot in Fig.\ \ref{FES-fig10} the function 
\begin{equation} \label{F}
F(\omega,\tau) 
= {\rm Im} \sum_{p > k_{F}}
\tilde{\Phi}_{{\bf p}}(\omega ,\tau)
\end{equation}
where $\tilde{\Phi}_{{\bf p}}(\omega ,\tau)$ is the
Fourier transform of $\tilde{\Phi}_{{\bf p}}(t,\tau)$ defined by
Eq.\ (\ref{factor}). Note that, in the presence of the
band dispersion renormalization, the
wave--function $\tilde{\Phi}_{{\bf p}}$ (which is  independent of 
$\Delta E_g(t)$) no longer coincides with $\Phi_{\bf p}^0$
as in Eq.\ (\ref{nla}). As can be seen in Fig.\ \ref{FES-fig10}(a), 
when the {\em e--h} correlations are taken into account,
 the  pump--induced redistribution
of oscillator strength between the states of the
continuum that contribute to the resonance manifests
itself as a dynamical redshift. This shift opposes
the rigid band blueshift $\Delta E_{g}(t)$ 
(when the latter is included). 
At the same time, the resonance strength is enhanced
significantly.  
The latter effect originates from the interplay between the 
transient increase in the effective mass/density of states 
of the photoexcited {\em e--h} pair
and the ``dressing'' of this pair  with the FS excitations
[described by the effective
potential $\tilde{V}({\bf p},t)$ in Eq.\ (\ref{phip})]. 
In contrast, such an oscillator strength enhancement is 
suppressed in the HFA (which neglects the {\em e--h} 
correlations), as seen in Fig.\ \ref{FES-fig10}(b), 
in which case  the main feature
is the redshift of the resonance
due to the  pump--induced increase of the binding energy
$E_M$ coming from the transient increase in the effective
mass. 

{In Fig.\ \ref{FES-fig11} the effect of the renormalization
of the band dispersion on the nonlinear absorption
spectrum is shown.} The optically--induced
increase in the {\em e--h} interactions
enhances significantly the strength of the FES
and compensates part of the bleaching induced by the
rigid band shift. A smaller enhancement is also seen
in the HFA, where the pump--induced increase in 
the binding energy $E_M$ competes with the effects
of the bandgap renormalization.

\subsection{Ultrafast Dynamics of the FES--Exciton Hybrid} 
\label{sec:hybrid-dynamics}

In this section we present the  results for
the evolution of the PP spectra of the FES--exciton
hybrid \cite{sha2000}. 
The spectra in Ref. \cite{sha2000} 
were obtained by the numerical solution 
of the coupled equations  (\ref{for phi})\ and\ (\ref{for s}), with
the time--dependent band dispersions 
$\epsilon_{ij\bf k}^c(t)$ and $\epsilon_{\bf k}^v(t)$.
In typical MDQW's, the intersubband 
scattering strength characterized by 
the screened potential $v_{12}$ is much smaller than 
the intrasubband scattering 
strength characterized by $v_{ii}$  
(a value $v_{12}/ v_{11}\simeq 0.2$ was deduced from the 
fit to the linear absorption spectrum in  \cite{mue90}).
As discussed in the previous section, 
in the absence of coupling ($v_{12}=0$),
the different nature of the exciton  and FES
leads to distinct dynamics under  ultrafast excitation.
In the presence of the coupling, one should expect 
new
effects coming from the interplay of this difference 
and the intersubband
scattering that hybridizes the two resonances. Indeed, 
it was demonstrated in Ref. \cite{sha2000} that, at negative time
delays, the PP spectrum undergoes a drastic transformation
due to a transient light--induced redistribution of the oscillator
strength between the FES and the exciton. We
discuss this at length in this section and show that
such a redistribution is a result of the dynamical FS response to
the pump pulse. In fact, the ultrafast PP spectra of
the FES--exciton hybrid can serve as an experimental test of the
difference between the FES and exciton dynamics.

\begin{figure}[hbt]
 \vspace{-10mm}

 \epsfxsize=3.0in
 \hspace{25mm}\epsffile{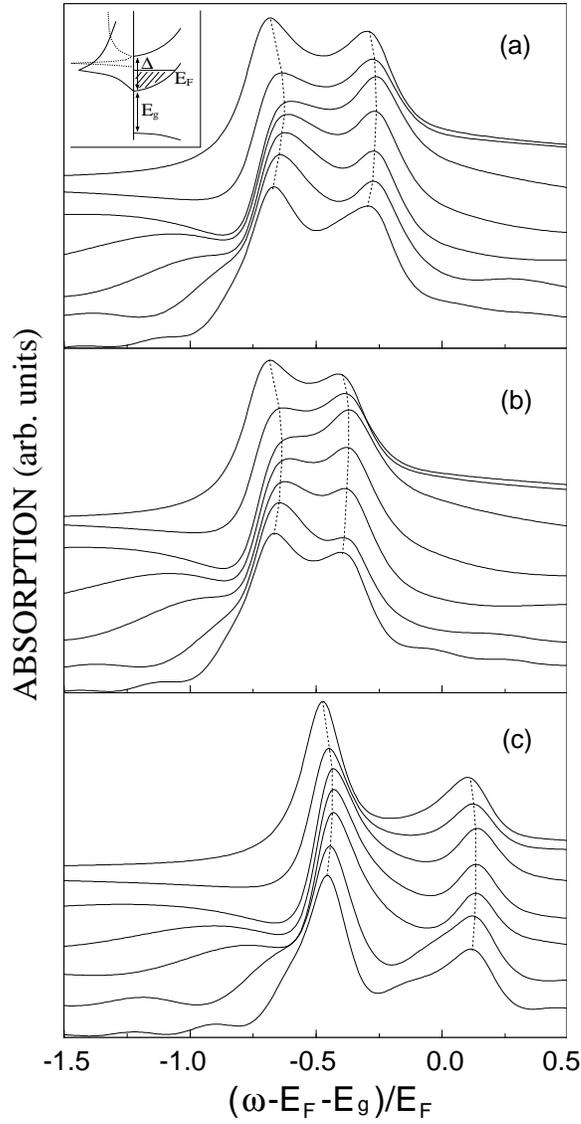}

 \vspace{5mm}
 \caption{
 (a) Calculated PP spectra with (a) $\Delta=1.7E_F$, 
 (b) $\Delta=1.6E_F$, and (c)  $\Delta=1.6E_F$ (HFA), 
 for short pump duration $t_pE_F/\hbar=2.0$, and negative time 
 delays $\tau\Gamma/\hbar=-2.0$ (lowest curve), $-1.2$,
 $-0.6$, $-0.4$, $-0.2$, 0, and linear absorption spectrum (upper curve).
 Inset: schematic plot of the energy spectrum of the two-subband
 QW (right) and absorption spectrum (left).
 }
\label{hybrid-fig}
 \end{figure}

The calculations in Ref. \cite{sha2000} 
 were performed at zero temperature for
below--resonant pump  with detuning $\Omega \sim E_F$
and duration $t_pE_F/\hbar=2.0$, and by 
adopting the  typical values of parameters
$v_{12}/v_{11}=0.2$, $\Gamma=0.1E_F$, and 
$v_{11} {\cal N} = 0.3$, ${\cal N}$ being the density of states, 
previously extracted from fits to the linear absorption spectra 
 \cite{mue90,mahbook}
($E_F\sim 15-20$ meV in typical GaAs/GaAlAs QW's \cite{kim92,bre95}).
Note, however, that similar results were also obtained for a broad 
range of parameter values. In Fig.\ \ref{hybrid-fig}(a) we plot the nonlinear
absorption spectra at different negative
time delays $\tau<0$. For better visibility, the curves are
shifted vertically with decreasing $|\tau|$ (the highest curve
represents the linear absorption spectrum). For the chosen value
of the subband separation 
$\Delta$,  the FES and excitonic components of the hybrid are
distinguishable in the linear absorption spectrum, with the FES peak
carrying larger oscillator strength. It can be seen that, at short
$\tau<0$, the oscillator strength is first transferred to the
exciton and then, with further increase in $|\tau|$, back to the FES. At
the same time, both peaks experience a blueshift, which is
larger for the FES than for the exciton peak because 
the ac--Stark effect   for the 
exciton is weaker due to the subband separation $\Delta$
and the correlation effects. 

The transient exchange of oscillator strength originates
from the different nature of the 
FES and exciton components of the hybrid. At negative
time delays, the time--evolution of the exciton is
governed by its dephasing time, which is essentially determined
by the homogeneous broadening $\Gamma$ (in doped systems the
exciton--exciton correlations do not play a significant role
due to the screening). The pump pulse first leads to a bleaching of
the exciton peak, which then recovers its strength at
$|\tau|\sim\hbar/\Gamma$. On the other hand, since the FES 
is a many--body  continuum resonance, (i) the bleaching
of the FES peak is stronger, and (ii) the polarization decay
of the FES is determined not by $\Gamma$, but by the scattering
with the low--lying FS excitations. 
This leads to much faster dynamics, roughly determined by the inverse 
Coulomb energy $E_{M}$  (see discussion in the previous 
section).
However, the time--evolution of the hybrid spectrum 
is not a simple superposition of the dynamics of its
components. Indeed, the pump-induced self-energies lead 
to the  flattening of the subbands or, to the first
approximation, to a time--dependent increase in the 
effective mass (and hence the density of states), which 
in turn increases the {\em e--h} scattering \cite{pri99}.  
Important is, however, that, due to the subband
separation and different nature of the resonances, such an
increase is stronger for the FES. Therefore, the effect of the pump
is to reduce the excitonic enhancement of the FES peak (coming from
the resonant scattering of the photoexcited electron by the exciton level)
as compared to the linear absorption case, 
resulting in the  oscillator strength transfer from the FES back
to exciton. In fact, such a transfer is strong even for smaller
$\Delta$ 
[see Fig.\ \ref{hybrid-fig}(b)]. It should be emphasized that the above
feature cannot be captured within the HFA.
Indeed, the latter approximates the FES by a bound 
state  and thus neglects the difference between the FES
and exciton dynamics originating from the unadiabatic response of
the FS to the change in the {\em e--h} correlations. 
This is demonstrated in Fig.\ \ref{hybrid-fig}(c) where we show 
the spectra obtained
without the FS dynamical response, i.e., by setting
$s_{ij}=0$. Although in that case both peaks show blue shift and
broadening,  there is no significant transfer of oscillator strength 

This concludes our discussion of the coherent nonlinear response of
the FES. In the rest of this article, we will review the 
role of size--dependent correlation effects, 
due to quasiparticle 
scattering via surface collective modes,  
on the ultrafast dynamics of the SP resonance 
in small metal nanoparticles. 

\clearpage

\section{ Quasiparticle 
Scattering with  Surface Collective Modes in 
Metal Nanoparticles}
\label{sec:decay}

\subsection{Electron--Electron Interactions in Metal Nanoparticles}
\label{sec:screen}

Before we proceed with  the ultrafast 
dynamics, we discuss in this section  the effect of the surface collective
excitations on the {\em e--e} interactions in a spherical metal
particle. In particular, we present a detailed derivation of the
dynamically screened Coulomb potential 
\cite{sha98,sha99} 
by generalizing
a method previously developed for calculations of 
local field corrections to the optical fields \cite{lus74}.

The potential 
$U(\omega;{\bf r},{\bf r}')$ at point ${\bf r}$ arising from 
an electron at point ${\bf r}'$ is determined by the equation \cite{mahbook}
\begin{eqnarray}\label{dyson}
U(\omega;{\bf r},{\bf r}')=u({\bf r}-{\bf r}')+
\int d{\bf r}_1 d{\bf r}_2u({\bf r}-{\bf r}_1)
\Pi(\omega;{\bf r}_1,{\bf r}_2)U(\omega;{\bf r}_2,{\bf r}'),
\end{eqnarray}
where $u({\bf r}-{\bf r}')=e^2|{\bf r}-{\bf r}'|^{-1}$ is the unscreened
Coulomb potential
and $\Pi(\omega;{\bf r}_1,{\bf r}_2)$ is the polarization
operator.
There are three contributions to $\Pi$, 
arising from the polarization of the 
conduction electrons, the $d$--electrons, and the  medium
surrounding the nanoparticles: 
$\Pi=\Pi_c+\Pi_d+\Pi_m$. It is useful to rewrite 
Eq.\ (\ref{dyson}) in the ``classical'' form 
\begin{equation}
\label{gauss}
\nabla\cdot({\bf E}+4\pi{\bf P})=4\pi e^2\delta({\bf r}-{\bf r}'),
\end{equation}
where ${\bf E}(\omega;{\bf r},{\bf r}')=
-\nabla U(\omega;{\bf r},{\bf r}')$ is the screened Coulomb
field
and 
${\bf P}={\bf P}_c+{\bf P}_d+{\bf P}_m$ is the electric
polarization vector, related to the potential $U$ as
\begin{equation}\label{delp}
\nabla {\bf P}(\omega;{\bf r},{\bf r}')
=-e^2\int d{\bf r}_1\Pi(\omega;{\bf r},{\bf r}_1)
U(\omega;{\bf r}_1,{\bf r}').
\end{equation}
In the random phase approximation (RPA),
the intraband polarization operator is given by  
\begin{eqnarray}\label{pol-sp}
\Pi_c(\omega;{\bf r},{\bf r}')=\sum_{\alpha\alpha'}
\frac{f(E_{\alpha}^c)-f(E_{\alpha'}^c)}
{E^c_{\alpha}-E^c_{\alpha'}+\omega+i0}
\psi^c_{\alpha}({\bf r})\psi^{c\ast}_{\alpha'}({\bf r})
\psi_{\alpha}^{c\ast}({\bf r}')\psi^c_{\alpha'}({\bf r}'),
\end{eqnarray}
where $E^c_{\alpha}$ and  $\psi^c_{\alpha}$ are the
single--electron
eigenenergies and eigenfunctions in the nanoparticle,
and $f(E)$ is the Fermi--Dirac 
 distribution (we
set $\hbar=1$). Since we are interested in
frequencies much larger than the single--particle level spacing, 
$\Pi_c(\omega)$ can be expanded in terms of
$1/\omega$. For the real part, 
$\Pi'_c(\omega)$, we obtain in the leading order
\begin{equation}\label{pol1-sp}
\Pi'_c(\omega;{\bf r},{\bf r}_1)=
-\frac{1}{m\omega^2}\nabla [n_c({\bf r})\nabla \delta({\bf r}-{\bf r}_1)],
\end{equation}
where $n_c({\bf r})$ is the conduction electron density. In the
following we assume, for simplicity, a step density  
profile,
$n_{c}({\bf r})=\bar{n}_{c}\,\theta(R-r)$, where 
$\bar{n}_{c}$ is the average density. The leading contribution to the
imaginary part, $\Pi''_c(\omega)$,  is proportional to
$\omega^{-3}$, 
so that   $\Pi''_c(\omega)\ll \Pi'_c(\omega)$.

By using Eqs.\ (\ref{pol1-sp}) and (\ref{delp}), we  obtain a familiar
expression for 
${\bf P}_c$ at high frequencies,
\begin{equation}\label{vecc}
{\bf P}_c({\omega};{\bf r},{\bf r}')
=\frac{e^2n_c({\bf r})}{m\omega^2}\nabla U(\omega;{\bf r},{\bf r}')
=\theta(R-r)\chi_c(\omega){\bf E}(\omega;{\bf r},{\bf r}'),
\end{equation} 
where $\chi_c(\omega)=-e^2\bar{n}_c/m\omega^2$ is the conduction
electron susceptibility. Note that, for a step density  
profile, ${\bf P}_c$ vanishes
outside the particle. The  $d$--band and dielectric
medium contributions to
${\bf P}$ are also given by similar relations,
\begin{eqnarray}
{\bf P}_d({\omega};{\bf r},{\bf r}')
=\theta(R-r)\chi_d(\omega)
{\bf E}(\omega;{\bf r},{\bf r}'),\label{vecd}
\\
{\bf P}_m({\omega};{\bf r},{\bf r}')
=\theta(r-R)\chi_m
{\bf E}(\omega;{\bf r},{\bf r}'),\label{vecm}
\end{eqnarray}
where $\chi_i=(\epsilon_i-1)/4\pi$, $i=d,m$ are the corresponding
susceptibilities and  the step functions account for the boundary
conditions \cite{boundary}. Using Eqs.\ (\ref{vecc})--(\ref{vecm}),
one can write a closed--form equation for $U(\omega;{\bf r},{\bf r}')$. 
Using Eq.\ (\ref{delp}), 
the second term of Eq.\ (\ref{dyson}) can be presented as
%
$
-e^{-2}\int d{\bf r}_1 u({\bf r}-{\bf r}_1)
\nabla \cdot {\bf P}(\omega;{\bf r}_1,{\bf r}').
$
%
Substituting the above expressions for ${\bf P}$, 
we then obtain after 
integration by parts
\begin{eqnarray}\label{self1}
\epsilon(\omega)
U(\omega;{\bf r},{\bf r}')=
&& \mbox{\hspace{-6mm}}
\frac{e^2}{|{\bf r}-{\bf r}'|}
+\int d{\bf r}_1 
\nabla_1\frac{1}{|{\bf r}-{\bf r}_1|}\cdot\nabla_1
[\theta(R-r)\chi(\omega)
\nonumber\\ && \mbox{\hspace{-6mm}}
+\theta(r-R)\chi_m]U(\omega;{\bf r}_1,{\bf r}')
\nonumber\\ && \mbox{\hspace{-6mm}}
+i\int d{\bf r}_1 d{\bf r}_2\frac{e^2}{|{\bf r}-{\bf r}_1|}
\Pi''_c(\omega;{\bf r}_1,{\bf r}_2)U(\omega;{\bf r}_2,{\bf r}'),
\end{eqnarray}
with
\begin{equation}
\label{eps}
\epsilon(\omega)\equiv 1+4\pi\chi(\omega)
=\epsilon_d(\omega)-\omega_p^2/\omega^2,
\end{equation}
$\omega_p^2=4\pi e^2\bar{n}_c/m$ being the
plasmon frequency in the conduction band.
The last term in the rhs of Eq.\ (\ref{self1}), proportional to
$\Pi''_c(\omega)$, can be regarded as a small correction.
To solve Eq.\ (\ref{self1}), we first eliminate the angular
dependence by expanding 
$U(\omega;{\bf r},{\bf r}')$ in spherical harmonics,
$Y_{LM}(\hat{\bf r})$, with coefficients $U_{LM}(\omega;r,r')$.
Using the corresponding expansion of 
$|{\bf r}-{\bf r}'|^{-1}$ with  coefficients 
$Q_{LM}(r,r')=\frac{4\pi}{2L+1}r^{-L-1}r'^{L}$ (for $r>r'$), 
we get the following equation for $U_{LM}(\omega;r,r')$:
\begin{eqnarray}\label{self2}
\epsilon(\omega)U_{LM}(\omega;r,r')=
&& \mbox{\hspace{-6mm}}
Q_{LM}(r,r')+
4\pi\left[\chi(\omega)-\chi_m\right]
\frac{L+1}{2L+1}\left({r\over R}\right)^L U_{LM}(\omega;R,r')
\nonumber\\ && \mbox{\hspace{-6mm}}
+ie^2\sum_{L'M'}\int dr_1dr_2r_1^2r_2^2Q_{LM}(r,r_1)
\Pi''_{LM,L'M'}(\omega;r_1,r_2)
\nonumber\\ && \mbox{\hspace{-6mm}}\times U_{L'M'}(\omega;r_2,r'),
\end{eqnarray}
where 
\begin{eqnarray}
\Pi''_{LM,L'M'}(\omega;r_1,r_2)=
\int d\hat{\bf r}_1d\hat{\bf r}_2 
Y_{LM}^{\ast}(\hat{\bf r}_1)
\Pi''_c(\omega;{\bf r}_1,{\bf r}_2)Y_{L'M'}(\hat{\bf r}_2),
\end{eqnarray}
are the coefficients of the multipole expansion of
$\Pi''_c(\omega;{\bf r}_1,{\bf r}_2)$. For $\Pi''_c=0$, 
the solution of 
Eq.\ (\ref{self2}) can be presented in the form
\begin{eqnarray}\label{sol}
U_{LM}(\omega;r,r')=a(\omega)e^2Q_{LM}(r,r')
+b(\omega)\frac{4\pi e^2}{2L+1}\frac{r^Lr'^L}{R^{2L+1}},
\end{eqnarray}
with frequency--dependent coefficients $a$ and $b$.
Since  $\Pi''_c(\omega)\ll\Pi'_c(\omega)$ for the relevant
frequencies, the solution of Eq.\ (\ref{self2}) in the
presence of the last term can be written in the
same form as Eq.\ (\ref{sol}),  but with modified $a(\omega)$ and
$b(\omega)$. Substituting  
Eq.\ (\ref{sol}) into Eq.\ (\ref{self2}), we obtain after lengthy
algebra in the lowest order in $\Pi''_c$
\begin{equation}
\label{ab}
a(\omega)=\epsilon^{-1}(\omega),
~~b(\omega)=\epsilon^{-1}_L(\omega)-\epsilon^{-1}(\omega),
\end{equation}
where
\begin{equation}\label{epsL}
\epsilon_L(\omega)={L\over 2L+1} 
\epsilon(\omega)
+{L+1\over 2L+1}\epsilon_m+i\epsilon''_{cL}(\omega),
\end{equation}
is the effective dielectric function, whose zero,
$\epsilon'_L(\omega_L)=0$, determines the frequency of the collective
surface excitation with angular momentum $L$ \cite{kreibig},
\begin{equation}
\label{omegaL}
\omega_L^2=\frac{L\omega_p^2}{L\epsilon'_d(\omega_L)+(L+1)\epsilon_m}.
\end{equation}
In Eq.\ (\ref{epsL}), $\epsilon''_{cL}(\omega)$ 
characterizes the damping of the $L$--pole collective mode by
single--particle excitations, and is given by
\begin{equation}\label{epscl}
\epsilon''_{cL}(\omega)=\frac{4\pi^2 e^2}{(2L+1)R^{2L+1}}
\sum_{\alpha\alpha'}|M^{LM}_{\alpha\alpha'}|^2
[f(E_{\alpha}^c)-f(E_{\alpha'}^c)]\delta(E^c_{\alpha}-E^c_{\alpha'}+\omega),
\end{equation}
where $M^{LM}_{\alpha\alpha'}$ are the matrix elements of 
$r^LY_{LM}(\hat{\bf r})$. 
Due to the momentum non--conservation in a nanoparticle, the matrix
elements are finite, which leads to the size--dependent width of
the $L$--pole mode  \cite{kaw66,lus74}:  
\begin{equation}
\label{gammaL}
\gamma_L=\frac{2L+1}{L}\frac{\omega^3}{\omega_p^2}\epsilon''_{cL}(\omega).
\end{equation}
For $\omega\sim\omega_L$, one can show that the width, 
$\gamma_L\sim v_F/R$, is independent of $\omega$.
Note that, in noble metal particles,
there is an additional {\em d}--electron contribution to 
the imaginary part of $\epsilon_L(\omega)$ at 
frequencies above the onset $\Delta$ of  the interband transitions.

Putting everything together, we arrive at the following expression
for the dynamically--screened interaction potential
in a nanoparticle:
\begin{eqnarray}\label{screen}
U(\omega;{\bf r},{\bf r}')
={u({\bf r}-{\bf r}')\over\epsilon(\omega)}
+ {e^2\over R}\sum_{LM}{4\pi\over 2L+1}
\frac{1}{\tilde{\epsilon}_L(\omega)}
\left({rr'\over R^2}\right)^L 
Y_{LM}(\hat{\bf r})Y^{\ast}_{LM}(\hat{\bf r}'),
\end{eqnarray}
with $\tilde{\epsilon}_L^{-1}(\omega)=
\epsilon_L^{-1}(\omega)-\epsilon^{-1}(\omega)$.
Equation (\ref{screen})
represents a generalization of the plasmon
pole approximation to spherical metal particles. 
The two contributions to the rhs originate from  two types of
dynamical screening. The first describes the usual  bulk-like
screening of the Coulomb potential by the electrons inside the
particle. The second contribution is a new effective interaction
induced by the {\em surface}: the potential of an electron  inside
the nanoparticle excites  high--frequency surface  collective modes,
which in turn act as image charges that interact with the second
electron. It should be emphasized that, unlike in the case of the
optical fields, the surface--induced dynamical screening
of the Coulomb potential is  {\em  size--dependent}.

Note that the excitation energies of the surface collective
modes are lower than the bulk plasmon energy, also given by 
Eq.\ (\ref{omegaL}) but with $\epsilon_m=0$.  
As discussed below, this opens up new channels of quasiparticle scattering.

\subsection{Conduction Electron Scattering}
\label{sec:cond}

In this section, we present the calculation of 
the  rates of electron scattering
in the conduction band accompanied by the emission of  surface
collective modes and discuss its possible experimental manifestations.
In the first order in the surface--induced potential, given by the
second term in the rhs of Eq.\ (\ref{screen}), the corresponding
scattering rate can be obtained from  the Matsubara
self--energy \cite{mahbook} 
\begin{eqnarray}\label{mselfc}
\Sigma_{\alpha}^{c}(i\omega)=
-\frac{1}{\beta}\sum_{i\omega'}\sum_{LM}\sum_{\alpha'}
\frac{4\pi e^2}{(2L+1)R^{2L+1}}
\frac{|M_{\alpha\alpha'}^{LM}|^2}{\tilde{\epsilon}_{L}(i\omega')}
G_{\alpha'}^{c}(i\omega'+i\omega),
\end{eqnarray}
where 
$G_{\alpha}^{c}=(i\omega-E_{\alpha}^c)^{-1}$ is the non-interacting
Green function of the conduction electron. Here the matrix elements
$M_{\alpha\alpha'}^{LM}$ are calculated with the one--electron 
wave functions $\psi_{\alpha}^c({\bf r})=R_{nl}(r)Y_{lm}(\hat{\bf r})$. 
Since $|\alpha\rangle$ and  $|\alpha'\rangle$ are the initial and
final states of the scattered electron,
the main contribution to  the $L$th term of the
angular momentum sum in Eq.\ (\ref{mselfc}) will come from electron
states with energy difference  
$E_{\alpha}-E_{\alpha'}\sim \omega_L$. Therefore, 
$M_{\alpha\alpha'}^{LM}$ can be expanded in terms of the small
parameter 
$E_0/|E_{\alpha}^c-E_{\alpha'}^c|\sim E_0/\omega_L$, where
$E_0=(2mR^2)^{-1}$ is the characteristic confinement energy.
The leading term can be
obtained by using the following procedure \cite{kaw66,lus74}.
We present $M_{\alpha\alpha'}^{LM}$ as
\begin{equation}
\label{melem}
M_{\alpha\alpha'}^{LM}
=\langle c,\alpha |r^LY_{LM}(\hat{\bf r})|c,\alpha'\rangle
=\frac{\langle c,\alpha |[H,[H,r^LY_{LM}(\hat{\bf r})]]|c,\alpha'\rangle}
{(E_{\alpha}^c-E_{\alpha'}^c)^2},
\end{equation}
where $H=H_0+V(r)$ is the Hamiltonian of an electron in a 
nanoparticle with confining potential $V(r)=V_0\theta(r-R)$. Since 
$[H,r^LY_{LM}(\hat{\bf r})]
=-\frac{1}{m}\nabla[r^LY_{LM}(\hat{\bf r})]\cdot\nabla$, the
numerator in Eq.\ (\ref{melem}) contains a term proportional to the
gradient of the confining potential, which peaks sharply at the 
surface. The corresponding 
contribution to the 
 matrix element describes the surface scattering of an
electron making 
the $L$--pole transition between the states $|c,\alpha\rangle$ and
$|c,\alpha'\rangle$, and gives the dominant  term of the
expansion. Thus, in the leading order in  
$|E_{\alpha}^c-E_{\alpha'}^c|^{-1}$, we obtain
\begin{eqnarray}
\label{melem1}
M_{\alpha\alpha'}^{LM}=
&& \mbox{\hspace{-6mm}}
\frac{\langle c,\alpha |
\nabla [r^LY_{LM}(\hat{\bf r})]\cdot\nabla V(r)
|c,\alpha'\rangle}
{m(E_{\alpha}^c-E_{\alpha'}^c)^2}
\nonumber\\ && \mbox{\hspace{-6mm}}
=\frac{LR^{L+1}}{m(E_{\alpha}^c-E_{\alpha'}^c)^2}
V_0R_{nl}(R)R_{n'l'}(R)\varphi_{lm,l'm'}^{LM},
\end{eqnarray}
with
$\varphi_{lm,l'm'}^{LM}=\int d\hat{\bf r}
Y_{lm}^{\ast}(\hat{\bf r})Y_{LM}(\hat{\bf r})Y_{l'm'}(\hat{\bf r})$.
Note that, for $L=1$, 
Eq.\ (\ref{melem1}) becomes exact.
For electron energies close to the Fermi level, $E_{nl}^c \sim E_F$,
the radial quantum  numbers are large, and the product 
$V_0 R_{nl}(R) R_{n'l'}(R)$ can be evaluated by using
semiclassical wave--functions.
In the limit $V_0\rightarrow \infty$, this product is given by 
 \cite{kaw66} 
$2\sqrt{E_{nl}^cE_{n'l'}^c}/R^3$, where
$E_{nl}^c=\pi^2(n+l/2)^2E_0$ is the 
electron eigenenergy for large $n$.
Substituting this expression into  
Eq.\ (\ref{melem1}) and then into  Eq.\ (\ref{mselfc}),
we obtain
\begin{eqnarray}\label{mselfc1}
\Sigma_{\alpha}^{c}(i\omega)=
 && \mbox{\hspace{-6mm}}
-\frac{1}{\beta}\sum_{i\omega'}\sum_{L}\sum_{n'l'}
C_{ll'}^{L}\,\frac{4\pi e^2 }{(2L+1)R}
\,\frac{E_{nl}^cE_{n'l'}^c}{(E_{nl}^c-E_{n'l'}^c)^4}
\nonumber\\ && \mbox{\hspace{-6mm}}\times
\frac{(4LE_0)^2}{\tilde{\epsilon}_{L}(i\omega')}
\,G_{\alpha'}^{c}(i\omega'+i\omega),
\end{eqnarray}
with
\begin{equation}
\label{C}
C_{ll'}^{L}=\sum_{M,m'}|\varphi_{lm,l'm'}^{LM}|^2
=\frac{(2L+1)(2l'+1)}{8\pi}\int_{-1}^{1}dxP_l(x)P_L(x)P_{l'}(x),
\end{equation}
where $P_l(x)$ are Legendre polynomials; we used properties
of the spherical harmonics in the derivation of Eq.\ (\ref{C}).
For  $E_{nl}^c\sim E_F$, the typical angular momenta are large, 
$l\sim k_{_F}R\gg 1$, and one can use the large--$l$ asymptotics of
$P_l$; for the low multipoles of interest, $L\ll l$, the
integral in Eq.\ (\ref{C}) can be
approximated by $\frac{2}{2l'+1}\delta_{ll'}$.
After performing the Matsubara summation, we
obtain for the imaginary part of the self--energy that 
determines the electron scattering rate 
\begin{eqnarray}
\label{imselfc}
\mbox{Im}\Sigma_{\alpha}^{c}(\omega)=
-\frac{16e^2}{R}E_0^2\sum_L\int dE \, g_l(E)
\frac{L^2E E_{\alpha}^c}{(E_{\alpha}^c-E)^4}
\mbox{Im}\frac{N(E-\omega)+f(E)}
{\tilde{\epsilon}_{L}(E-\omega)},
\end{eqnarray}
where  $N(E)$ is the Bose distribution and $g_l(E)$ is the
density of states of a conduction electron with angular momentum
$l$,
\begin{equation}
\label{gl}
g_l(E)=2\sum_{n}\delta(E_{nl}^c-E)\simeq
\frac{R}{\pi}\sqrt{\frac{2m}{E}},
\end{equation}
where we replaced the sum over $n$
by an integral (the factor of 2 accounts for spin).

Each term in the sum in the rhs of Eq.\ (\ref{imselfc}) represents
a channel of electron scattering mediated by a 
collective surface mode with angular momentum $L$. For low $L$,
the difference between the energies  of  modes
with successive values of $L$ is larger than their
widths, so the different channels are well separated. 
Note that since all $\omega_L$ are smaller than the
frequency of the bulk plasmon, one can replace 
$\tilde{\epsilon}_{L}(\omega)$ by  $\epsilon_{L}(\omega)$  in the
integrand of Eq.\ (\ref{imselfc}) for frequencies 
$\omega \sim \omega_{L}$.

Consider now the $L=1$ term in Eq.\ (\ref{imselfc}), which
describes the SP--mediated scattering channel. 
The main contribution to the integral comes from the SP pole in 
$\epsilon_1^{-1}(\omega)=3\epsilon_{s}^{-1}(\omega)$, where
$\epsilon_{s}(\omega)$ is the same as in Eq.~(\ref{epseff}). To
evaluate the integral in Eq.\ (\ref{imselfc}), we can in the first
approximation replace $\mbox{Im}\epsilon_{s}^{-1}(\omega)$ by  a
Lorentzian,
\begin{eqnarray}
\label{lor}
\mbox{Im}\,\frac{1}{\epsilon_{s}(\omega)}=
&&\mbox{\hspace{-6mm}}
-\frac{\gamma_s\omega_p^2/\omega^3+\epsilon''_d(\omega)}
{[\epsilon'(\omega)+2\epsilon_m]^2+
[\gamma_s\omega_p^2/\omega^3+\epsilon''_d(\omega)]^2}
\nonumber\\ &&\mbox{\hspace{-6mm}}
\simeq
-\frac{\omega_s^2}{\epsilon'_d(\omega_s)+2\epsilon_{m}}
\,
\frac{\omega_s\gamma}
{(\omega^2-\omega_s^2)^2+\omega_s^2\gamma^2},
\end{eqnarray}
where 
$\omega_s\equiv \omega_1=\omega_p/\sqrt{\epsilon'_d(\omega_s)+2\epsilon_m}$
and $\gamma=\gamma_s+\omega_s\epsilon''_d(\omega_s)$ are the SP
frequency and width, respectively. For  typical widths
$\gamma\ll\omega_s$, the integral in Eq.\ (\ref{imselfc}) can be easily
evaluated, yielding
\begin{equation}
\label{imselfc1}
\mbox{Im}\Sigma_{\alpha}^{c}(\omega)=
-\frac{24e^2\omega_sE_0^2}{\epsilon'_d(\omega_s)+2\epsilon_m}
\frac{E_{\alpha}^c\sqrt{2m(\omega-\omega_s)}}
{(\omega-E_{\alpha}^c-\omega_s)^4}
[1-f(\omega-\omega_s)]. 
\end{equation}
Using the relation
$e^2k_F[\epsilon'_d(\omega_s)+2\epsilon_m]^{-1}=3\pi\omega_s^2/8E_F$,
the SP--mediated scattering rate, 
$\gamma_e^s(E_{\alpha}^c)=-\mbox{Im}\Sigma_{\alpha}^{c}(E_{\alpha}^c)$,
takes the form  
\begin{equation}
\label{gammae}
\gamma_e^s(E)=9\pi\frac{E_0^2}{\omega_s}
\frac{E}{E_F}\left(\frac{E-\omega_s}{E_F}\right)^{1/2}
[1-f(E-\omega_s)].
\end{equation}
Recalling that $E_0=(2mR^2)^{-1}$, we see that the scattering rate of a 
conduction electron is {\em size--dependent}:
$\gamma_e^s\propto R^{-4}$. At $E=E_F+\omega_s$, the scattering
rate jumps to the value $9\pi(1+\omega_s/E_F)E_0^2/\omega_s$,
and then {\em increases} with energy as $E^{3/2}$ 
(for $\omega_s\ll E_F$). This should be 
contrasted with the usual (bulk) plasmon--mediated  scattering,
originating from the first term in Eq.\ (\ref{screen}), in which
case the rate decreases as $E^{-1/2}$ above the onset \cite{mahbook}.

Note that the total electron scattering rate is the sum, 
$\gamma_e+\gamma_e^s$, of the SP--mediated ($\gamma_e^s$) and the
bulk--like ($\gamma_e$) scattering rates. In order to be observable,
the former should exceed the latter. The typical size at which
$\gamma_e^s$ becomes important can be estimated by equating
$\gamma_e^s$ and the Fermi liquid {\em e--e} scattering
rate \cite{pines},   
$\gamma_e(E)=\frac{\pi^2q_{_{TF}}}{16k_{_F}}\frac{(E-E_F)^2}{E_F}$.
For energies $E\sim E_F+\omega_s$, the two rates
become comparable for
\begin{equation}
\label{size}
(k_{_F}R)^2\simeq
12\frac{E_F}{\omega_s}\left(1+\frac{E_F}{\omega_s}\right)^{1/2}
\left(\frac{k_{_F}}{\pi q_{_{TF}}}\right)^{1/2}.
\end{equation}
In the case of a copper 
nanoparticle with $\omega_s\simeq 2.2$ eV, we obtain 
$k_{_F}R\simeq 8$, which corresponds  to a  radius of $R\simeq 3$ nm. 
At the same time, in this energy range, the width $\gamma_e^s$ exceeds
the mean level spacing $\delta$, so that the energy spectrum is
still continuous. The strong size dependence of $\gamma_e^s$
indicates that, although $\gamma_e^s$ increases with energy slower
than $\gamma_e$, the SP--mediated scattering should dominate for
nanometer--sized particles. Note also that the size and energy
dependences of the scattering in the different channels are similar:
the rate of scattering via the $L$th channel is given by
Eq.\ (\ref{gammae}) with $\omega_s$ replaced by  $\omega_L$, 
Eq.\ (\ref{omegaL}), and the numerical factor 9 replaced by
$3L(2L+1)$. 

Concluding this section, we have shown that the SP--mediated
scattering is the dominant scattering mechanism of conduction
electrons in nanometer-sized nanoparticles for energies larger than
$\omega_s$ but smaller than $\omega_p$. The scattering rate in
the $L$th channel, $\gamma_e^L$, increases with energy, in sharp
contrast with the bulk--plasmon--mediated scattering rate.
The total scattering rate as a function
of energy represents a series of steps at $E=\omega_L$,  on
top of a smooth energy increase. We expect that this new  effect
should be observable experimentally by measuring {\em e--e}
scattering rate in size--selected cluster beams in time--resolved
two--photon photoemission spectrum \cite{og97}.

\subsection{$d$--Band Hole Scattering}

\label{sec:d-band}

We now turn to the interband processes in noble metal particles 
and consider the scattering of a $d$--hole into the conduction band.
The corresponding surface--induced
potential, given by the  $L$th term in Eq.\ (\ref{screen}), has the form 
\begin{eqnarray}
\label{spscreen}
U_{LM}(\omega;{\bf r},{\bf r}')={4\pi\over 2L+1}
\frac{1}{\epsilon_L(\omega)}{e^2\over R}
\left({rr'\over R^2}\right)^L 
Y_{LM}(\hat{\bf r})Y^{\ast}_{LM}(\hat{\bf r}'),
\end{eqnarray}
where $\epsilon_L(\omega)$ is given by Eq.\ (\ref{epsL}) 
With this potential, the $d$--hole self--energy is given by
\begin{eqnarray}\label{mselfd}
\Sigma_{\alpha}^{d}(i\omega)=
-{e^2\over R^{2L+1}}\sum_{\alpha'}|{\tilde M}_{\alpha\alpha'}^{LM}|^2
{1\over \beta}\sum_{i\omega'}
\frac{G_{\alpha'}^{c}(i\omega'+i\omega)}{\epsilon_{L}(i\omega')},
\end{eqnarray}
where ${\tilde M}_{\alpha\alpha'}^{LM}=
\langle c,\alpha |r^LY_{LM}(\hat{\bf r})|d,\alpha'\rangle$
is the {\em interband} transition matrix element 
[compare with  Eq.\ (\ref{melem})].
Since the final state energies in the conduction band are high 
(in the case of interest here, they are close to
the Fermi level), the matrix element can be 
approximated by a bulk--like expression
${\tilde M}_{\alpha\alpha'}^{LM}=
\delta_{\alpha\alpha'}{\tilde M}_{\alpha\alpha}^{LM}$,
the corrections due 
to  surface scattering 
being suppressed by a factor of $(k_{_F}R)^{-1}\ll 1$. 

The largest contribution to the self energy Eq.\ (\ref{mselfd})
comes from the dipole channel, $L=1$, mediated by the SP. In this case, 
after performing the 
frequency summation, we obtain for Im$\Sigma_{\alpha}^{d}$ 

\begin{eqnarray}\label{imselfd}
{\rm Im}\Sigma_{\alpha}^{d}(\omega)
=-{9e^2 \mu^2\over m^2(E^{cd}_{\alpha})^2R^3}
\mbox{Im}\frac{N(E^c_{\alpha}-\omega)+f(E^c_{\alpha})}
{\epsilon_{s}(E^{c}_{\alpha}-\omega)},
\end{eqnarray} 
where $E^{cd}_{\alpha}=E^{c}_{\alpha}-E^{d}_{\alpha}$ and $\mu$ is
the interband dipole matrix element \cite{sha98}.
We see that the scattering rate of a $d$-hole with energy $E^{d}_{\alpha}$, 
$\gamma_{h}^s(E^{d}_{\alpha})=\mbox{Im}\Sigma_{\alpha}^{d}(E^{d}_{\alpha})$,
has a strong $R^{-3}$ dependence on the nanoparticle size, which is,
however, different from that of the intraband scattering, 
Eq.\ (\ref{gammae}). 

The most important difference between the interband
and the intraband SP--mediated scattering rates
lies in their dependence on energy.
Since the surface--induced  potential, Eq.\ (\ref{spscreen}),
only allows for vertical (dipole) interband single--particle
excitations, the phase space available for the scattering of a
$d$--hole with energy $E^{d}_{\alpha}$ is restricted to a 
{\em single} final state in the conduction band, with energy
$E^{c}_{\alpha}$. As a result of this restriction, the $d$--hole
scattering rate, $\gamma_{h}^s(E^{d}_{\alpha})$, exhibits a {\em peak} 
as the difference between the energies of final and initial states,  
$E^{cd}_{\alpha}=E^{c}_{\alpha}-E^{d}_{\alpha}$,
approaches the SP frequency $\omega_{s}$ 
[see Eq.\ (\ref{imselfd})].  
In contrast, the energy dependence of $\gamma_e^s$ is smooth due
the larger phase space available for scattering
within the conduction band. 
This leads to the additional integral over final state energies in
Eq.\ (\ref{imselfc}), which smears out the SP resonant enhancement
of the intraband scattering. 

As we show 
later,
the fact that the  scattering rate of a $d$--hole is 
dominated by the SP resonance affects strongly the nonlinear
optical dynamics in small nanoparticles. 
This is the case, in particular, when the SP frequency,
$\omega_{s}$, is close to the onset of interband transitions,
$\Delta$, as, e.g., in copper and gold  
nanoparticles \cite{kreibig,big95,per97}.
Indeed, if the optical pulse excites an  {\em e--h} pair with
excitation energy $\omega$ close to $\Delta$, the $d$--hole can
subsequently scatter into the conduction band by emitting a
SP. According to Eq.~(\ref{imselfd}), for $\omega\sim\omega_{s}$, 
such a scattering process should be resonantly enhanced. In order to
have an observable effect on the absorption spectrum,  the 
scattering rate of the photoexcited $d$--hole should be comparable
(or larger) than that of the photoexcited electron.
Close to $E_F$, the electron scattering in the conduction band
comes from a two--quasiparticle process; the corresponding
rate in noble--metals is estimated 
as \cite{og97} $\gamma_{e}\sim 10^{-2}$ eV. 
If one assumes the bulk value for $\mu$ ($2\mu^2/m\sim 1$ eV near
the L-point \cite{eir62}), then $\gamma_{h}^s$ exceeds $\gamma_{e}$ for 
$R< 2.5$\ nm. In fact, one would expect
that, in nanoparticles, $\mu$ is larger than in the bulk due to the
localization of the conduction electron  wave--functions \cite{kreibig}.

\section{Surface Plasmon Nonlinear Optical dynamics}
\label{sec:optics}

In this section, we  discuss the effect of the SP--mediated
{\em interband} scattering on the ultrafast optical dynamics
in noble metal nanoparticles.
We are interested in the situation when the hot electron
distribution has already thermalized and the electron gas is cooling
to the lattice (stage {\bf III}). 
In this case the transient response of a
nanoparticle can be described by the time--dependent absorption
coefficient $\alpha(\omega,t)$, given by  Eq.\ (\ref{absor}) with
time--dependent temperature. In noble--metal
particles, the temperature dependence of $\alpha$ originates
from two different sources. First is the phonon--induced
correction to $\gamma_s$, 
 which is proportional to the {\em lattice}
temperature $T_l(t)$. 
As mentioned in section \ref{sec:nano}, 
for small
nanoparticles this effect is relatively weak.
Note that, as  discussed e.g. in Ref. \cite{inou98}, 
the contribution to  $\gamma_s$ coming from the 
{\em e--e} interaction
\cite{fly98,halas1,halas2,per97,gurzhi,kaveh}, which depends 
on the electron temperature, plays a minor role when 
the interband transitions are resonantly excited. 
 Second, near the onset
of the interband transitions, $\Delta$, the absorption coefficient
depends on the {\em electron} temperature $T(t)$ via the interband
dielectric function $\epsilon_d(\omega)$ 
[see Eqs.\ (\ref{absor}) and (\ref{epseff})].
In fact, in copper or gold  nanoparticles, $\omega_s$ can be tuned close to
$\Delta$, so   the SP damping by {\em interband} {\em e--h}
excitations leads to an additional broadening of the absorption
peak \cite{kreibig}. In this case, it is the temperature dependence of
$\epsilon_d(\omega)$ that dominates the pump--probe 
dynamics. Below we show that, near the SP resonance, both the
temperature and frequency dependence of
$\epsilon_d(\omega)=1+4\pi\chi_d(\omega)$ are strongly affected by
the SP--mediated interband scattering.

We start with the RPA expression for 
the interband susceptibility \cite{mahbook}, 
$\chi_d(i\omega)=\tilde{\chi}_d(i\omega)+\tilde{\chi}_d(-i\omega)$,
\begin{equation}\label{susc}
\tilde{\chi}_d(i\omega)=
-\sum_{\alpha}{e^2\mu^2\over m^2(E^{cd}_{\alpha})^2}
{1\over \beta}\sum_{i\omega'}
G_{\alpha}^{d}(i\omega')
G_{\alpha}^{c}(i\omega'+i\omega),
\end{equation}
where $G_{\alpha}^{d}(i\omega')$ is the Green function of a
$d$--electron. With the $d$-band fully occupied,
the only allowed SP--mediated interband scattering 
is that of the  $d$--hole. We assume here, for simplicity, a
dispersionless $d$--band with energy $E^d$.
Substituting  $G_{\alpha}^{d}(i\omega')= 
[i\omega'-E^{d}+E_F-\Sigma_{\alpha}^{d}(i\omega')]^{-1}$,
with $\Sigma_{\alpha}^{d}(i\omega)$ given by Eq.~(\ref{mselfd}),
and  performing the frequency summation, we  obtain
\begin{equation}\label{interband}
\tilde{\chi}_d(\omega)=
{e^2 \mu^2\over m^2}\int {dE^c\, g(E^c)\over (E^{cd})^2}
{f(E^c)-1\over \omega-E^{cd}+i\gamma_h^s(\omega,E^c)},
\end{equation}
where $g(E^c)$ is the density of states of
conduction electrons. The scattering rate of a $d$-hole, 
$\gamma_h^s(\omega,E^c)={\rm Im}\Sigma^{d}(E^c-\omega)$,
is  obtained from  Eq.\ (\ref{imselfd}) with $E_d=E^c-\omega$:
\begin{equation}\label{gamhole}
\gamma_h^s(\omega,E^c)
=-{9e^2 \mu^2\over m^2(E^{cd})^2R^3}
f(E^c)\mbox{Im}{1\over\epsilon_{s}(\omega)},
\end{equation}
where $N(\omega)$ is negligible for frequencies
$\omega\sim\omega_s\gg k_BT$ \cite{sha98}.
The rate $\gamma_h^s(\omega,E^c)$ exhibits a 
sharp peak  as a function of the frequency of the 
probe optical field. The reason for this is that the scattering rate
of a $d$--hole with energy $E$ depends explicitly on the  {\em
difference} between the final and initial states, $E^c-E$,
as discussed above; 
therefore, for a $d$--hole
with energy $E=E^c-\omega$, the dependence on the final 
state energy, $E^c$, cancels out in $\epsilon_s(E^c-E)$ 
[see Eq.\ (\ref{imselfd})].
In other words, the optically--excited 
$d$--hole scatters resonantly into the conduction band as 
$\omega$ approaches $\omega_s$. It is important to note that 
$\gamma_h^s(\omega,E^c)$ is, in fact, proportional to the 
absorption coefficient $\alpha(\omega)$ [see Eq.\ (\ref{absor})].
Therefore, $\alpha$ and $\gamma_h^s$ should be calculated 
self--consistently from
Eqs.\ (\ref{absor}), (\ref{epseff}), (\ref{interband}), and
(\ref{gamhole}).

It should be emphasized that  the effect of $\gamma_h^s$ on
$\epsilon_d''(\omega)$  increases with temperature.
Indeed, it can be seen from Eq.\ (\ref{gamhole}) that the value of
$\gamma_h^s$ is appreciable only if $E^c-E_F > k_B T$. Since
the main contribution to $\tilde{\chi}''_d(\omega)$ comes from
energies $E^c\sim \omega-\Delta+E_F $, the effect of 
$d$--hole scattering on the absorption becomes important only for
elevated electron temperatures: $k_B T> \omega_{s}-\Delta$. 
As a result, near the SP resonance, the time evolution of the
differential absorption, which is governed by the temperature
dependence of $\alpha$, becomes strongly {\em size--dependent}.

The numerical results discussed  below are taken from 
Refs.  \cite{sha98,sha99}. 
In the experiment of Bigot {\em et. al.} \cite{big95},
the pump--probe measurements were
performed on  $R\simeq 2.5$ nm copper nanoparticles. The SP frequency,  
$\omega_{s}\simeq 2.22$ eV, was
slightly above the onset of the
interband transitions, $\Delta\simeq 2.18$ eV.
In order to describe  the time--evolution
of the differential absorption spectra, one first needs to
determine the time--dependence of the electron temperature, $T(t)$,
due to the relaxation of the electron gas to the lattice. 
For this, a simple 
two--temperature model is employed,
defined by heat equations for 
$T(t)$ and the lattice temperature $T_l(t)$:
\begin{eqnarray}
\label{TT}
C(T)\frac{\partial T}{\partial t}&=&-G(T-T_l),
\nonumber\\ 
C_l\frac{\partial T_l}{\partial t}&=&G(T-T_l),
\end{eqnarray}
where $C(T)=\gamma T$ and $C_l$ are the electron and lattice heat
capacities, respectively, and $G$ is 
the {\em e--p} coupling \cite{tt,anis75,allen}.
The parameter values used in Refs. \cite{sha98,sha99} 
were $G=3.5\times 10^{16}$ Wm$^{-3}$K$^{-1}$, 
$\gamma=70$ Jm$^{-3}$K$^{-2}$, and $C_l=3.5$ Jm$^{-3}$K$^{-1}$,
and the initial condition was taken as $T_0=1000$ K.
The time--dependent absorption
coefficient $\alpha(\omega,t)$ was calculated self--consistently; 
the differential transmission is proportional to
$\alpha_r(\omega)-\alpha(\omega,t)$, where $\alpha_r(\omega)$ was
calculated at the room temperature.

\begin{figure}
\vspace{-10mm}

\epsfxsize=2.7in
\hspace{-1mm}\epsffile{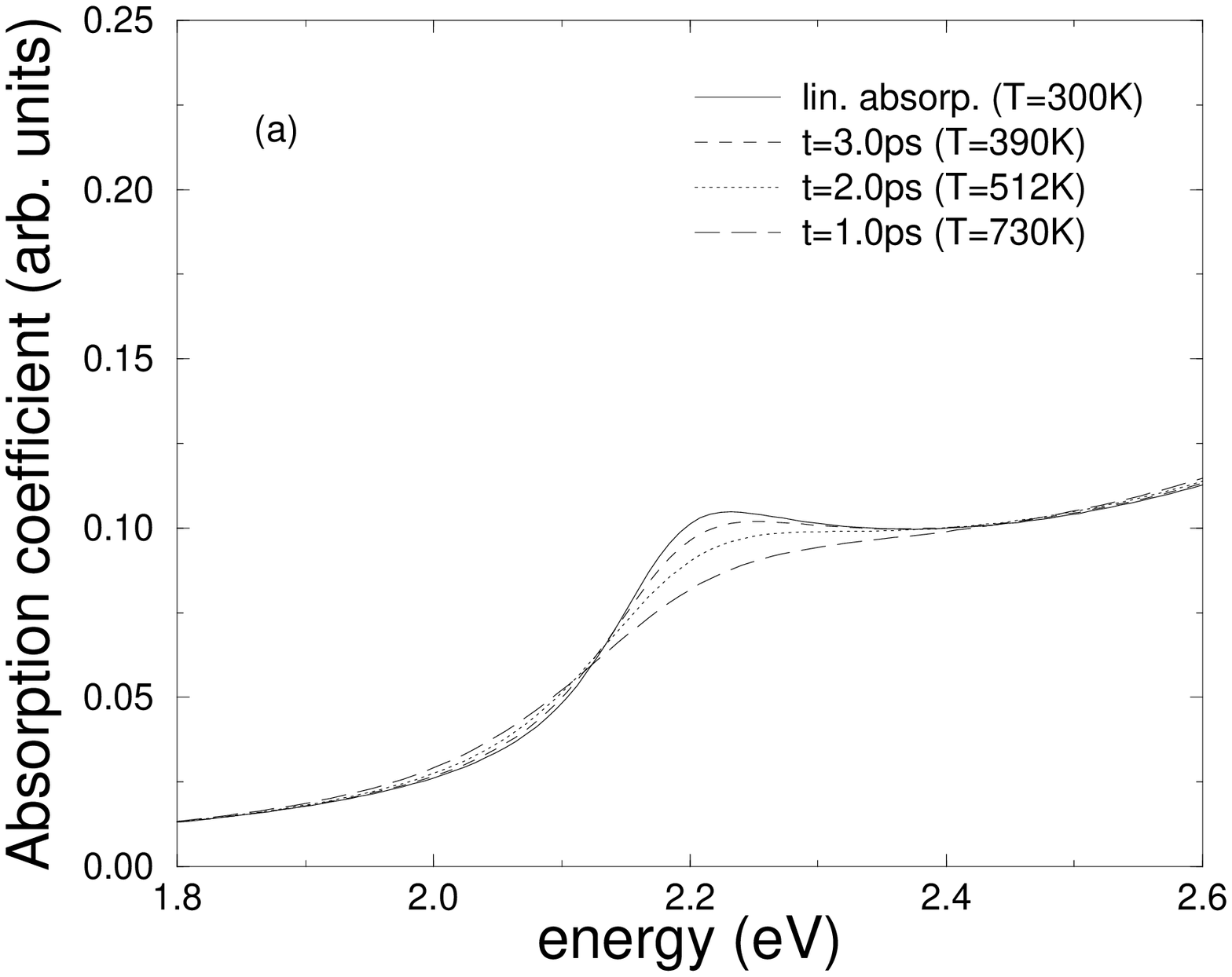}
\epsfxsize=2.7in
\hspace{-1mm}\epsffile{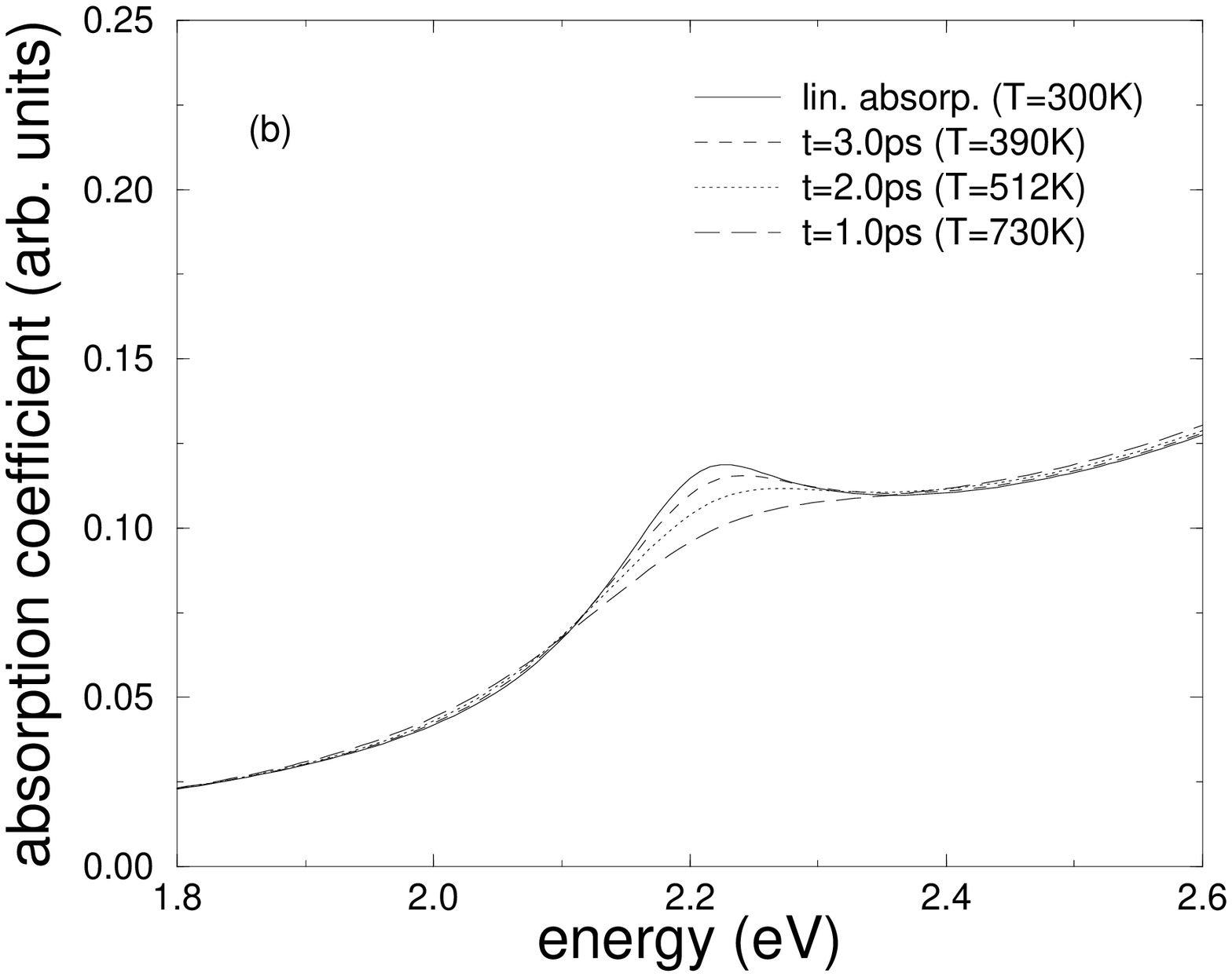}

\epsfxsize=2.7in
\hspace{-1mm}\epsffile{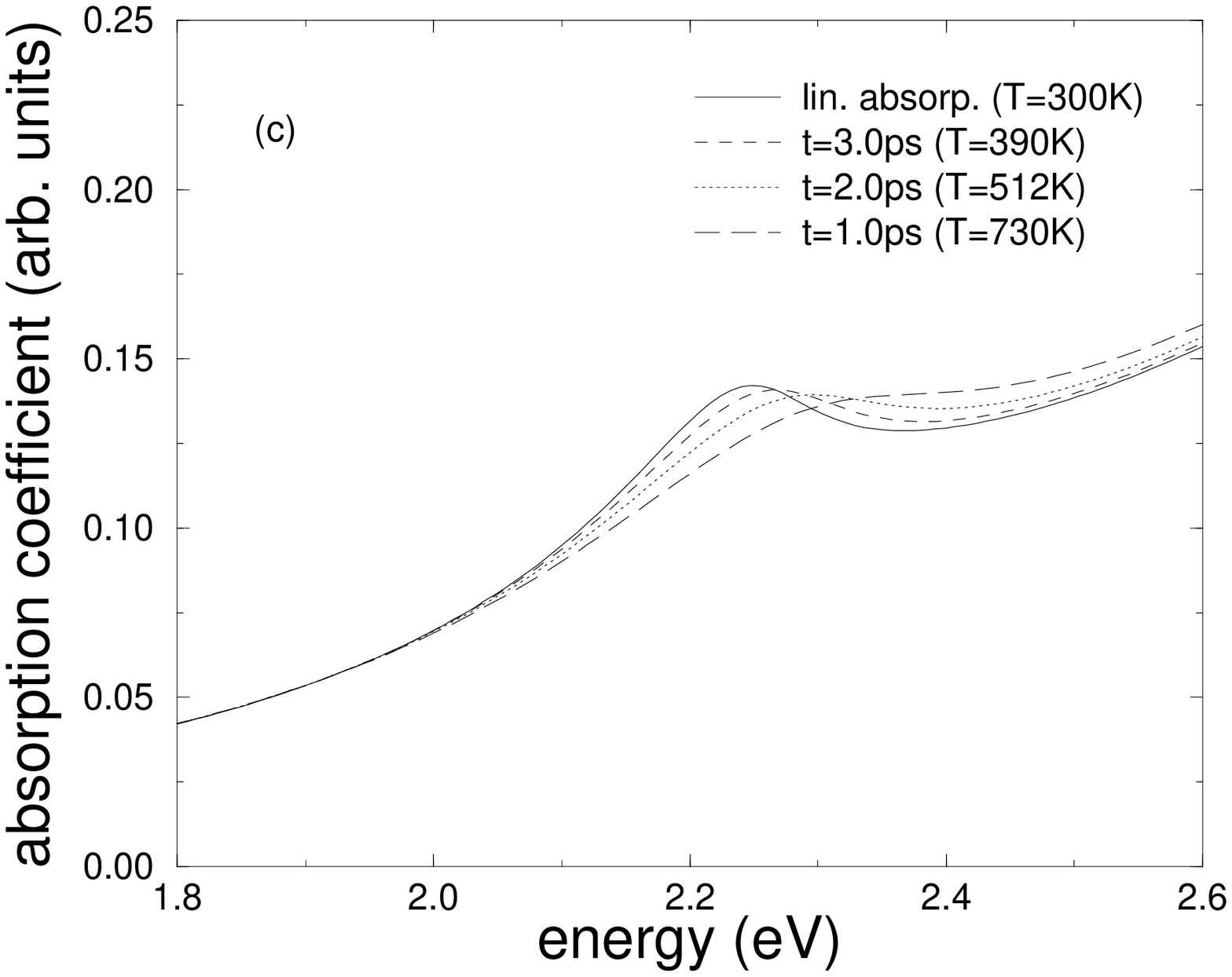}

\vspace{5mm}
\caption{
 Calculated absorption spectra at positive time
 delays for nanoparticles with  
 (a) $R=5$ nm, (b) $R=2.5$ nm, and (c) $R=1.2$ nm.
 }
\label{chem-fig}
 \end{figure}

\begin{figure}[htb]
 \vspace{-8mm}

 \epsfxsize=2.9in
\hspace{-4mm} \epsffile{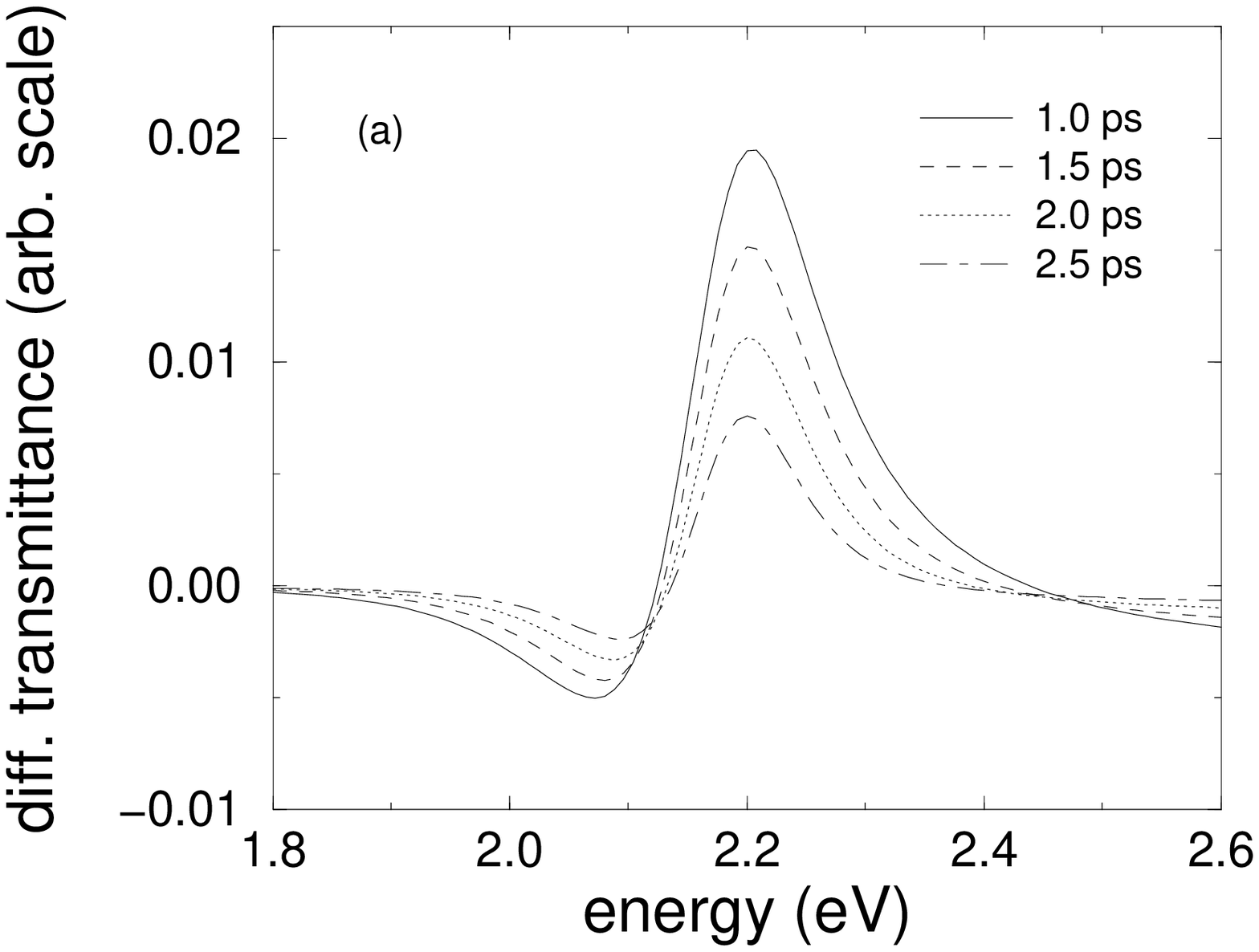}
 \epsfxsize=2.9in
\hspace{-4mm} \epsffile{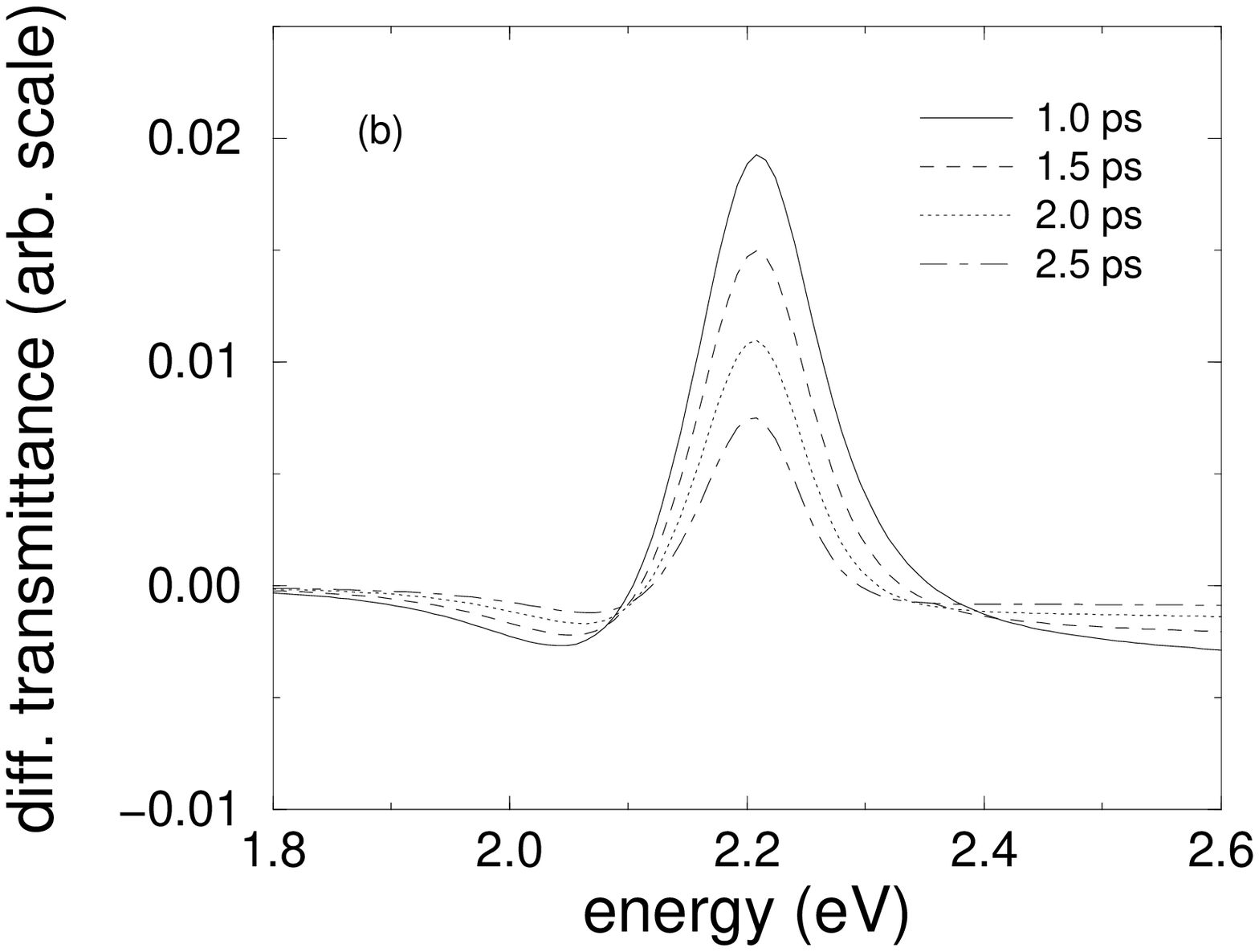}

 \epsfxsize=2.9in
\hspace{-4mm} \epsffile{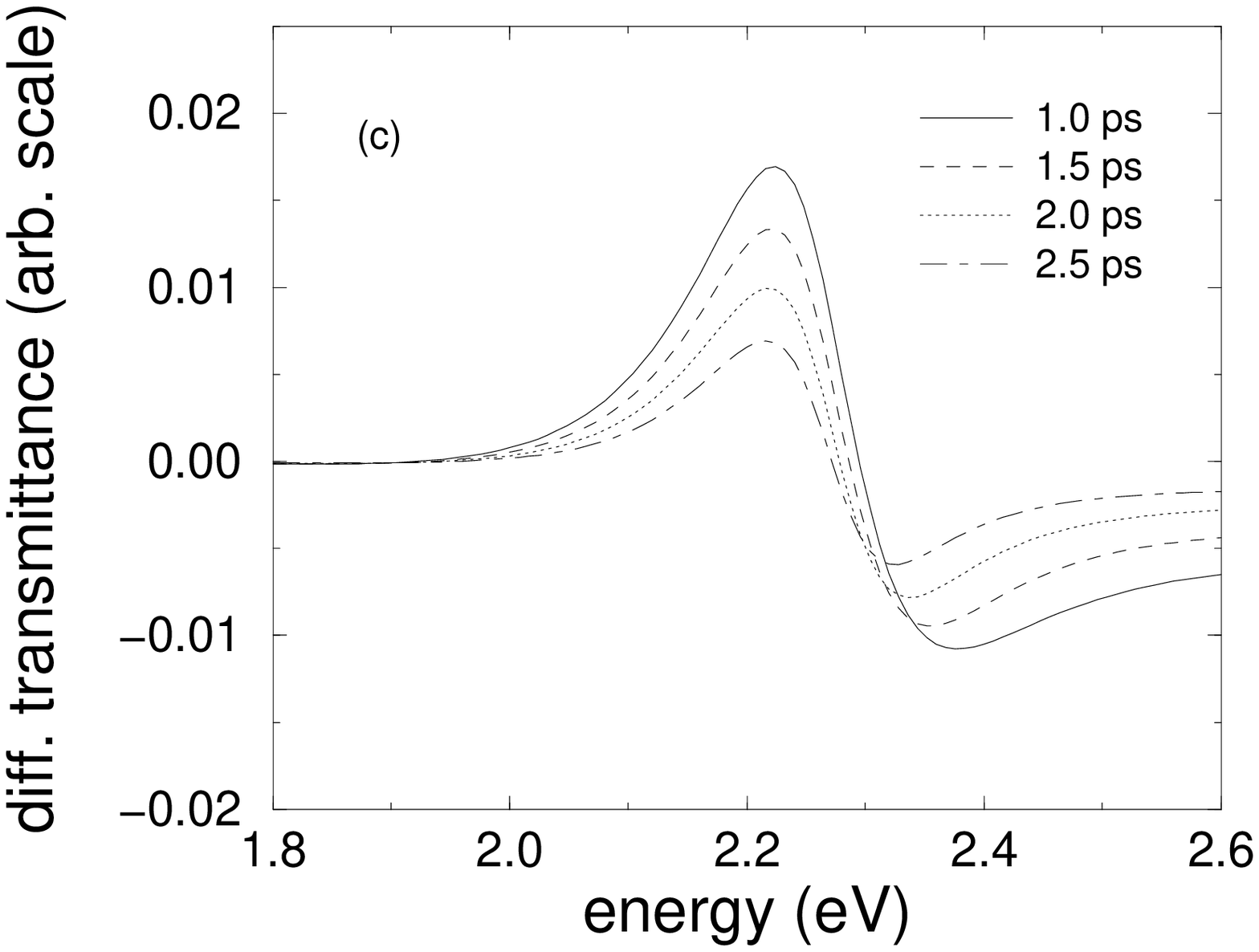}

 \vspace{5mm}
 \caption{
 Calculated differential transmission spectra at positive time
 delays for nanoparticles with  
 (a) $R=5$ nm, (b) $R=2.5$ nm, and (c) $R=1.2$ nm.
 }
\label{bspd-fig2}
 \end{figure}

In Fig.\ \ref{chem-fig}, the calculated nonlinear 
absorption spectra are 
shown for various nanoparticle sizes.  
Fig.\ \ref{chem-fig}(a) shows the spectra
at several time delays for $R=5.0$\ nm; for this size, the
SP--mediated {\em d}--hole scattering has no effect. 
With decreasing nanoparticle size, the {\em linear}
 absorption spectra are
not significantly altered, as can be seen in Figs.\ \ref{chem-fig}(b) and (c).
However, the change in the nonlinear 
absorption spectra becomes pronounced at
short time delays corresponding to higher temperatures 
[see  Figs.\ \ref{chem-fig}(b) and (c)]. 
This effect is more clearly seen in the
differential transmission spectra, shown in Fig.\ \ref{bspd-fig2}, 
which undergo a qualitative transformation with decreasing size.

Note that it is necessary to  include the intraband
{\em e--e} scattering in order to reproduce  
the differential transmission 
lineshape observed in the experiment \cite{big95}. 
For optically excited electron energy close to $E_F$, this  can be 
achieved  by adding the {\em e--e} scattering rate \cite{pines} 
$\gamma_e(E^c)\propto [1-f(E^c)][(E^c-E_F)^2+(\pi k_B T)^2]$ to
$\gamma_h^s$ in Eq.\ (\ref{interband}). The difference in 
$\gamma_e(E^c)$ for $E^c$ below and above $E_F$ leads to a
lineshape similar to that expected from the combination of
red--shift and broadening [see Fig.\ \ref{bspd-fig2}(a)].

In Figs.\ \ref{bspd-fig2}(b)\ and\ (c) the differential transmission
spectra  are shown with decreasing nanoparticle size.  
For $R=2.5$ nm, the apparent red--shift is reduced 
[see Fig.\ \ref{bspd-fig2}(b)]. This change can be explained  as follows.  
Since here $\omega_{s}\sim\Delta$, the SP is damped by the interband
excitations. This broadens the spectra for $\omega>\omega_{s}$, so
that the absorption peak is  
{\em asymmetric}. The $d$--hole scattering with the SP
enhances the damping; since the $\omega$--dependence of 
$\gamma_h^s$ follows that of $\alpha$, this effect is larger above
the resonance. On the other hand, the efficiency of the scattering
increases with temperature, as discussed above. Therefore, for short
time delays, the relative increase in the absorption is larger for
$\omega>\omega_{s}$.
With decreasing size, the strength  of this effect increases further,
leading to an apparent blue--shift [see Fig.\ \ref{bspd-fig2}(c)].  
Such a strong change in the absorption dynamics
originates from the $R^{-3}$ dependence of the $d$--hole scattering
rate; reducing the size by the factor of two results in an
enhancement of $\gamma_h^s$  by an order of magnitude.

\begin{figure}[htb]
\vspace{-10mm}

 \epsfxsize=2.9in
\hspace{-5mm} \epsffile{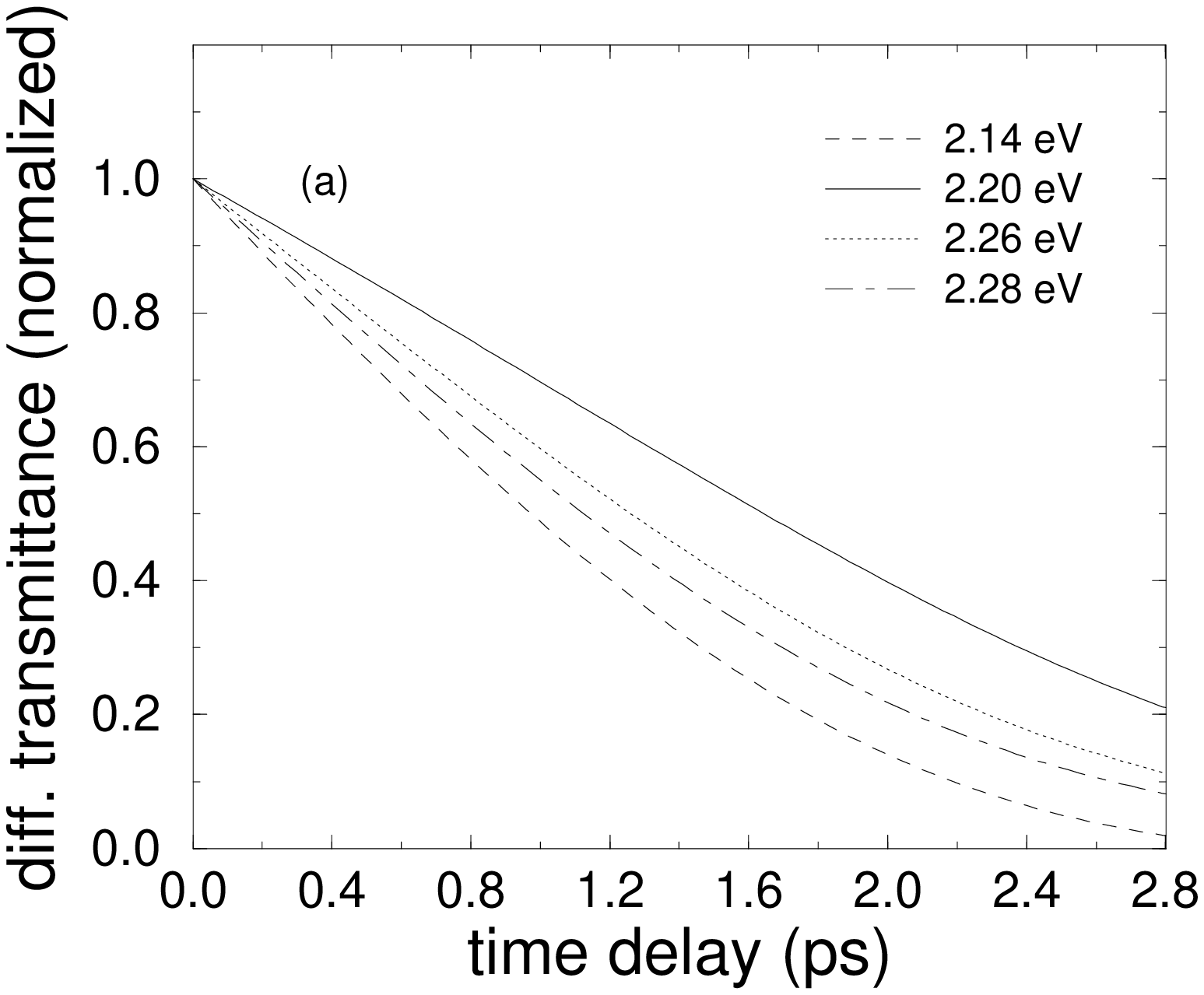}
 \epsfxsize=2.9in
\hspace{-5mm} \epsffile{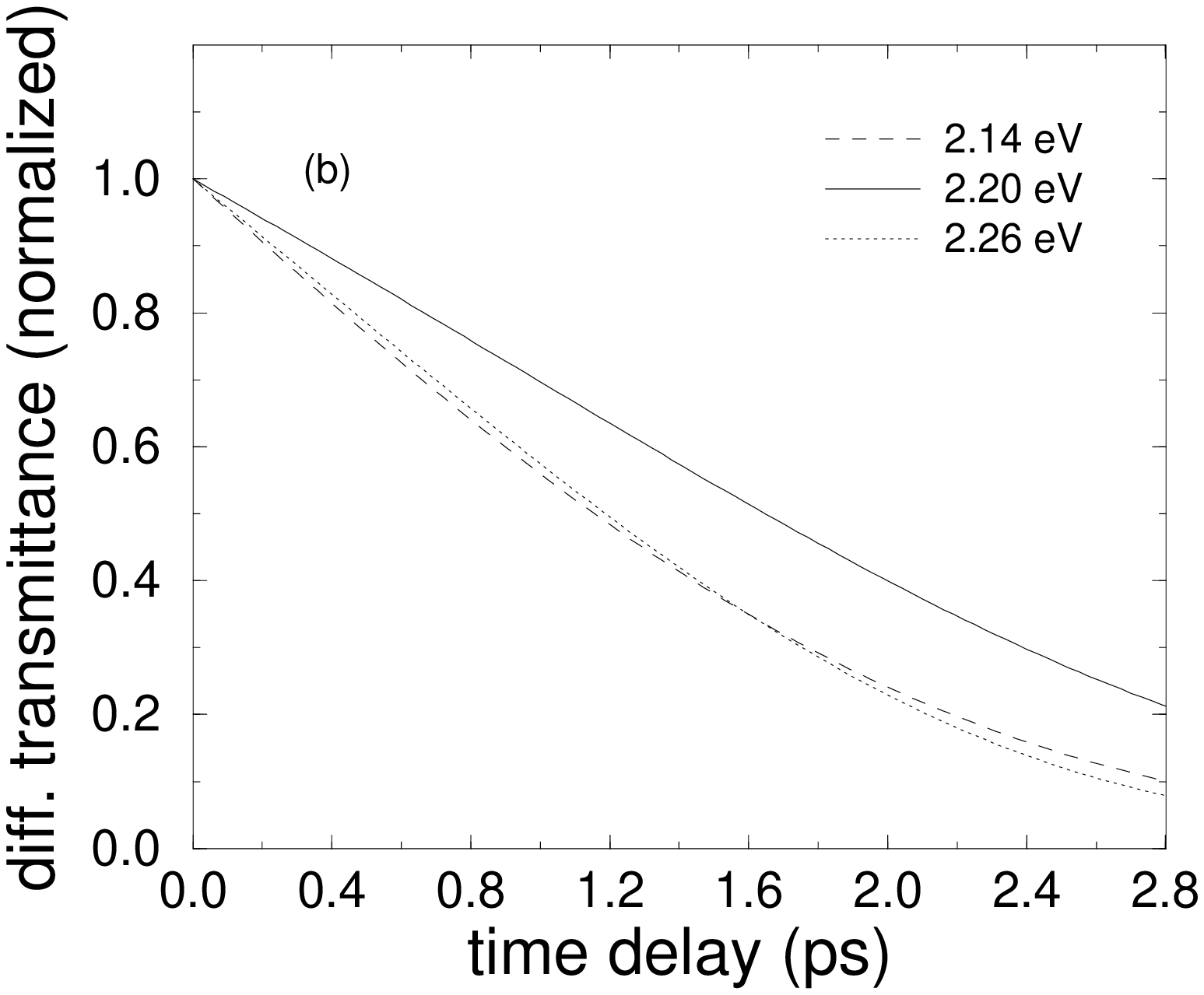}

 \epsfxsize=2.9in
\hspace{-5mm} \epsffile{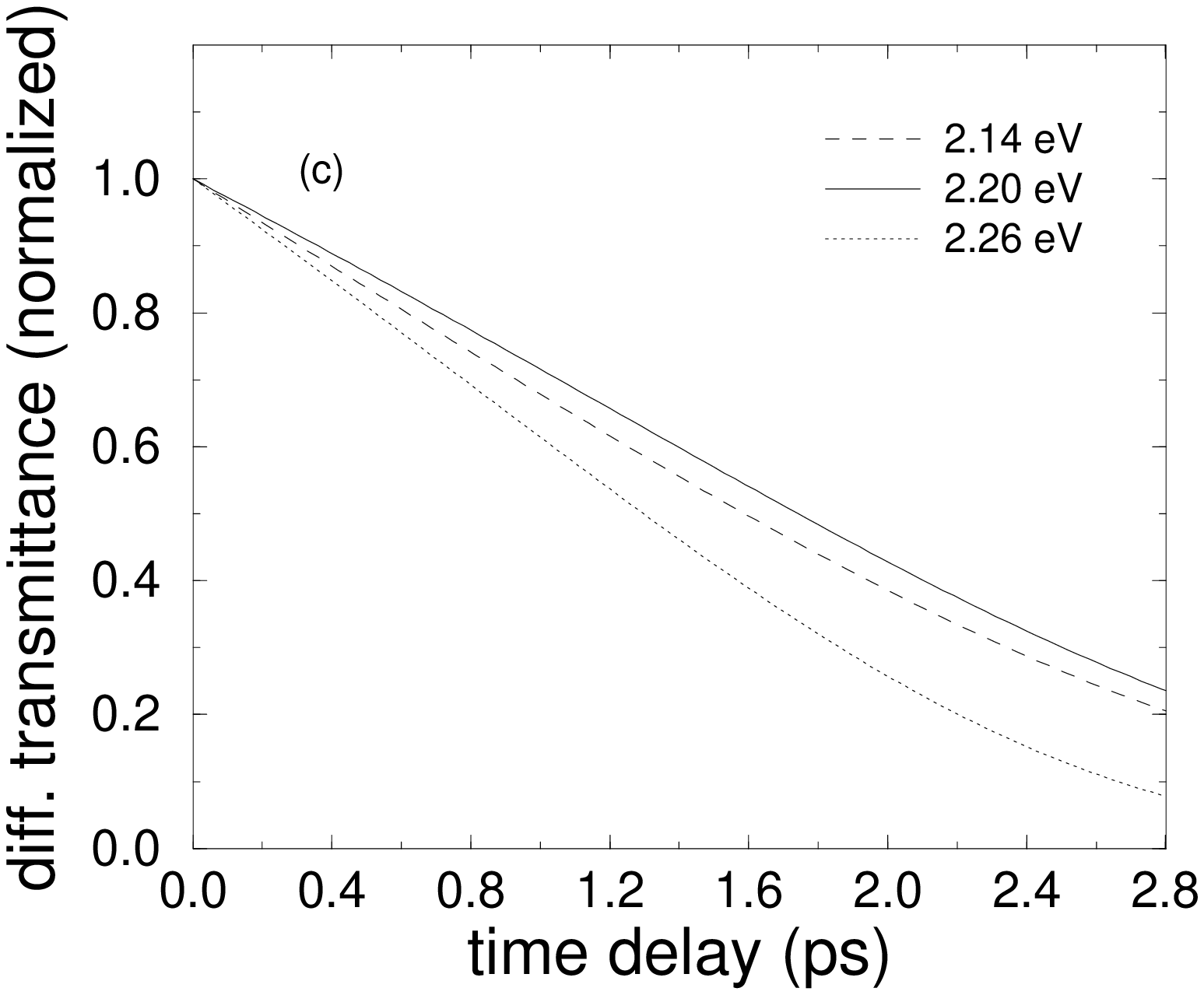}
 \epsfxsize=2.9in
\hspace{-5mm} \epsffile{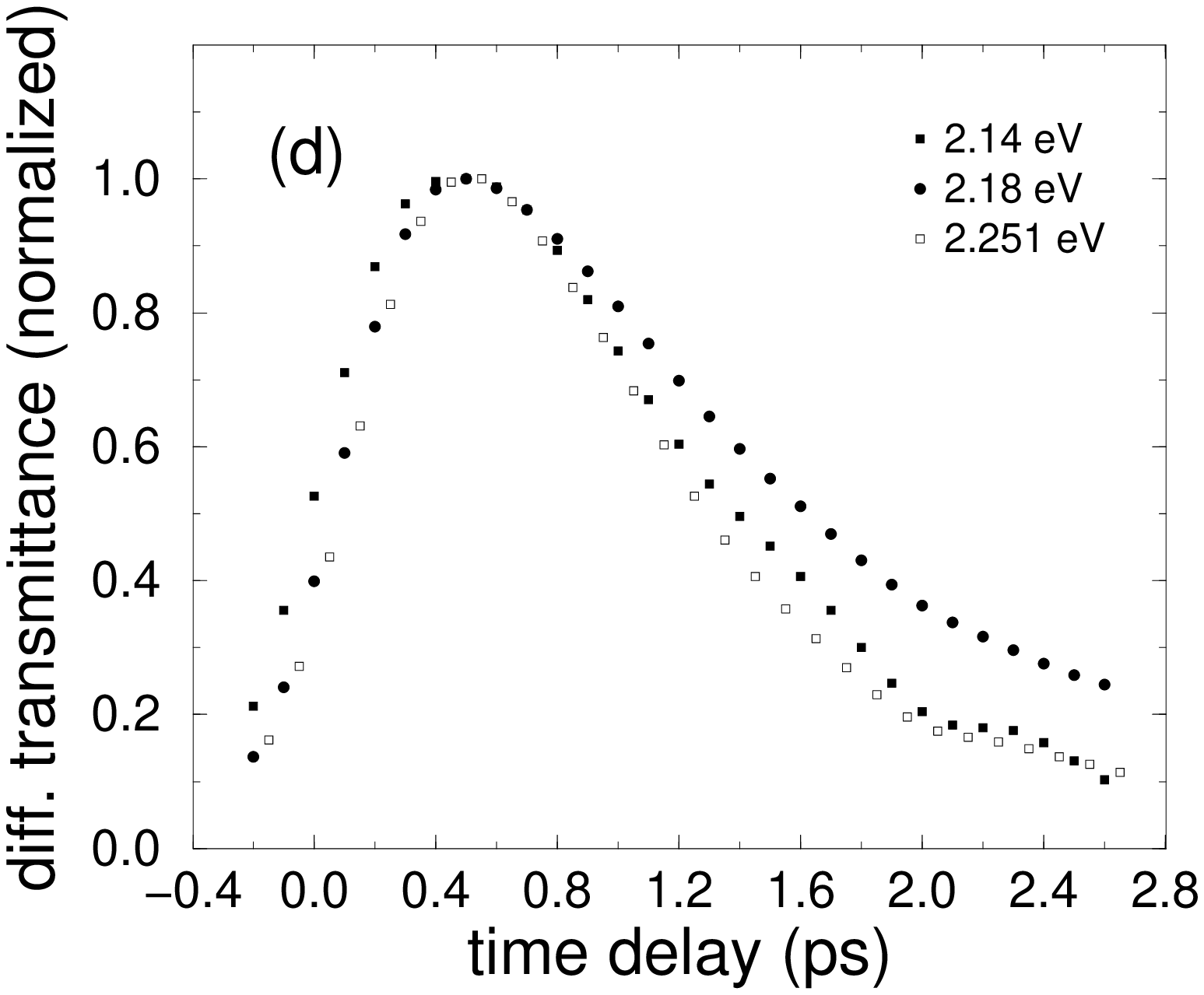}

 \vspace{5mm}
 \caption{
 Temporal evolution of the differential transmission 
 at frequencies close the SP resonance for nanoparticles
  with  
 (a) $R=5$ nm, (b) $R=2.5$ nm, and (c) $R=1.2$ nm. 
 (d) Measured time--resolved pump--probe signal.
 }
\label{bspd-fig3}
 \end{figure}

In   Fig.\ \ref{bspd-fig3}, the time evolution of the differential
transmission are shown for several frequencies close to $\omega_s$. 
It can be seen that the relaxation is slowest at the SP resonance;
this characterizes the robustness of the collective mode, which
determines the peak position, versus the single--particle
excitations, which determine the resonance width. 
For larger sizes, at which $\gamma_h^s$ is small, the change in  
the differential transmission decay rate with {\em frequency} 
is smoother above the resonance [see Fig.\ \ref{bspd-fig3}(a)]. 
This stems from the asymmetric lineshape of the
absorption peak, mentioned above: the  absorption is
larger for $\omega>\omega_{s}$, so that its {\em relative} change 
with temperature is weaker.
For smaller nanoparticle size, the decay rates become similar above
and below  $\omega_s$  [see Fig.\ \ref{bspd-fig3}(b)]. This change in the
frequency dependence is related to the stronger SP damping for
$\omega>\omega_{s}$ due to the $d$--hole scattering, as discussed
above. Since this additional damping is reduced with decreasing
temperature, the relaxation is faster above the
resonance.
This rather ``nonlinear'' relation between the
time--evolution of the pump--probe signal and that of the temperature
becomes even stronger for smaller sizes 
[see Fig.\ \ref{bspd-fig3}(c)]. In this
case, the frequency dependence of the differential transmission
decay  below and above $\omega_s$ is reversed. Note that a
frequency dependence consistent with our calculations presented 
in Fig.\ \ref{bspd-fig3}(b) was, in fact, observed in the experiment of
Ref. \cite{big95}, shown in  Fig.\ \ref{bspd-fig3}(d).

\section{Conclusions}

\label{sec:conclusion}

In this  article we 
presented an overview of ultrafast nonlinear 
optics experiments that probe the dynamics 
of modulation--doped quantum wells and metal 
nanoparticles.
We also discussed 
some examples where the ultrafast dynamics 
of such confined Fermi seas cannot be described in terms of the 
usual dephasing and relaxation time approximations. 
These examples  show the  important role 
of dynamical and non--equilibrium 
many--body correlations during femtosecond time
scales. 

More specifically, 
we discussed a recent theory 
for the ultrafast nonlinear optical response of the FES.
We focussed on coherent effects, 
which dominate the  pump--probe spectra
during negative time delays 
and off--resonant excitation conditions. 
We  demonstrated  that the dynamical FS response 
leads to qualitatively different coherent  dynamics
of the FES  as compared to the Hartree--Fock
approximation.
In the latter case, the time evolution of the
resonance bleaching is governed by the dephasing time. 
In contrast, in the former case,  polarization interference
and {\em e--h} correlation 
effects 
dominate. This results in an initial fast FES dynamics, with a 
response time determined by the characteristic 
inverse Coulomb energy $E_{M}$
as well as an apparent resonance enhancement during
negative time delays,
followed by a long--time decay determined by the dephasing time. 
Such dynamical features  should be observable 
in ultrafast PP experiments.
Using a simple model, we showed that the different
dynamics of the FES and Hartree-Fock treatment can
be attributed to the non--Lorentzian broadening of
the HFA bound state due to its interactions 
with the gapless FS excitations, a process which is,
of course, beyond the dephasing time approximation.
In addition, we showed that the pump excitation 
directly affects the strength of the {\em e--h} 
scattering processes, which changes the frequency
dependence of the resonance. The latter can be
thought of as an excitation--induced dephasing
effect that leads to a transient enhancement of
the FES. These  results indicate that ultrafast 
spectroscopy provides a powerful tool to study
the role of  correlations in the nonlinear response
of a Fermi liquid during
time scales shorter than the dephasing times. 
The dynamical features discussed above can also be used 
as an experimental signal  to probe the
crossover from FES to exciton bound state 
(exciton Mott transition) as a function of the  FS density.

The above {\em e--h} correlation effects 
and the fundamental differences in the dynamics
between an exciton state and a FES many--body resonance 
can be best observed experimentally in asymmetric modulation--doped 
quantum wells in the case where the lowest  occupied subband is in close
proximity  to the second unoccupied subband and is coupled strongly
to it  via intersubband {\em e--h} scattering. 
In such a system there is  a strong redistribution of the
oscillator strength between the FES and exciton peaks
which is caused by the different dynamics of the FES and exciton
components of the hybrid as well as by their coupling due to the 
{\em e--h} correlations. This originates from the dynamical Fermi
sea response and leads to a strong transient changes in the
PP spectra.

We also  discussed the role of
size--dependent correlations in the electron relaxation in small 
metal particles. We identified a new mechanism of quasiparticle
scattering, mediated by collective surface excitations, which
originates from the surface--induced dynamical
screening of the {\em e--e} interactions. The behavior of the
corresponding scattering rates with varying energy and temperature
differs substantially from that in the bulk metal. We showed that
the conduction electron scattering rate increases with energy, in
sharp contrast to the bulk behavior, which could be observed in
two--photon photoemission measurements. We also 
found  that in
noble metal particles, the resonant energy dependence of the $d$--hole
scattering rate affects strongly the differential absorption. 

An important aspect of the SP--mediated scattering is its strong
dependence on size. Our estimates show that it
becomes comparable to the usual Fermi--liquid scattering in
nanometer--sized particles. This size regime is, in fact,
intermediate between ``classical'' particles with sizes larger
than 10 nm, where the bulk--like behavior dominates, and
very small clusters with only dozens of atoms, where the metallic
properties are completely lost. Although the static properties
of nanometer--sized particles are also size--dependent,
the deviations from their bulk values do not change the qualitative
features of the electron {\em dynamics}. In contrast, the
size--dependent {\em many--body} effects discussed here {\em do}
affect the dynamics in a significant way during time scales comparable
to the relaxation times. 
The SP--mediated interband scattering
reveals itself in the size--dependence of the transient pump--probe
spectra. In particular, as the nanoparticle size decreases, 
the calculated time--resolved  differential absorption lineshape
shows a transition from an apparent redshift to a blueshift. This
transition, absent in the RPA, comes from the correlations between
collective surface and single--particle excitations.
At the same time, near the SP resonance, these correlation leads to
significant size--dependent changes in the frequency dependence of
the relaxation time of the pump--probe signal. 
These results indicate the need for a systematic
experimental studies of   the size--dependence of the transient
nonlinear optical  response, as we approach the transition 
from boundary--constrained nanoparticles to molecular 
clusters. 
We expect that in the coming years ultrafast nonlinear 
optical spectroscopy will provide new insight 
into  many--body effects in strongly correlated, 
and magnetic Fermi sea systems \cite{sc1,sc2,sc3,sc4,sc5,sc6}.

\section*{Acknowledgements}
\addcontentsline{toc}{section}{\numberline{}Acknowledgements}
 
This work was supported by the NSF grant ECS-9703453
and by  HITACHI Ltd. 
Many of the calculations discussed 
in this review were performed by N. Primozich.
It is a pleasure to acknowledge
numerous
fruitful discussions with D. S. Chemla and J.--Y. Bigot.

\appendix
%
%
%
\section{APPENDIX A }

In this appendix we briefly outline the formalism of Ref. 
\cite{per99}.
The pump--probe signal is determined by the polarization,
\begin{equation}\label{polar}
P(t) = \mu e^{-i\omega_{p}t}\langle \Psi(t) | U |
\Psi(t) \rangle, 
\end{equation}
where the state $|\Psi(t)\rangle$ satisfies the
time-dependent Schr\"{o}dinger equation,
\begin{equation}\label{schro-1}
\left[i \frac{\partial}{\partial t} - H_{tot}(t)\right]
|\Psi(t) \rangle = \left[i \frac{\partial}{\partial t} -
H -H_{p}(t)-H_{s}(t)\right] |\Psi(t) \rangle =0 \, , 
\end{equation}
with the Hamiltonians $H$ and $H_{p,s}(t)$,
given by Eqs. (\ref{H0})--(\ref{Hp}).
The Hilbert
space of the bare semiconductor, i.e. in the absence
of optical fields,
consists of disconnected subspaces
$\zeta \{ \nu_{eh} \}$ which are labeled by the
number of (interband) {\em e--h} pairs, $\nu_{eh}$.
The corresponding bare Hamiltonian, $H$, conserves
the number of {\em e--h} pairs in each band separately
and in the $\zeta \{ \nu_{eh} \}$-basis has a
block--diagonal form. The Hamiltonians $H_{p,s}(t)$
couple the different subspaces $\zeta \{ \nu_{eh} \}$
by causing interband transitions.

In the description of PP experiments, we
are interested only in the polarization component
propagating along the probe direction ${\bf k}_{s}$.
For a weak probe, the nonlinear signal
then arises from the linear response of the
system in the presence of the pump, 
described
by the time--dependent Hamiltonian $H + H_{p}(t)$,
to the probe--induced  perturbation $H_s(t)$.
Within $\chi^{(3)}$, the above is  true 
even for comparable pump and probe  amplitudes.
However, since, in contrast to $H$, the Hamiltonian $H + H_{p}(t)$ 
does not conserve the number of carriers in each band
separately,  the calculation of the linear response
function is not practical. Therefore, we seek 
to replace $H + H_{p}(t)$ by
an effective Hamiltonian $\tilde{H}(t)$
that {\em does} conserve the number of {\em e--h} pairs in
each of its Hilbert subspaces (i.e., is
``block--diagonal''). 
As derived in Ref.  \cite{per99}, 
this can be accomplished
in any given order
in the pump field ${\cal E}_p(t)$
and for any pulse duration 
by using  
a time--dependent Schrieffer--Wolff /Van Vleck
canonical transformation  \cite{s-w,cohen}.
Here it is sufficient to 
``block--diagonalize''
the Hamiltonian $H+ H_{p}(t)$ up to the second order
in  ${\cal E}_p(t)$. 
The  transformation that 
achieves this 
has the form $e^{-\hat{T}_{2}}
e^{-\hat{T}_{1}}[H+H_{p}(t)]e^{\hat{T}_{1}}e^{\hat{T}_{2}}$,
where the anti--Hermitian operators $\hat{T}_{1}(t)$
and $\hat{T}_{2}(t)$ create/annihilate one and two
{\em e--h} pairs, respectively.

We proceed with the first step and eliminate the
single-pair pump-induced transitions in the
time--dependent Schr\"{o}dinger equation of the
pump/bare--semiconductor system,
\begin{equation} 
\left[i \frac{ \partial}{ \partial t} -H-H_{p}(t)\right]
|\Psi(t) \rangle=0 \, .
\label{schro-2}
\end{equation}
This is achieved by substituting
$|\Psi(t) \rangle = e^{\hat{T}_1(t)}|\chi(t) \rangle$
and acting with the operator $e^{-\hat{T}_1(t)}$
on the lhs of Eq.\ (\ref{schro-2}),
\begin{equation} 
e^{-\hat{T}_1(t) }
\left[ i \frac{\partial}{\partial t}
- H \right] e^{\hat{T}_1(t) } | \chi(t) \rangle 
=  e^{-\hat{T}_1(t) } \left[ H_{p}(t) \right] 
 e^{\hat{T}_1(t) } | \chi(t) \rangle.
\label{schroed}
\end{equation}
The anti--Hermitian operator $\hat{T}_{1}(t)$ has
a decomposition 
\begin{equation} 
\hat{T}_{1}(t) = \hat{\cal P}(t) 
e^{-i {\bf k}_{p} \cdot {\bf r}} -
\hat{\cal P}^{\dag}(t) e^{i {\bf k}_{p} \cdot {\bf r}},
\label{unit} 
\end{equation}
where $\hat{\cal P}^{\dag}(t)$ and $\hat{\cal P}(t)$
create and annihilate single {\em e--h} pairs,
respectively.
The effective Hamiltonian
can be found from the condition that the terms
describing single--pair interband transitions
cancel each other
in  Eq.\ (\ref{schroed}). 
In Ref.  \cite{per99}, 
it was shown that the multiple commutators 
of $\hat{\cal P}(t)$ with its time derivatives can be eliminated 
from  Eq.\ (\ref{schroed}) 
to all orders. 
By expanding Eq.\ (\ref{schroed})
and neglecting third or higher 
order terms in $\hat{\cal P}$, 
we obtain the following equation: 
\begin{equation}
i\frac{\partial \hat{\cal P}^{\dag}(t)}{\partial t} = 
\left[ H,\hat{\cal P}^{\dag}(t)\right] 
+\mu {\cal E}_p(t) U^{\dag},
\label{sigma1}
\end{equation}
with initial condition
$\hat{\cal P}^{\dag}(-\infty)=0$. The formal
solution is
\begin{equation} 
\hat{\cal P}^{\dag}(t)
= i\mu  \int_{-\infty}^{t} dt'{\cal E}_{p}(t')
e^{ -i H (t-t')} U^{\dag} e^{iH(t-t')}.
\label{sigop}
\end{equation}
Note that, since the Hamiltonian $H$ conserves the
number of {\em e--h} pairs and the optical transition
operator $U^{\dag}$ creates a single {\em e--h} pair, 
$\hat{\cal P}^{\dag}(t)$ also creates a single
{\em e--h} pair. Furthermore, since both $H$ and
$U^{\dag}$ conserve momentum, so does 
$\hat{\cal P}^{\dag}(t)$. 
Eq.\ (\ref{schroed}) then takes the form
\begin{eqnarray}
\left[i   \frac{\partial}{\partial t} 
-\tilde{H}(t)\right]|\chi(t) \rangle  = 
+ \frac{\mu}{2}
\left[{\cal E}_p(t)
e^{2 i {\bf k}_{p} {\bf r}} 
\left[U^{\dag}, \hat{\cal P}^{\dag}(t) \right]
+ {\rm H.c.} \right]
| \chi(t) \rangle,
\label{first}
\end{eqnarray}
where 
\begin{equation} 
\tilde{H}(t) = H +\frac{\mu}{2}\,\left({\cal E}_p(t)
\left[\hat{\cal P}(t),U^{\dag} \right]
+ {\rm H.c.} \right)
\label{eff1} 
\end{equation} 
is the sought time--dependent effective
Hamiltonian that  conserves the number of {\em e--h}
pairs and $\hat{\cal P}^{\dag}(t)$ is given by 
Eq.\ (\ref{sigma1}). Note that, since $\hat{\cal P}^{\dag}(t)$
is linear in the 
pump field ${\cal E}_p$, the pump--induced
term in $\tilde{H}(t)$ [second term in Eq.\ (\ref{eff1})] is quadratic
($\propto {\cal E}_p{\cal E}_p^{\ast}$). The rhs
of Eq.\ (\ref{first}) describes the pump--induced
two--pair transitions. These can be eliminated as well
by performing a second canonical transformation, 
$| \chi(t)\rangle = e^{\hat{T}_{2}(t)} |\Phi(t) \rangle$.
Following the same procedure, 
we use the anti--Hermiticity of $\hat{T}_{2}(t)$ to decompose it as 
\begin{equation} 
\hat{T}_{2}(t) = \hat{\cal P}_{2}(t) 
e^{-2 i {\bf k}_{p} \cdot {\bf r}} -
{\cal P}^{\dag}_{2}(t) e^{ 2i {\bf k}_{p} \cdot {\bf r}},
\label{unit2} 
\end{equation}
where ${\cal P}^{\dag}_{2}(t)$ and $\hat{\cal P}_{2}(t)$
create and annihilate two {\em e--h} pairs, respectively.
Substituting the above expression for $|\chi(t)\rangle$
into Eq.\ (\ref{first}) and requiring that all two--pair
transitions  cancel out, we obtain the following
equation for ${\cal P}^{\dag}_{2}(t)$,
\begin{equation}
i\frac{\partial {\cal P}^{\dag}_{2}(t)}{\partial t} = 
\left[ \tilde{H}(t),{\cal P}^{\dag}_{2}(t) \right] +\frac{\mu}{2}\,
{\cal E}_p(t) [\hat{\cal P}^{\dag}(t),U^{\dag}].
\label{sig2}
\end{equation}
Note that $\hat{\cal P}_{2}^{\dag}(t)$ only affects
the PP polarization via  higher order
(${\cal E}_{p}^{4}$) corrections, which are neglected here.
However, it does determine the four--wave--mixing
(FWM) polarization (see below).

To obtain the condition of validity of this approach,
it is useful to write down a formal solution
(\ref{sigop}) of Eq.\ (\ref{sigma1}) in the basis
of the N--hole many--body eigenstates,
$|\alpha N \rangle$, with energies $E_{\alpha N}$, 
of the Hamiltonian $H$. Here the index $\alpha$
labels all the other quantum numbers, so that
$N$=0 corresponds to the semiconductor ground state
$| 0 \rangle$, $|\alpha 1\rangle$ denotes the one--pair states
(exciton eigenstates in the undoped case, with $\alpha$ labeling
both bound and scattering states), $|\alpha 2\rangle$ denotes the
two--pair (biexciton in the undoped case) eigenstates, etc. 
In this basis, the solution of Eq.\ (\ref{sigma1}) can be written as
\begin{equation}
\frac{  \langle \beta N+1
|\hat{\cal P}^{\dag}(t) |\alpha N \rangle}{
\langle \beta N+1| U^{\dag} |\alpha N \rangle }
=-i\mu\int_{-\infty}^{t}dt' {\cal E}_{p}(t')
e^{-i(t-t') \left( \Omega
+ \Delta E_{\alpha \beta}^{N} \right)}
e^{-\Gamma(t-t')},  \label{sigm1}
\end{equation}
where we separated out the detuning $\Omega$ and denoted
$\Delta E_{\alpha \beta}^{N}=E_{\beta N+1}-E_{\alpha N}$.
It can be seen that for resonant excitation (small
$\Omega$) the rhs of Eq.\ (\ref{sigm1}) is of the order
of $\mu {\cal E}_{p} t_{p}$. Thus, for short pulses,
this parameter justifies the expansion in terms of
the optical fields. For off-resonant excitation, this
expansion is valid for any pulse duration provided
that $\mu {\cal E}_p/\Omega < 1$. Similar
conditions can be obtained for the two--pair transition
described by $\hat{\cal P}_{2}$.

The nonlinear polarization Eq.\ (\ref{polar}) can now
be expressed in terms of the linear response
to the probe field:
\begin{equation}\label{polar-p}
P(t) = \mu e^{-i\omega_{p}t}\langle \Psi(t) | U |\Psi(t)\rangle
= \mu e^{-i\omega_{p}t}\langle \Phi(t)|U_{T}(t) |\Phi(t)\rangle,
\end{equation}
where, in the first order in ${\cal E}_{s}(t)$, the
state $| \Phi(t)\rangle$ is given by
\begin{equation} 
|\Phi(t) \rangle = |\Phi_0(t)\rangle 
- i\mu\int_{-\infty}^{t}dt'{\cal K}(t,t') 
\left[  {\cal E}_{s}(t')e^{i  {\bf k}_{s}
{\bf r}+i\omega_{p}\tau} U^{\dag}_{T}(t') 
+ {\rm H.c.} \right] |\Phi_0(t')\rangle.
\label{phipert}
\end{equation}
Here ${\cal K}(t,t')$ is the time-evolution
operator satisfying
\begin{equation}
i \frac{\partial}{\partial t}{\cal K}(t,t')=
\tilde{H}(t){\cal K}(t,t'), \label{evol} 
\end{equation}
and
$U^{\dag}_{T}(t) = e^{-\hat{T}_{2}(t) }
e^{-\hat{T}_{1}(t) } U^{\dag}
e^{\hat{T}_{1}(t)}e^{\hat{T}_{2}(t) }$
is the (transformed) optical transition
operator. In Eq.\ (\ref{phipert}),
$|\Phi_0(t)\rangle = {\cal K}(t,-\infty)
|0 \rangle$ is the time--evolved ground state 
$|0 \rangle$; since $\tilde{H}(t)$ conserves the
number of {\em e--h} pairs,  
$|\Phi_0(t)\rangle$ contains no interband {\em e--h} pairs (in
undoped semiconductors, it coincides with the ground state, 
$|\Phi_0(t)\rangle= |0\rangle$). 
>From  Eqs.\ (\ref{phipert}) and (\ref{polar-p}),
the polarization $P(t)$ takes the form 
\begin{eqnarray} \label{pol}
P(t) =
i\mu^2 e^{-i\omega_{p}t}
&&\mbox{\hspace{-6mm}}
\int_{-\infty}^{t}dt'
\biggl[\langle\Phi_0(t)| U_{T}(t){\cal K}(t,t')
[{\cal E}_{s}(t') e^{i{\bf k}_{s}\cdot{\bf r}
+i\omega_{p}\tau} U^{\dag}_{T}(t')
\nonumber\\ &&\mbox{\hspace{-6mm}}
+{\cal E}_{s}^{\ast}(t')e^{-i{\bf k}_{s}\cdot{\bf r}
-i\omega_{p}\tau} U_{T}(t')] |\Phi_0(t')\rangle
\nonumber\\ &&\mbox{\hspace{-6mm}}
-\langle\Phi_0(t')|
\bigl[{\cal E}_{s}(t')e^{i{\bf k}_{s}\cdot{\bf r}
+i\omega_{p}\tau} U^{\dag}_{T}(t') 
\nonumber\\ &&\mbox{\hspace{-6mm}}
+{\cal E}_{s}^{\ast}(t')e^{-i{\bf k}_{s}\cdot{\bf r}
-i\omega_{p}\tau} U_{T}(t')\bigr]
{\cal K}(t',t)U_{T}(t)|\Phi_0(t)\rangle
\biggr]. 
\end{eqnarray}
The above expression for the {\em total} polarization
contains contributions propagating in various
directions. To obtain the polarization propagating
in a specific direction, one has to expand the
effective--transition operator $U_{T}(t)$ in terms
of $\hat{T}_{1}$ and $\hat{T}_{2}$. Using 
Eqs.\ (\ref{unit})\ and\ (\ref{unit2}) and
keeping only terms contributing to PP
and FWM polarizations, we obtain \cite{per99}
\begin{eqnarray}\label{U}
U^{\dag}_{T}(t) = 
U^{\dag}_{{\rm 1}}(t)
+U^{\dag}_{{\rm 2}}(t) e^{i {\bf k}_{p}{\bf r}}
+ U_{FWM}(t) e^{- 2 i {\bf k}_{p}{\bf r}} +\cdots,
\end{eqnarray}
where (to lowest order in the pump field)
\begin{eqnarray}\label{Upp}
U^{\dag}_1(t)&=&U^{\dag}
+ \frac{1}{2} \left[ \hat{\cal P}(t), \left[U^{\dag},
\hat{\cal P}^{\dag}(t)\right] \right] +
\frac{1}{2} \left[ \hat{\cal P}^{\dag}(t),
\left[ U^{\dag}, \hat{\cal P}(t)\right] \right],
\nonumber\\
U^{\dag}_2(t)&=&
\left[\hat{\cal P}^{\dag}(t), U^{\dag}\right],
\nonumber\\
U_{FWM}^{\dag}(t)&=&
\frac{1}{2}\left[\left[ U,\hat{\cal P}^{\dag}(t)
\right] ,\hat{\cal P}^{\dag}(t)\right]
-\left[U,\hat{\cal P}_2^{\dag}(t)\right]. 
\end{eqnarray}
Here operators $U^{\dag}_1(t)\equiv\tilde{U}(t)$ and
$U_{FWM}^{\dag}(t)$ create one  
{\em e--h} pair, while $U^{\dag}_2(t)$ creates two {\em e--h} pairs
(note that $U_{FWM}$ in Eq.\ (\ref{Upp}) annihilates an {\em e--h} pair).

\subsection*{Pump--probe polarization}

In order to extract the PP polarization
from Eq.\ (\ref{pol}), one should retain only 
terms that are proportional to
$e^{i {\bf k}_{s}\cdot{\bf r}}$. Substituting Eqs.
(\ref{U}) into Eq.\ (\ref{pol}), we obtain
$P_{{\bf k}_{s}}(t)=e^{i{\bf k}_{s} \cdot {\bf r}- i\omega_{p}
(t - \tau)} 
[\tilde{P}^{(1)}(t)+\tilde{P}^{(2)}(t)]$,
where 
\begin{eqnarray}\label{PP1}
 \tilde{P}^{(1)}(t) =i\mu^2  
\int_{-\infty}^{t} dt' {\cal E}_{s}(t') 
\langle\Phi_0(t)|\tilde{U}(t) {\cal K}(t,t')
\tilde{U}^{\dag}(t') |\Phi_0(t')\rangle,
\end{eqnarray}
and
\begin{eqnarray}\label{PP2}
\tilde{P}^{(2)}(t) = i\mu^2   
\int_{-\infty}^{t} dt' {\cal E}_{s}(t')
\langle\Phi_0(t)| U_2(t) {\cal K}(t,t')
U^{\dag}_2(t') |\Phi_0(t')\rangle. 
\end{eqnarray}
Note that the above formulae apply to both undoped
and doped semiconductors. 

Eqs.\ (\ref{PP1})--(\ref{PP2}) express the nonlinear
PP polarization in terms of the linear
response to the probe field of a system described
by the time--dependent effective Hamiltonian
(\ref{eff1}).
Such a form for the nonlinear response allows one to
distinguish between two physically distinct contributions
to the optical nonlinearities.
Assuming that a  short probe pulse arrives at $t=\tau$, consider
the first term, Eq.\ (\ref{PP1}), which gives the
single--pair (exciton for undoped case)  
contribution to the PP polarization. At
negative time delays, $\tau<0$, the probe excites
an {\em e--h} pair, described by state
$\tilde{U}^{\dag}(\tau)|\Phi_0(\tau)\rangle
\simeq U^{\dag}|0\rangle$; since the probe arrives
before the pump, the effective transition operator
coincides with the ``bare''one [see Eqs.\ (\ref{Upp})\
and\ (\ref{sigop})]. The first contribution to the
optical nonlinearities comes from the effective
Hamiltonian, $\tilde{H}(t)$, governing the propagation
of that interacting {\em e--h} pair in the interval
$(\tau, t)$ via the time--evolution operator
${\cal K}(t,\tau)$. Note that since the pump pulse
arrives at $t=0$, for  $|\tau| \gg \Gamma^{-1}$, the
negative time--delay signal vanishes. At $t>0$,  the
{\em e--h} pair (exciton in the undoped case) 
``feels'' the effect of the pump via mainly
the transient bandgap shift, leading, e.g., to
ac--Stark effect, and the change in the band dispersions
(increase in effective mass/density of states), leading to enhanced 
{\em e--h} scattering (exciton binding energy
in the undoped case). Note that  $\tilde{H}(t)$
also contains a  contribution coming from 
the interactions
between probe-- and pump--excited {\em e--h} pairs,
which  are however perturbative in the doped case
for  short pulses or off--resonant excitation
and  lead to subdominant corrections.
Importantly, 
the response of the system to the optically--induced 
corrections in $\tilde{H}(t)$
takes into account {\em all} orders in the pump field, which is
necessary for the adequate description, e.g., of the
ac--Stark effect and the pump--induced changes in
the {\em e--h} correlations. Indeed, although the pump--induced
term in Eq.\ (\ref{eff1}) is quadratic in ${\cal E}_p$,
the time--evolution of the interacting {\em e--h}
pair is  described without expanding
${\cal K}(t,\tau)$ in the pump field. On the other hand,
the third--order
polarization ($\chi^{(3)}$) can be obtained by expanding 
${\cal K}(t,\tau)$ to the lowest order. The second
contribution to the optical nonlinearities comes from
the matrix element of the final state,
$\langle \Phi_0(t)|\tilde{U}(t)$
in Eq.\ (\ref{PP1}). The latter, given by Eq.\ (\ref{Upp}), 
contains the lowest order (quadratic)
pump--induced terms which describe the Pauli blocking, 
pair--pair, and pair--FS interaction effects (exciton--exciton
interactions in the undoped case \cite{per99}).
 Note that the matrix element of the {\em initial} state
contributes for positive time delays, i.e., if the
probe arrives after the pump pulse. In this case,
however, the pump--induced term in the effective
Hamiltonian (\ref{eff1}) vanishes (since it lasts
only for the duration of the pulse) so that for
positive $\tau> t_p$ the PP signal is
determined by the matrix elements rather than by
$\tilde{H}(t)$. If the probe arrives during the
interaction of the system with the pump pulse 
($\tau\sim t_p$), both  the effective Hamiltonian and
 the matrix elements contribute to the polarization. In
this case, there is also a biexcitonic contribution
[given by Eq.\ (\ref{PP2})] coming from a simultaneous
excitation of two {\em e--h} pairs by the pump
{\em and} the probe. However, such a biexciton state
does {\em not} contribute to negative ($\tau<0$)
time delays. 
As can be seen from the above discussion, 
our theory separates out a number of
contributions that play a different role 
for different 
time delays and excitation conditions.

\subsection*{FWM polarization}

By extracting from Eq.\ (\ref{pol}) all the terms propagating
in the the FWM direction, $2{\bf k}_p-{\bf k}_s$,
we obtain \cite{per99} 
(for delta--function probe ${\cal E}_s(t)={\cal E}_s\delta(t-\tau)$)

\begin{eqnarray}\label{FWM}
 P_{FWM}(t) = 
&&\mbox{\hspace{-6mm}}
i 
e^{i(2{\bf k}_p-{\bf k}_s) \cdot {\bf r}- i\omega_{p}
(t +\tau)}\theta(t-\tau)\mu^2{\cal E}_{s}^{\ast}
\nonumber\\&&\mbox{\hspace{-6mm}}\times
\biggl[\langle\Phi_0(t)|U 
{\cal K}(t,\tau)U_{FWM}^{\dag}(\tau)|\Phi_0(\tau)\rangle 
-(t\leftrightarrow \tau)\biggr],
\end{eqnarray}
where the FWM transition operator $U_{FWM}^{\dag}(t)$
is given by Eq.\ (\ref{Upp}). It is convenient to
express $U_{FWM}^{\dag}(t)$ in terms of the ``irreducible''
two--pair operator $W^{\dag}(t)=\frac{1}{2}
\hat{\cal P}^{\dag 2}-\hat{\cal P}_2^{\dag}$, satisfying  
\begin{equation}\label{W}
i\frac{\partial W^{\dag}(t)}{\partial t} = 
\left[\tilde{H}(t),W^{\dag}(t)\right] 
+ \mu {\cal E}_p(t)U^{\dag}\hat{\cal P}^{\dag}(t).
\end{equation}
In terms of $W^{\dag}(t)$, the state
$U_{FWM}^{\dag}(t)|\Phi_0(t)\rangle$ in Eq.\
(\ref{FWM}) can be presented as a sum of
two-- and one--pair contributions:
\begin{eqnarray}
\label{Ufwm}
U_{FWM}^{\dag}(t)|\Phi_0(t)\rangle
=UW^{\dag}(t)|\Phi_0(t)\rangle
-\hat{\cal P}^{\dag}(t)U\hat{\cal P}^{\dag}(t)
|\Phi_0(t)\rangle.
\end{eqnarray}
For undoped semiconductors, $|\Phi_0(t)\rangle$ represents the 
ground state $|0\rangle$ of $H$.
The operator $W^{\dag}(t)$, being quadratic in
the pump field ($\propto {\cal E}_p^2$), describes
the simultaneous excitation of two interacting {\em e--h}
pairs by the pump pulse and includes the biexciton and exciton--exciton
scattering effects. 
Introducing the amplidutes $\chi_{\bf k}$ and $\Phi_{\bf k}$ as
\begin{equation}
\label{Ufwm1}
U_{FWM}^{\dag}(t)|0\rangle
=\sum_{\bf k}\chi_{\bf k}(t)a_{\bf k}^{\dag}b_{\bf -k}^{\dag}|0\rangle,
\end{equation}
and
\begin{equation}
\tilde{\cal K}(t,t')U_{FWM}^{\dag}(t')|0\rangle
=\sum_{\bf k}\Phi_{\bf k}(t,t')a_{\bf k}^{\dag}b_{\bf -k}^{\dag}|0\rangle,
\end{equation}
where $\Phi_{\bf k}(t,t')$ satisfies
\begin{equation}
\label{eqPhi}
i\partial_t\Phi_{\bf k}(t,t')=\sum_{{\bf q}}\langle 0|b_{\bf -k}a_{\bf k} 
\tilde{H}(t)a_{{\bf q}}^{\dag}b_{{\bf -q}}^{\dag}|0\rangle
\Phi_{{\bf q}}(t,t'),
\end{equation}
with initial condition $\Phi_{\bf k}(t,t)=\chi_{\bf k}(t)$, 
one finally obtains for the FWM polarization
\begin{equation}
\label{pol-gen}
 P_{FWM}(t) = 
i e^{i(2{\bf k}_p-{\bf k}_s) \cdot {\bf r}- i\omega_{p}
(t +\tau)}\theta(t-\tau)\mu^2{\cal E}_{s}^{\ast}
\sum_{\bf k}[\Phi_{\bf k}(t,\tau)-\Phi_{\bf k}(\tau,t)].
\end{equation}
Note that the third order polarization 
[corresponding to $\chi^{(3)}$] is obtained
by replacing  $\tilde{H}$ with $H$.

\section{APPENDIX B}
In this appendix we clarify our convention for
the time delay $\tau$ and relate it to the most
commonly used conventions in PP
and FWM. In the generic
experimental configuration two laser pulses
$E_{1}(t)e^{i{\bf k}_1 \cdot {\bf r} - i\omega (t-t_1)}$,
and
$E_{2}(t)e^{i{\bf k}_2 \cdot {\bf r} - i\omega (t-t_2)}$,
respectively centered at time $t= t_1$ and $t= t_2$ are
incident on a sample. Let us define $\Delta t$ as,
\begin{equation}
\Delta t =  t_1 - t_2
\end{equation}
and consider a FWM experiment where the signal is
measured in the direction
$2{\bf k}_2 - {\bf k}_1$. Then
for a two-level-atom, the signal vanishes for
$\Delta t <0$, while  for $\Delta t>0$  its amplitude,
which decays with time as $e^{-t/T_2}$, is determined by 
the Pauli blocking. In a system with Coulomb interactions 
(such as  a semiconductor) a FWM signal is observed both
for $\Delta t<0$ and $\Delta t>0$. The $\Delta t <0$
signal is entirely due to the Coulomb interaction.

In PP experiments, one usually chooses 
one pulse (the ``pump'') to have an amplitude $E_{p}$
much larger than that of the other pulse (the
``probe''),  $E_{s}$. As discussed in the text, a weak 
probe measures the linear
response of the system (bare or dressed by the
pump). If we choose that $E_{p} = E_{2}$,  the pump
induces coherent and incoherent populations when
it arrives in the sample {\em before} the probe
i.e. for $t_2<t_1$. This is usually defined as
``positive'' time delay $\tau =  t_2 - t_1 >0$ in
the PP literature. Note that
$\tau =  - \Delta t$, i.e., the ``regular'' sequence
in PP experiments is the {\em reverse} of
that of FWM experiments. For $\tau <0$, the origin
of the PP signal is that the probe creates
a linear polarization in the sample which lasts for
time $\sim T_2=\Gamma^{-1}$ and, consequently, is
scattered by polarization 
excited by the pump field. The signal
observed for $\tau <0$ is therefore due to coherent
effects.

In FWM experiments,
there is no restriction on the magnitude of the two
incident fields $E_{2}$ and $E_{1}$, which are often
chosen to have amplitudes of the same order.
 Note however that, at the $\chi^{(3)}$ level, the FWM
and pump--probe polarizations are linear
in the $E_{1}(t)$ field
and thus the above linear response calculation 
applies even for comparable pump and probe amplitudes.

\section{ APPENDIX C}

In this Appendix  we present the explicit expressions 
for the renormalized transition matrix elements in the presence of
the pump excitation. 
The direct transition matrix element is given by
\begin{eqnarray}
M_{{\bf p}}(t) 
&&\mbox{\hspace{-6mm}}
= 1 - \left| {\cal P}_{eh}({\bf p},t)\right|^{2} 
+ \Biggl[ {\cal P}^{*}_{{\rm eh}}({\bf p},t) 
\sum_{k'<k_{F}} {\cal P}_{eh}^{{\rm e}}({\bf pk';k'};t)
+ {\rm H.c.} \Biggr] 
\nonumber\\&&\mbox{\hspace{-6mm}}
-\frac{1}{2} \sum_{p'>k_{F}} {\cal P}_{eh}^{*}({\bf p'},t) 
\Biggl[ {\cal P}_{eh}^{{\rm e}}({\bf p' p; p'};t)
-{\cal P}_{eh}^{{\rm h}}({\bf p' p;p'}; t)
\nonumber\\&&\mbox{\hspace{-6mm}} 
-{\cal P}_{eh}^{{\rm h}}({\bf p' p ; p};t)   
+  {\cal P}_{eh}^{{\rm e}}({\bf p' p; p};t) \Biggr] 
\nonumber\\&&\mbox{\hspace{-6mm}}
+ \frac{1}{2} \sum_{p'>k_{F}}
\Biggl[{\cal P}_{eh}({\bf p},t)  + 
{\cal P}_{eh}({\bf p'},t)\Biggr] 
\nonumber\\&&\mbox{\hspace{-6mm}}\times
\Biggl[ {\cal P}_{eh}^{{\rm e} }({\bf p p'; p};t)
+ {\cal P}_{eh}^{{\rm h}}({\bf p p'; p};t)\Biggr]^{*} \label{me1}.
\end{eqnarray}
The first term on the rhs of the above equation describes the phase
space filling contribution, 
while the rest of the terms are due to  the 
mean field  pair--pair and pair--FS interactions. 

The pump--induced indirect transition matrix element is given by
\begin{eqnarray}
M_{{\bf p p' k}}(t)=
\mbox{\hspace{-6mm}}&&
\Biggl[ {\cal P}_{eh}({\bf k},t) - {\cal P}_{eh}
({\bf p + p' - k},t) \Biggr]^{\ast}  
{\cal P}_{eh}^{{\rm e}} ({\bf p  p';k};t)
\nonumber\\&&\mbox{\hspace{-6mm}}
-\Biggl[{\cal P}_{eh}({\bf p},t)+{\cal P}_{eh}({\bf
p'},t)\Biggr]^{\ast}
\nonumber\\&&\mbox{\hspace{-6mm}}  \times
\Biggl[{\cal P}_{eh}^{{\rm e}} ({\bf p  p';p};t) 
+ {\cal P}_{eh}^{{\rm h} *}({\bf p  p';p+p'-k};t)\Biggr] 
\nonumber\\&&\mbox{\hspace{-6mm}}
+ {\cal P}_{eh}({\bf p+p'-k},t) 
\Biggl[ {\cal P}_{eh}^{{\rm e}}({\bf k,p+p'-k;p'};t) 
\nonumber\\&&\mbox{\hspace{-6mm}}
- {\cal P}_{eh}^{{\rm e} }({\bf k,p+p'-k;p};t)\Biggr]^{\ast}  
\nonumber\\&&\mbox{\hspace{-6mm}}
+ {\cal P}_{eh}^{\ast}({\bf k},t) 
\Biggl[ {\cal P}_{eh}^{{\rm h}}({\bf k,p+p'-k;p'};t)
+ {\cal P}_{eh}^{{\rm e}}({\bf p ,p';k};t)
\nonumber\\&&\mbox{\hspace{-6mm}}
- {\cal P}_{eh}^{{\rm e}}({\bf p , p'; p+p'-k};t)
-{\cal P}_{eh}^{{\rm h}}({\bf k,p+p'-k;p};t) \Biggr]
\nonumber\\&&\mbox{\hspace{-6mm}}
+ {\cal P}_{eh}({\bf p},t)  {\cal P}_{eh}^{{\rm e}*}({\bf k, p+p'-k;
p'};t)
\nonumber\\&&\mbox{\hspace{-6mm}}
- {\cal P}_{eh}({\bf p'},t)  {\cal P}_{eh}^{e *}({\bf k, p+p'-k;p};t) 
\label{me2}.
\end{eqnarray} 

The effective {\em e--h} potential is given by
\begin{eqnarray}\label{veh}
\upsilon_{eh}({\bf q; kk'};t)= 
\upsilon({\bf q})  -  \frac{\mu}{2}\,{\cal E}_p(t)
&&\mbox{\hspace{-6mm}} 
\Biggl[ {\cal P}_{eh}^{e}({\bf k+q,k';k'+q};t)
\nonumber\\&&\mbox{\hspace{-6mm}} 
+{\cal P}_{eh}^{h}({\bf k , k'+q;k'};t) 
\nonumber\\&&\mbox{\hspace{-6mm}}
+{\cal P}_{eh}^{e\ast}({\bf k,k'+q;k'};t) 
\nonumber\\&&\mbox{\hspace{-6mm}}
+ {\cal P}_{eh}^{h\ast}({\bf k+q, k';k'+q};t)
\Biggr],
\end{eqnarray} 
and the effective {\em e--e} potential is given by
\begin{eqnarray} \label{vee}
\upsilon_{ee}({\bf q;k k'};t) =  
\upsilon({\bf q})  + \frac{\mu}{4}\, {\cal E}_p(t)
&&\mbox{\hspace{-6mm}}
\Biggl[ {\cal P}_{eh}^{e}({\bf k+q,k'-q;k'};t) 
\nonumber\\&&\mbox{\hspace{-6mm}}
- {\cal P}_{eh}^{e}({\bf k+q,k'-q;k};t)
\nonumber\\&&\mbox{\hspace{-6mm}}
+ {\cal P}_{eh}^{e\ast}({\bf k,k';k'-q};t)
\nonumber\\&&\mbox{\hspace{-6mm}}
-{\cal P}_{eh}^{e\ast}({\bf k,k';k+q;}t) \Biggr].
\end{eqnarray} 

\addcontentsline{toc}{section}{\numberline{}References}

\newpage

\clearpage

\clearpage

\end{document}